\documentstyle[12pt]{article}

\advance\textheight by \topskip
\begin{document}
\def\lsim{\mathrel{\lower2.5pt\vbox{\lineskip=0pt\baselineskip=0pt
\hbox{$<$}\hbox{$\sim$}}}}
\def\gsim{\mathrel{\lower2.5pt\vbox{\lineskip=0pt\baselineskip=0pt
\hbox{$>$}\hbox{$\sim$}}}}

\def\gs{SU(2)_{\rm L} \times U(1)_{\rm Y}}
\def\ql{{\tilde{q}_{{i}_{\scriptscriptstyle L}}}}
\def\qr{{\tilde{q}_{{i}_{\scriptscriptstyle R}}}}
\def\tl{{\tilde{t}_{{\scriptscriptstyle L}}}}
\def\tr{{\tilde{t}_{{\scriptscriptstyle R}}}}
\def\bl{{\tilde{b}_{{\scriptscriptstyle L}}}}
\def\br{{\tilde{b}_{{\scriptscriptstyle R}}}}
\def\mt{{\tilde{m}^2}_{{\scriptscriptstyle t}}}
\def\mb{{\tilde{m}^2}_{{\scriptscriptstyle b}}}
\def\mz{{m}_{\scriptscriptstyle Z}^{2}}
\def\at{{A_{\scriptscriptstyle t}}}
\def\ab{{A_{\scriptscriptstyle b}}}
\def\qt{{Q_{\scriptscriptstyle t}}}
\def\qb{{Q_{\scriptscriptstyle b}}}
\def\tu{{\tilde{t}_{{\scriptscriptstyle 1}}}}
\def\td{{\tilde{t}_{{\scriptscriptstyle 2}}}}
\def\bu{{\tilde{b}_{{\scriptscriptstyle 1}}}}
\def\bd{{\tilde{b}_{{\scriptscriptstyle 2}}}}
\def\taul{{\tilde{\tau}_{{\scriptscriptstyle L}}}}
\def\taur{{\tilde{\tau}_{{\scriptscriptstyle R}}}}
\def\tauu{{\tilde{\tau}_{{\scriptscriptstyle 1}}}}
\def\taud{{\tilde{\tau}_{{\scriptscriptstyle 2}}}}
\def\mnu{{\tilde{m}^2_{{\scriptscriptstyle \nu}}}}
\def\mtau{{\tilde{m}^2}_{{\scriptscriptstyle \tau}_{\scriptscriptstyle 1}}}
\def\mtad{{\tilde{m}^2}_{{\scriptscriptstyle \tau}_{\scriptscriptstyle 2}}}
\def\w{{\tilde{W}^{\pm}}}
\def\h{{\tilde{H}^{\pm}}}
\def\chiu{{\tilde{\chi}^{+}}_{{\scriptscriptstyle 1}}}
\def\chid{{\tilde{\chi}^{+}}_{{\scriptscriptstyle 2}}}
\def\sw{s_{{\scriptscriptstyle W}}}
\def\cw{c_{{\scriptscriptstyle W}}}
\def\mfa{{\tilde{m}^2}_{{\scriptscriptstyle fa}}}
\def\mfb{{\tilde{m}^2}_{{\scriptscriptstyle fb}}}
\def\mi{{\tilde{M}^2}_{{\scriptscriptstyle i}}}
\def\mj{{\tilde{M}^2}_{{\scriptscriptstyle j}}}

\newcommand{\sfe}{{\tilde f}}
\newcommand{\sg}{{\tilde \chi}}
\newcommand{\sne}{{\tilde \chi^o}}
\newcommand{\scp}{{\tilde \chi^+}}
\newcommand{\dx}{dx}  
\newcommand{\dy}{dy}  
\newcommand{\slas}[1]{\rlap/ #1}
\newcommand{\diag}{{\rm diag}}
\newcommand{\Tr}{{\rm Tr}}
\vspace*{-1cm}
\begin{flushright}
{\large FTUAM 97/12}\\
\vspace{0.02cm}
{\large FT-UCM 5/97}\\
{{\it hep-ph}/9710313}\\
{Oct.-1997}\\
\end{flushright}

\begin{center}
\begin{large}
\begin{bf}
DECOUPLING OF SUPERSYMMETRIC PARTICLES \\
\end{bf}
\end{large}
\vspace{0.4cm}
ANTONIO DOBADO$^{\star}$\\
\vspace{0.08cm} 
{\em  Departamento de F{\'\i}sica Te{\'o}rica\\
  Universidad Complutense de Madrid\\
 28040-- Madrid,\ \ Spain} \\
\vspace{0.2cm} 
 MARIA J. HERRERO$^{\dagger}$\\
\vspace{0.08cm}
and\\
\vspace{0.08cm}
SIANNAH PE{\~N}ARANDA$^{\dagger}$\\
\vspace{0.08cm}
{\em  Departamento de F{\'\i}sica Te{\'o}rica\\
  Universidad Aut{\'o}noma de Madrid\\
  Cantoblanco,\ \ 28049-- Madrid,\ \ Spain}
\end{center}
\begin{center}
{\bf ABSTRACT}
\end{center}
\begin{quotation}
\noindent
The possibility of a heavy supersymmetric spectrum at the Minimal Supersymmetric Standard Model is
considered and the decoupling from the low energy electroweak scale is analyzed in detail. The 
formal proof of decoupling of supersymmetric particles from low energy physics is stated in terms 
of the effective action for the particles of the Standard Model that results by integrating out 
all the sparticles in the limit where their masses are larger than the electroweak scale. The 
computation of the effective action for the standard electroweak gauge bosons $W^{\pm}, Z$ and 
$\gamma$ is performed by integrating out all the squarks, sleptons, charginos and neutralinos to 
one-loop. The Higgs sector is not considered in this paper. The large sparticle masses limit is 
also analyzed in detail. Explicit analytical formulae for the two-point functions of the electroweak gauge 
bosons to be valid in that limit are presented. Finally, the decoupling of sparticles in the $S, T$ and $U$ 
parameters is studied analitically. A discussion on how the decoupling takes place in terms of both the 
physical sparticle masses and the non-physical mass parameters as the $\mu$-parameter and the 
soft-breaking parameters is included.
\end{quotation}
${\star}${\em e-mail: dobado@eucmax.sim.ucm.es}\\
${\dagger}${\em e-mail: herrero@delta.ft.uam.es}\,\,,\,\,
{\em siannah@delta.ft.uam.es}\\

\vspace{0.5cm}
\begin{large}
{\bf Introduction}
\end{large}

The Standard Model (SM) of particle physics provides a highly successful description of all  
particle physics phenomena occurring at present day accelerators. No experimental  
deviations from the SM have yet been found, even at LEP where the 
measurement of precision observables provides a very sensitive test to possible 
deviations through radiative corrections. Therefore, it is clear that the SM
is an extremely accurate effective low energy theory at energy scales
up to, at least, 100 GeV. In spite of this success there is, however, the 
strong  belief among particle physicists that the SM cannot be the 
ultimate theory  but it must be the low energy effective theory  of some other
more fundamental one which will probably incorporate gravitational interactions. 
Among the various extensions of the SM, the Supersymmetric (SUSY)
theories are the most claimed ones \cite{IS}, \cite{HC}. The motivations have been principally
theoretical. The idea of supersymmetry relating fermions  and bosons is
certainly beautiful itself, but one of the most used arguments in favour of supersymmetry
is that it guarantees the absence of quadratic divergences in scalar self-energies which, in 
turn, keeps the wanted stability of widely separated electroweak and grand unified scales. In 
practice, the supersymmetric theories are mostly built within same grand unification 
scenario and have as a common prediction the existence of, at least, one light elementary 
scalar Higgs boson with a mass close to the electroweak scale. At present, both
aspects the idea of grand unification and the existence of a light Higgs boson
are still waiting for experimental confirmation.

However, the  greatest  dilemma for supersymmetry is the unknown mechanism  for 
supersymmetry breaking. Since no supersymmetric partners of the known  particles   
have yet been observed, one must not only explain why supersymmetry is broken, 
but also why it is broken in such a way that all the supersymmetric partners 
are heavier than the known particles. The present experimental limits on
the sparticle masses \cite{DATA} indicate that supersymmetry must be indeed strongly 
broken. In practice, when building a supersymmetric theory for low energy physics   
one typically avoids the issue of the exact mechanism  of supersymmetry breaking
and parametrizes this phenomenon by introducing a set of supersymmetry breaking 
parameters which characterize the scale of the sparticle masses. These explicit 
SUSY breaking terms are required, in addition, to be of soft type, meaning that 
no new quadratic divergences are generated in the broken SUSY theory \cite{GG}. 
 In other to keep the wanted shielding of the electroweak scale against potentially 
large corrections from the grand unified scale, at least at a phenomenologically
acceptable  level, these soft SUSY breaking mass parameters shouldn't be larger than 
say $O(1TeV)$. This condition translates  into a restriction 
on the largest allowed mass splitting between  the known particles and their 
superpartners of this same order of magnitude and one concludes that the 
supersymmetric particles cannot be heavier than say $O(1TeV)$. The simplest 
low energy supersymmetric model that includes these soft SUSY breaking terms   
is the Minimal Supersymmetric Standard Model (MSSM) \cite{HAMSSM}, which is the one we have 
chosen to work with in this paper. 

One interesting aspect that arises in these softly broken SUSY theories, and 
in particular in the MSSM, is the question of decoupling of heavy sparticles 
from the low energy  SM and how does it really occurs if it occurs    
at all. This is the main subject of the present paper. It is a common belief that  when the 
spectrum of supersymmetric particles is considered much  heavier  than the low energy 
electroweak scale they decouple from the low energy physics, even at the quantum level, and 
the resulting low energy effective theory is the SM  itself. It is true that there are 
indications that this will happen, but a rigorous proof 
of decoupling is still lacking. On one hand there are numerical studies of 
observables that measure electroweak radiative corrections, like $\Delta r$ and $\Delta \rho$
 \cite{CHA}, or the $S, T$ and $U$ parameters \cite{HA2} as well as in the $Z$ boson, top quark and Higgs 
decays \cite{SOLA}, which indicate  that the one loop corrections from supersymmetric particles 
decrease up to negligible values in the limit of very heavy sparticle masses. 
 On the other hand there are some analytical studies, in certain asymptotic limits, of these and 
 related observables \cite{CHA}-\cite{GL1}, as well as some computations of the effective potential
 for the scalar sector~\cite{QUI}, where one of the mass parameters like the so-called $\mu$-parameter 
or the soft-SUSY breaking parameters are pushed to infinity. They 
also indicate some kind of decoupling but they are not completely general since 
they mostly assume, for simplicity, the restrictive hypothesis of universality of the soft    
SUSY breaking parameters. Besides, one must be aware that not all these asymptotic limits  
in terms  of the mass scale parameters of the MSSM lead to a complete heavy SUSY
spectra. In order to study the true decoupling, one must be careful in taking   
the proper asymptotic limits such that all the sparticles, and not just some of them, 
are got heavier than their superpartners. In other words, the choices of SUSY
parameters that lead to some heavy and some light SUSY particles are not the proper
ones to study  the decoupling hypothesis, since the presence of new light spectra  
at low energies is by itself contradictory  with  the idea of decoupling. Therefore
to explore the decoupling phenomenon it is more convenient to study the asymptotic behaviour of 
the theory in the large sparticle physical masses limit  themselves rather than in the large MSSM mass 
parameters limit. In this spirit there are some works which deal with  
the Higgs sector of the MSSM and study the consequences of taking one of the Higgs 
particles infinitely heavy. In particular, when the mass of the pseudoscalar boson $A^{o}$ is 
pushed to infinity it is proved in \cite{HA3} that, up to one loop level, all the effects of the heavy Higgs 
particles decouple from the electroweak precision observables. More specifically,  
in this limit four of the Higgs particles, $H^{\pm}, H^{o}$ and $A^{o}$ become infinitely heavy  
and the fifth one, $h^{o}$ , remains close to the electroweak scale. The unique trace  
of the SUSY Higgs sector that is left to low energies is  the unavoidable 
presence of the light Higgs particle $h^{o}$ itself which turns out to be, in the
above limit, indistinguishable from the Higgs particle of the SM. In this sense, it is concluded
in \cite{HA3} that there is indeed decoupling in the Higgs sector of the MSSM. Notice, however, that
this behaviour is generic of the MSSM, but in more extended Higgs sectors and in non-supersymmetric
models with two Higgs doublets it is known that the decoupling of the heavy Higgs particles may not 
occur~\cite{AM}.

A formal proof of decoupling must be driven, however, along the lines stated in the famous
Decoupling Theorem of Appelquist and Carazzone \cite{OJO}. The theorem states that under certain 
conditions in a given Quantum Field Theory with light and heavy particles, if the 
heavy particles are integrated out to all orders in perturbation theory, the remaining
effective action to be valid at energies much lower that the heavy particle masses does not show
any trace of these heavy particles. In that case they are said to decouple from the low energy 
theory. More specifically, all the quantum effects of the heavy particles that are left in the effective action  
can be either absorbed into a redefinition of the parameters of the original theory or wave
function renormalization  
referring to  the  light fields, or they are suppressed by inverse powers of the  
heavy masses and, therefore, vanish in the infinite mass limit. Among the conditions 
on the starting field theory to be sure that the theorem holds are the renormalizability
and the absence of spontaneous symmetry breaking and chiral fermions. Examples  where the 
Decoupling Theorem does not hold are well known. Particularly interesting 
are the cases of the Higgs particle and the top quark in the SM which are known
not to decouple from low energy physics \cite{VE}. For instance, by integrating out
the top quark at one loop one finds new contributions in the effective action  
which do not vanish in the infinite top mass limit \cite{CHI}. The same occurs when integrating out
the Higgs particle at one-loop.\cite{MJ} This can also be seen in the observable $\Delta \rho$ 
where the dominant contribution from the top quark loops in the large  top mass limit  goes quadratically 
with the top mass and the dominant contribution from the Higgs particle loops in the large Higgs mass
limit goes logarithmically with
the Higgs mass. In these examples one understands the departure from the Decoupling 
Theorem because the SM is a gauge theory with both spontaneous symmetry breaking and chiral 
fermions~\cite{ANT}. 

 The question whether the Decoupling Theorem applies or not in the case of heavy 
sparticles  in the MSSM is not obvious at all, in our opinion. The MSSM is 
a model which incorporates the SM particle content, the SM gauge 
interactions and it is built such that the wanted spontaneous breaking pattern, 
$\gs \rightarrow U(1)_{\rm em}$, takes place as in the Electroweak Theory. Even more, 
the soft-SUSY breaking terms are required not just to generate the needed SUSY particle 
masses but in addition to allow for this spontaneous electroweak symmetry     
breaking~\cite{LL}. In  few words, the MSSM is a 
gauge theory, as the SM, with spontaneous symmetry breaking and chiral fermions and  
therefore, the direct application of the Decoupling Theorem should, in the
principle, be questioned. 
   
In our opinion, a formal proof of decoupling must involve the explicit 
computation of the effective action by integrating out one by one all the 
sparticles in the MSSM to all orders in perturbation theory, and by considering 
the heavy sparticle masses limit. The proof will be conclusive if the remaining   
effective action, to be valid at energies much lower than the SUSY particle 
masses, turns out to be that of the SM with all the SUSY effects being absorbed into  
a redefinition of the SM parameters or else they are suppressed by inverse powers  
of the SUSY particle masses and vanish in the infinite masses limit.

In the present paper we give a first step in this direction and compute the two-point functions part 
of the effective action which results by integrating out all the SUSY particles of the MSSM,  
except the Higgs sector, at the one loop level. The integration of the Higgs sector presents 
interesting features by itself and we have preferred to consider it separately in a forthcoming 
work \cite{DMS}. Of course, the complete proof of decoupling must involve the computation of all
the n-point Green functions. The part of the effective action we have chosen to start with is 
the one for the electroweak gauge bosons, $W^{\pm}$, Z and $\gamma$ and, in 
particular, we have derived here the two point functions. The three and four point functions with 
external $W^{\pm}$, Z and $\gamma$ will be analized in a different work \cite{YO}. From the two-point
functions we will derive, in addition, the contributions  
from the SUSY particles to the $S, T$ and $U$ parameters in the large SUSY masses limit. 
From these analytical expressions we will be able to conclude how the decoupling
really occurs in these parameters. In order to keep our computation of the heavy 
SUSY particle quantum effects in a general form we have chosen to work with the masses 
themselves. They are the proper parameters of the large mass expansions instead of another more model 
dependent choices as the $\mu$-parameter or the soft-SUSY-breaking parameters. We have
considered the physically plausible situation where all the sparticle masses are large as compared to 
the electroweak scale but they are allowed, in principle, to be different from each other. 
We will explore the interesting question of to what extent the usual hypothesis of SUSY masses being  
generated by soft-SUSY- breaking terms and the universality of the mass parameters
do or do not play a relevant role  in getting decoupling. In fact, we will show in this
paper, that the basic requirement of $\gs$ gauge invariance on the SUSY breaking 
terms is sufficient to obtain decoupling in the MSSM.

Finally, we have dedicated special attention and have been very careful in evaluating 
analytically the large SUSY masses limit of the Green functions. For this purpose, 
we have applied the so-called m-Theorem \cite{GMR} which provides a rigorous technique to compute 
Feynman integrals with both large and small masses in the asymptotic limit 
$\tilde{m}_{\scriptscriptstyle i} \rightarrow \infty$ for the large masses. 

The paper is organized as follows: The first section contains a short review of 
the MSSM and the supersymmetric spectra and fixes the notation. The large sparticle masses limit is described
in section two. A discussion on how to get large mass
values for all the squarks, sleptons, neutralinos and charginos in the MSSM is also included. 
The third section is devoted to the computation of the effective action for the 
electroweak gauge bosons  $W^{\pm}$, Z and $\gamma$ in the MSSM that results by integrating 
out, in the path integral, squarks, sleptons, charginos and neutralinos to one-loop.
The exact results to one-loop for the two-point functions of the electroweak gauge bosons 
are presented in the fourth section. A discussion on the application of the m-Theorem and the 
analytical results for the self-energies of the electroweak gauge bosons in the large 
SUSY masses limit are also included and commented in
this section. The decoupling of heavy sparticles in the $S, T$ and $U$ parameters  
is analyzed in section five. Explicit formulae for these parameters in the large 
SUSY masses limit as well as a discussion on these results are also presented 
in this section. The conclusions are summarized in section six. Appendix A contains
the exact results to one-loop of the self-energies, $\Sigma^{{\scriptscriptstyle XY}}$, 
and the part of the two-point functions that is proportional to $k_{\mu}k_{\nu}$, $R^{{\scriptscriptstyle XY}}$.
The evaluation of the relevant Feynman integrals in the large masses 
limit and the application of the m-Theorem are performed in Appendix B. The asymptotic
results in the above limit for the transverse and longitudinal parts of the two-point 
functions, $\Sigma_{\scriptscriptstyle T}^{{\scriptscriptstyle XY}}$ and 
$\Sigma_{\scriptscriptstyle L}^{{\scriptscriptstyle XY}}$, are collected in Appendix C.


\section{Some remarks on the MSSM}
\label{sec:mssm}
\renewcommand\baselinestretch{1.3}

\hspace*{0.5cm} The Minimal Supersymmetric extension of the Standard Model consists of 
taking the Standard Model of electroweak and strong interactions as it is known
today, including the as yet undiscovered Higgs boson particle, and adding the
corresponding supersymmetric partners~\cite{HC},~\cite{HAMSSM}.

The supersymmetric partners of the quarks and leptons are the spin-zero squarks and
sleptons. For a given fermion $f$, there are two supersymmetric partners $\tilde
f_L$ and $\tilde f_R$ which are the scalar partners of the corresponding left and
right-handed fermions. Correspondingly, in the case of sneutrinos there is no
$\tilde \nu_R$. For simplicity we shall ignore intergenerational mixing.

The supersymmetric partners of the gauge bosons are the spin-one-half fermions
called gauginos. The partners of the gluon $g_\mu$, and the four weak bosons
$W_\mu^a$ (a=1,2,3), $B_\mu$ are, correspondingly, the gluino $\tilde g$, the
winos $\tilde W^a$ and the bino $\tilde B$.

In addition, the MSSM must possess two complex scalar Higgs doublets
$H_1$, $H_2$ in order to give masses to down and up type fermions in a manner
consistent with supersymmetry. The corresponding fermionic superpartners are the
Higgsinos $\tilde H_1$, $\tilde H_2$ which are also needed in pairs, in order to
avoid gauge anomalies. Supersymmetry, on the other hand, imposes strong
constraints on the form of the Higgs potential. In particular, the quartic
self couplings of the Higgs fields are fixed in terms of the $SU(2) \times U(1)$
gauge couplings and, therefore, the Higgs sector of the MSSM is always weakly
interacting.

The MSSM Lagrangian consists of two parts, a supersymmetry-conserving Lagrangian
and a supersymmetry-breaking Lagrangian:
\begin{equation}
  \label{lmssm}
  {\cal L}_{\rm MSSM}={\cal L}_{\rm SUSY}+{\cal L}_{\rm SUSY}^{\rm break}
\end{equation}

Usually ${\cal L}_{\rm MSSM}$ is required to contain no interaction terms of
higher dimensions than 4 so that the theory is renormalizable.

${\cal L}_{\rm SUSY}$ contains the SM Lagrangian for the gauge
bosons and fermions and the corresponding supersymmetrized Lagrangian for the
gauginos and sfermions. It also includes all possible renormalizable interactions among sparticles
themselves and among particles and sparticles which are compatible with
$SU(3)_{\rm c} \times \gs$ gauge invariance and with supersymmetry. In 
addition, there are the Lagrangian of the Higgs sector and the corresponding supersymmetrized 
Lagrangian for the Higgsino sector. This Higgs/Higgsino part includes the self-interaction 
terms and also the interactions with the other sectors. In particular it contains the Yukawa interactions with
the standard fermions and the supersymmetric Higgs potential:
\begin{equation}
  \label{higgspot}
  V_{\rm Higgs}^{\rm SUSY}= |\mu|^2 \left(|H_1|^2+|H_2|^2\right)
  +\frac{1}{8}(g^2+{g^\prime}^2)\left(|H_1|^2-|H_2|^2\right)^2
  +\frac{1}{2}g^2|H_1^* H_2|^2
\end{equation}

The supersymmetry-conserving Lagrangian, ${\cal L}_{\rm SUSY}$, is completely
general and model-independent since it is based just in $SU(3)_{\rm c} \times \gs$ gauge 
symmetry, renormalizability and supersymmetry. The parameters of ${\cal L}_{\rm SUSY}$ can be
summarized as follows:
\begin{itemize}
\item the gauge couplings $g_s$, $g$ and $g^\prime$ corresponding to the
  SM gauge group $SU(3)_{\rm c} \times \gs$ respectively,
\item the Higgs Yukawa couplings $\lambda_e$, $\lambda_u$ and $\lambda_d$ (which
  are $3 \times 3$ matrices in flavour space), and
\item  the supersymmetry-conserving mass parameter $\mu$.
\end{itemize}

Concerning the SUSY-breaking part, ${\cal L}_{\rm SUSY}^{\rm break}$, it is not
of a general form but its detailed form and the involved parameters depend on
the particular mechanism that is chosen in order to produce the SUSY breaking. However, 
we are forced to introduce some kind of explicit SUSY breaking at low energies since no 
supersymmetric partners of the known particles have been observed.

In order to study the decoupling of SUSY particles we would like to consider
this term in the most model independent way as possible. For this purpose, the minimal requirements
that  ${\cal L}_{\rm SUSY}^{\rm break}$ must fulfil are: 1) $SU(3)_{\rm c} \times \gs$
 gauge invariance, and  2) it must transmit masses to all the supersymmetric particles in a way 
that these result being considerably heavier than their standard partners. This last point is needed 
since we wish to study the consequences at low energies of having a gap between the SUSY 
spectra and the SM spectra.

Here we shall adopt the low-energy point of view in which one skirts the issue of the
exact mechanism of SUSY breaking, and parametrizes one's ignorance by introducing general 
mass-SUSY-breaking parameters that give the scale of the masses of the SUSY particles without 
inquiring into its origins. Usually, these explicit SUSY breaking terms are required, in addition to be of
soft type, meaning that no new quadratic divergences are generated by the
explicitly broken SUSY theory. These soft breaking terms were classified by
Girardello and Grisaru~\cite{GG} and are of four types: gaugino Majorana mass terms,
scalar mass terms, scalar-scalar-scalar trilinear interaction terms, and
scalar-scalar bilinear interaction terms.

In the case of the MSSM the set of all possible soft terms that respect
$SU(3)_{\rm c} \times SU(2)_L \times U(1)_Y$ gauge invariance are listed below:
\begin{eqnarray}
  \label{vsoft}
  V_{\rm soft}&=&m_1^2 |H_1|^2+m_2^2 |H_2|^2 -m_{12}^2 \left(\epsilon_{ij} H_1^i
  H_2^j+{\rm h.c.}\right)+M_{\tilde Q}^2 \left[\tilde t_L^* \tilde t_L+\tilde  b_L^* \tilde b_L\right] \nonumber\\
&+& M_{\tilde U}^2 \tilde t_R^* \tilde t_R +  M_{\tilde  D}^2 \tilde b_R^* \tilde b_R+ 
M_{\tilde L}^2 \left[\tilde \nu_L^* \tilde \nu_L+ \tilde \tau_L^* \tilde \tau_L\right]+
  M_{\tilde E}^2 \tilde \tau_R^* \tilde \tau_R \nonumber\\
&+&\frac{g}{\sqrt{2}m_W} \epsilon_{ij}\left[\frac{m_\tau A_\tau}{\cos\beta}H_1^i \tilde l_L^j \tilde \tau_R^*+
 \frac{m_b A_b}{\cos\beta}H_1^i\tilde q_L^j \tilde b_R^*-
  \frac{m_t A_t}{\sin\beta}H_2^i\tilde q_L^j \tilde t_R^*\right] \nonumber\\
&+&\frac{1}{2}\left[M_3 \bar{\tilde g} \tilde g + M_2 \bar{\tilde W}^a \tilde W^a +
  M_1 \bar{\tilde B} \tilde B \right]
\end{eqnarray}
where, 
\begin{equation}
\label{leftfields}
\tilde l_L \equiv \left(
  \begin{array}{c}
  \tilde \nu_L\\ \tilde \tau_L  
  \end{array}
  \right)\, ,\, \tilde q_L \equiv \left(
  \begin{array}{c}
  \tilde t_L\\ \tilde b_L  
  \end{array}
  \right)\,\, ,
\end{equation}
$\epsilon_{ij}$ ($i$,$j$=1,2) is the antisymmetric tensor to combine two $SU(2)_L$
doublets ($\epsilon_{12}=1$) and the third generation notation is used. The
generation labels that are omitted for brevity in this formula must be assumed.

The parameters of $V_{\rm soft}$ can be summarized as follows:
\begin{itemize}
\item three scalar Higgs mass parameters $m_1^2$, $m_2^2$, and $m_{12}^2$, with
  $m_{12}^2\equiv B \mu$ where $\mu$ is the mass parameter introduced in
  eq.~(\ref{higgspot}) and $B$ is a soft breaking parameter.
  These can be re-expressed in terms of the two Higgs vacuum expectation values,
  $v_1$ and $v_2$, and one physical Higgs mass. Here, $v_1$ ($v_2$) is the vacuum
  expectation value of the Higgs field which couples exclusively to down-type
  (up-type) quarks and leptons. Since $v_1^2+v_2^2 = (246 {\rm GeV})^2$ is fixed
  by the $W$ mass, we are left with just two independent parameters in the Higgs
  sector. These are usually chosen to be $\tan\beta=v_2/v_1$ and the mass of
  the pseudoscalar $m_{A^0}^2=m_{12}^2 (\tan\beta+\cot\beta)$,
\item scalar masses for the squarks and sleptons of each generation:
$M_{\tilde Q}$, $M_{\tilde U}$, $M_{\tilde D}$, $M_{\tilde L}$ and  $M_{\tilde E}$,
\item the trilinear soft breaking parameters for sleptons and squarks of
 each generation $A_e$, $A_u$ and $A_d$, and finally,
\item \label{un} gaugino Majorana masses $M_3$, $M_2$ and $M_1$ associated with the
  $SU(3)_{\rm c}$, $SU(2)_{\rm L}$ and $U(1)_{\rm Y}$ subgroups of the SM respectively.
\end{itemize}

At this point we would like to make a short comment that we think is relevant
for the discussion about decoupling which will be presented in this
paper. Notice that the mass terms for gauginos and scalar particles above are
the unique possible explicit mass terms that are compatible with $SU(3)_{\rm c} \times
\gs$ gauge invariance. For instance, we have chosen not to put
different mass terms for the two members of a scalar $SU(2)_{\rm L}$ doublet, and we
have not included Dirac fermion mass terms. Any of these two types of terms would
have broken the weak isospin symmetry and consequently the $SU(2)_L$ gauge symmetry
of the Lagrangian. Thus, the above mass terms are completely general and could have been 
introduced without mentioning the soft breaking requirement, but just the $SU(3)_{\rm c} \times \gs$ 
gauge invariance. In the following we will refer to these general terms simply as SUSY breaking 
mass terms. On the other hand, it is known that the SUSY breaking part of the Lagrangian is needed 
in order to incorporate the wanted spontaneous breaking of $\gs$ into $U(1)_{\rm em}$. It is not 
possible to break spontaneously the $\gs$ symmetry if the theory has an exact supersymmetry. The 
reason is that the supersymmetric Higgs potential in eq.~(\ref{higgspot}) is definite positive and
its minimum is at the symmetric configuration $H_1=H_2=0$. Once the SUSY-breaking Higgs 
mass parameters, $m_1^2$, $m_2^2$ and $m_{12}^2$, are included in the Higgs potential it reads:
\begin{eqnarray}
  \label{higgspotbr}
  V_{\rm Higgs}&=& m_{1H}^2 |H_1|^2+m_{2H}^2|H_2|^2-m_{12}^2 \left(\epsilon_{ij} H_1^i
  H_2^j+ {\rm h.c.}\right) \nonumber \\
&+&\frac{1}{8}(g^2+{g^\prime}^2)\left(|H_1|^2-|H_2|^2\right)^2+\frac{1}{2}g^2|H_1^* H_2|^2
\end{eqnarray}
where $m_{iH}^2\equiv |\mu|^2+m_{i}^2$ (i=1,2), and the real parameters $m_{i}^2$ can 
be either positive or negative, therefore allowing $V_{\rm Higgs}$ to develop non-trivial vacua. The parameter
region in which $\gs $ breaks down to $U(1)_{\rm em}$ at tree level is:\\
\\
\hspace*{3.4cm}$m_{1H}^2+ m_{2H}^2 \geq 2|m_{12}^2|$ (required for stability of $V_{\rm Higgs}$)\\ 
\hspace*{3cm}${|m_{12}^2|}^{2} > m_{1H}^2 m_{2H}^2$ (required for $\gs $ breaking) \\
\\
We will proceed on the assumption that these conditions are satisfied.

In the following we consider the mass eigenstates of the MSSM. Any set of particles of 
a given spin, baryon number, lepton number and the same $SU(3)_{\rm c} \times U(1)_{\rm em}$ quantum numbers can mix. 
Therefore, in principle, there can be mixing in all the sectors of the MSSM and one
must diagonalize mass matrices to obtain the mass eigenstates and the corresponding 
eigenvalues \cite{HC}, \cite{HAMSSM}, \cite{DCH}. We consider here all the sectors, except the Higgs sector that we prefer 
to analyze elsewhere \cite{DMS}.

\subsection{Squarks}
\hspace*{0.5cm}One must diagonalize $6 \times 6$ matrices corresponding to the weak eigenstate basis
$\{{\ql,\qr}\}$ where $i=1,2,3$ are the generation labels. We will ignore mixing between 
sfermions of different generations to avoid unacceptable large flavor changing
neutral currents and give here the $2 \times 2$ matrix corresponding to the one generation 
case. As above, we choose to use the notation of the third family. The mixing mass matrices 
for the stop and sbottom squarks in the $(\tl,\tr)$ and $(\bl,\br)$ bases respectively are as follows: \\
\begin{equation}
\label{eq:massmatrix}
{\hat M}_{\tilde{t}}^2 =\left(\begin{array}{cc}
 { L} & {\varepsilon}
\\ {\varepsilon} &{R} \,.
\end{array} \right)\,,\ \ \ \ \
{\hat M}_{\tilde{b}}^2 =\left(\begin{array}{cc}
 { L'} & {\varepsilon'}
\\ {\varepsilon'} &{ R'} \,.
\end{array} \right)\,,
\end{equation}
Notice that these are generic matrices with, in principle, unconstrained and unrelated
 matrix elements. The only constraint comes from $\gs$ gauge invariance which imposes the 
equality of the explicit breaking $\tl^{*} \tl $ and $\bl^{*} \bl $ mass terms. Thus, if the SUSY
breaking is just the explicit one, one  has  $L = L'$. This equality can be distorted by mass terms whose 
origin is not explicit SUSY breaking but spontaneous breaking of the $\gs$ symmetry. In the large 
SUSY masses limit that we are interested in, $L$ and $L'$ being of the order of $(1TeV)^{2}$, this 
distortion is relatively small since it goes as $(m_{t}^{2}-m_{b}^{2})$ and/or it is proportional to $\mz$.

We give here their specific expressions for completeness but we will keep the generic form,
eq.~(\ref{eq:massmatrix}), through most of this paper:
\begin{eqnarray}
\label{eq:inputssq}
{ L} &=& M_{{\scriptscriptstyle {\tilde Q}}}^{2}+m_{t}^{2}+ \mz (\frac{1}{2}-\qt 
 s_{{\scriptscriptstyle W}}^{2}) \cos{2\beta}\,,\nonumber\\
{R} &=& M_{{\scriptscriptstyle {\tilde U}}}^{2} +  
m_{t}^{2}+ \mz \qt s_{{\scriptscriptstyle W}}^{2} \cos{2\beta}\,,\nonumber\\
{\varepsilon} &=& m_{\scriptscriptstyle t}(\at-\mu \cot{\beta})\,,\nonumber\\
{L'} &=& M_{{\scriptscriptstyle {\tilde Q}}}^{2}+ m_{b}^{2} - \mz (\frac{1}{2}+\qb 
s_{{\scriptscriptstyle W}}^{2}) \cos{2\beta}\,,\nonumber\\
{R'} &=& M_{{\scriptscriptstyle {\tilde D}}}^{2} +  
m_{b}^{2} + \mz \qb s_{{\scriptscriptstyle W}}^{2} \cos{2\beta}\,,\nonumber\\
{\varepsilon'} &=&  m_{\scriptscriptstyle b}(\ab-\mu \tan{\beta})\,,
\end{eqnarray}
where $\qt=2/3$, $\qb=-1/3$, $s_{{\scriptscriptstyle W}}^{2} \equiv 
{\sin}^{2}\theta_{\scriptscriptstyle W}$ and $\tan\beta\equiv v_2/v_1$.

The mass eigenstates are denoted by $\tu,\td,\bu,\bd$ and are related to the weak eigenstates
$\tl ,\tr ,\bl ,\br$ by orthogonal matrices: \\
\begin{equation}
\left(\begin{array}{c}
\tl \\ \tr
\end{array} \right)= R_{t}
\left(\begin{array}{c}
\tu \\ \td
\end{array} \right)\,,\ \ \ \ \ 
\left(\begin{array}{c}
\bl \\ \br 
\end{array} \right)= R_{b}
\left(\begin{array}{c}
\bu \\ \bd
\end{array} \right)\,.
\end{equation}

\begin{equation}  
\label{eq:rotatmass}
R_{t} =\pmatrix{ c_{t} & -s_{t} \cr 
         s_{t} & c_{t}}\,,\ \ \ \ \ 
R_{b} =\pmatrix{ c_{b} & -s_{b} \cr 
         s_{b} & c_{b}}\,.
\end{equation}
\\
where $c_{q}\equiv \cos{\phi_{q}}$ ,$s_{q}\equiv \sin{\phi_{q}}$, $q=t,b$. 

The corresponding squared-mass eigenvalues and the mixing angles are given 
in terms of the generic matrix elements as follows: 
\begin{eqnarray}
\label{eq:autov}
\tilde{m}^2_{{\scriptscriptstyle t}_{1,2}} &=& \frac{1}{2}({L}+
{R})\pm\frac{1}{2}{\left[({L}-{R})^{2}+ 4{\varepsilon}^{2}\right]}^{\frac{1}{2}}\,,\nonumber\\
\tilde{m}^2_{{\scriptscriptstyle b}_{1,2}} &=& \frac{1}{2}({L'}+
{R'})\pm\frac{1}{2}{\left[({L'}-{R'})^{2}+4{\varepsilon'}^{2}\right]}^{\frac{1}{2}}\,,
\end{eqnarray}
\begin{equation}  
\tan{2 \phi_{t}}= \frac{2{\varepsilon}}{{L}-{R}}\,;\ \ \ \ \ 
\tan{2 \phi_{b}}= \frac{2{\varepsilon'}}{{L'}-{R'}}\,.
\end{equation}

\subsection{Sleptons}

Similar formulae hold for the sleptons sector but replacing $\tl\rightarrow \tilde{\nu}$,
$\tr\rightarrow 0$, $\bl\rightarrow \taul$, $\br\rightarrow \taur$ and the corresponding 
changes in the rotation matrices $R_{\nu}$ and $R_{\tau}$ which relate the weak eigenstates 
$(\tilde{\nu})$, $(\taul,\taur)$ with the mass eigenstates $(\tilde{\nu})$, 
$(\tauu,\taud)$, respectively: \\
\begin{equation}
\label{eq:Rl}  
R_{\nu} =\pmatrix{ 1 & 0 \cr 
         0 & 1}\,,\ \ \ \ \ 
R_{\tau} =\pmatrix{ c_{\tau} & -s_{\tau} \cr 
         s_{\tau} & c_{\tau}}\,.
\end{equation}
The mass squared eigenvalues are $\mnu$, $\mtau$, $\mtad$ respectively. The 
specific formulae (\ref{eq:inputssq}) for the mass matrix elements must also be correspondingly replaced: \\
\begin{equation}
\label{eq:inputssl}
\begin{array}{l}
{L}\rightarrow M_{{\scriptscriptstyle {\tilde L}}}^{2}+\frac{1}{2} \mz \cos{2\beta}\,,\ \ \ \ \
{R}\rightarrow 0\,,\ \ \ \ \ 
{\varepsilon}\rightarrow 0\,,\\
\\
{L'}\rightarrow M_{{\scriptscriptstyle {\tilde L}}}^{2}
+m^{2}_{{\scriptscriptstyle \tau}} 
-\mz (\frac{1}{2} - s_{{\scriptscriptstyle W}}^{2}) \cos{2\beta}\,,\\
\\
{R'}\rightarrow M_{{\scriptscriptstyle {\tilde E}}}^{2}
+m^{2}_{{\scriptscriptstyle \tau}} 
-\mz s_{{\scriptscriptstyle W}}^{2} \cos{2\beta}\,,\\
\\
{\varepsilon'}\rightarrow m_{\scriptscriptstyle \tau}(A_{\scriptscriptstyle \tau}-\mu \tan{\beta})\,.
\end{array}
\end{equation}
From the above formulae (\ref{eq:inputssq}) and (\ref{eq:inputssl}), we see that the $\tilde{f}_{\scriptscriptstyle L} -
\tilde{f}_{\scriptscriptstyle R}$ mixing is unimportant for most of the sfermions
except for the stop. In the case $\tan{\beta} \gg 1$, the mixing in the sbottom 
sector may also be non-negligible. 

\subsection{Charginos}
\hspace*{0.5cm}The charged gauginos, $\w$, and the charged Higgsinos, $\h$, can mix; the resulting mass 
eigenstates are the charginos. The 4-component Dirac fermions that represent these two 
charginos are denoted here by $\chiu$ and $\chid$. The mass matrix in the 
$(\tilde{W}^{+},\tilde{H}^{+})$ basis can be written generically as: \\
 \begin{equation}
\label{eq:chargemass}
X =\left(\begin{array}{cc}
 M_{2} & {\varepsilon}_{1}
\\ {\varepsilon}_{2} & \mu \,
\end{array} \right)\,.
\end{equation}
\hspace*{0.5cm}As in the case of squarks and sleptons there are three types of contributions. There
are contributions to the diagonal elements that come from explicit SUSY breaking, 
namely the Majorana mass $M_{2}$, and contributions that preserve SUSY, namely, the 
$\mu -$term. Both diagonal terms preserve $\gs$ invariance. The off-diagonal terms come from 
the spontaneous breaking of the $\gs$ symmetry. More specifically, when the Higgs field in the 
SUSY invariant interaction terms with a gaugino and a Higgsino is replaced by its {\it vev}, it gives 
rise to the following mixing entries, 
\begin{equation}  
{\varepsilon}_{1} = \sqrt{2} m_{\scriptscriptstyle W} \sin{\beta}\,,\ \ \ \ \ 
{\varepsilon}_{2} = \sqrt{2} m_{\scriptscriptstyle W} \cos{\beta}\,.
\end{equation}
\hspace*{0.5cm}Notice that $X$ is not symmetric unless $\tan{\beta}=1$. Therefore, two 
different unitary $2 \times 2$ matrices $U$ and $W$ are required to diagonalize the 
chargino mass matrix: 
\begin{equation} 
\label{eq:chargediag}
{\tilde{M}}^{+} = \diag(\tilde{M}^{+}_{\scriptscriptstyle 1},\tilde{M}^{+}
_{\scriptscriptstyle 2})
 = U^{*}XW^{-1}\,. 
\end{equation}
In principle, the diagonal elements can be either positive or negative. We choose 
$M_{2}$ to be positive and $\mu$ can be either positive or negative. The physical chargino
masses, $|{\tilde{M}^{+}_{\scriptscriptstyle 1}}|, |\tilde{M}^{+}_{\scriptscriptstyle 2}|$, 
are defined to be positive. 

In case of negative eigenvalues we follow the procedure of the second paper in~\cite{HC}. We define
a new matrix $V$ such that, 
\begin{equation} 
U^{*}XV^{-1} = \diag(|{\tilde{M}^{+}_{\scriptscriptstyle 1}}|,
|\tilde{M}^{+}_{\scriptscriptstyle 2}|)\,,
\end{equation}
and express all the interactions in terms of $U$ and $V$. $V$ can be trivially obtained from $W$
by changing all the signs in the file corresponding to the negative eigenvalue, i.e,
$V_{\scriptscriptstyle kl}=\eta_{k}W_{\scriptscriptstyle kl}$ (no sum over $k$) where 
$\eta_{k}$ is the sign of the eigenvalue $\tilde{M}_{k}^{+}$ $(k,l=1,2)$.

The corresponding squared mass eigenvalues are: \\
\begin{equation}
\tilde{M}_{\scriptscriptstyle 1,2}^{+{2}}=\frac{1}{2}(M^{2}_{\scriptscriptstyle 2}+
\mu^{2}+2 m_{{\scriptscriptstyle W}}^{2})\pm\frac{1}{2}{\left[{(M_{\scriptscriptstyle 2}+\mu)}^2
[{(M_{\scriptscriptstyle 2}-\mu)}^2+4m_{{\scriptscriptstyle W}}^{2}]\right]}^{\frac{1}{2}} 
\end{equation}
Notice that to reach the large SUSY masses limit that we are interested in, it is necessary to 
consider the mass parameters in the range $M_{\scriptscriptstyle 2},\mu \gg m_{\scriptscriptstyle W}$
and therefore, to a very good approximation, the off diagonal elements of $X$ in eq.~(\ref{eq:chargemass}) are
negligible as compared to the diagonal elements. The mixing is small and $\chiu $ will
be predominantly gaugino with a mass close to $M_{\scriptscriptstyle 2}$, whereas $\chid$ 
will be predominantly Higgsino with a mass close to $|\mu|$. In this case,
\begin{equation}  
\label{eq:uw}
U=W=\pmatrix{ 1 & 0 \cr 0 & 1}\,,\nonumber
\end{equation} 
and if  $\mu \geq 0$ or $\mu <0$ $V$ will be respectively: \\
\begin{equation} 
\label{eq:v} 
V =\pmatrix{ 1 & 0 \cr 0 & 1}\,,\ \ \ \ \ {\rm or}\ \ \ \ \  
V =\pmatrix{ 1 & 0 \cr 0 & -1 }\,.
\end{equation}
The corrections to these matrices are suppressed by powers of $(m_{\scriptscriptstyle W}/M)$
where $M$ is the largest of $M_{\scriptscriptstyle 2}$ and $\mu$.

\subsection{Neutralinos}

The neutral gauginos $\tilde B$ and ${\tilde W}_{\scriptscriptstyle 3}$ and the neutral Higgsinos 
${\tilde H}_{1}^{\scriptscriptstyle 0}$, ${\tilde H}_{2}^{\scriptscriptstyle 0}$ can mix;
the resulting mass eigenstates are the neutralinos. The 4-component Majorana fermions which 
represent these 4 neutralinos are denoted here by 
${\tilde {\chi}}_{1}^{\scriptscriptstyle 0}, {\tilde {\chi}}_{2}^{\scriptscriptstyle 0},
{\tilde {\chi}}_{3}^{\scriptscriptstyle 0}$ and 
${\tilde {\chi}}_{4}^{\scriptscriptstyle 0}$, following the standard notation given 
in~\cite{HC}.

The mass matrix in the $(\tilde B,{\tilde W}_{\scriptscriptstyle 3},
{\tilde H}_{1}^{\scriptscriptstyle 0}, {\tilde H}_{2}^{\scriptscriptstyle 0})$ basis can be
written generically as:\\
\begin{equation}
\label{eq:neumass}  
Y =\pmatrix{ M_{\scriptscriptstyle 1} & 0 & 
\varepsilon'_{\scriptscriptstyle 1} & \varepsilon'_{\scriptscriptstyle 2} \cr 
         0 & M_{\scriptscriptstyle 2} & 
\varepsilon'_{\scriptscriptstyle 3} & \varepsilon'_{\scriptscriptstyle 4} \cr
\varepsilon'_{\scriptscriptstyle 1} & \varepsilon'_{\scriptscriptstyle 3} & 0 & -\mu \cr
\varepsilon'_{\scriptscriptstyle 2} & \varepsilon'_{\scriptscriptstyle 4} & -\mu & 0 }\,, 
\end{equation}

Similarly to the chargino case, there are the explicit SUSY breaking mass terms in the 
diagonal, $M_{\scriptscriptstyle 1}$ and $M_{\scriptscriptstyle 2}$, and the SUSY 
preserving $\mu$-mass terms connecting ${\tilde H}_{1}^{\scriptscriptstyle 0}$ and
${\tilde H}_{2}^{\scriptscriptstyle 0}$. The rest of the off-diagonal terms come from the
spontaneous breaking of $\gs$ and are specifically given by:
\begin{equation}
\begin{array}{l}  
\varepsilon'_{\scriptscriptstyle 1} = -m_{\scriptscriptstyle Z} 
\cos{\beta}\sin{\theta_{\scriptscriptstyle W}}\,,\ \ \ \ \ 
\varepsilon'_{\scriptscriptstyle 2} = m_{\scriptscriptstyle Z} 
\sin{\beta}\sin{\theta_{\scriptscriptstyle W}}\,, \\
\varepsilon'_{\scriptscriptstyle 3} = m_{\scriptscriptstyle Z} 
\cos{\beta}\cos{\theta_{\scriptscriptstyle W}}\,,\ \ \ \ \ 
\varepsilon'_{\scriptscriptstyle 4} = -m_{\scriptscriptstyle Z} 
\sin{\beta}\cos{\theta_{\scriptscriptstyle W}}\,.\\
\end{array}
\end{equation}
Since $Y$ is symmetric, only one $4 \times 4$ unitary matrix, $Z$, is required to diagonalize it,
\begin{equation} 
\label{eq:neutdiag}
{\tilde{M}}^{\scriptscriptstyle 0} = \diag(\tilde{M}_{\scriptscriptstyle 1}^{\scriptscriptstyle 0}
,\tilde{M}_{\scriptscriptstyle 2}^{\scriptscriptstyle 0},
\tilde{M}_{\scriptscriptstyle 3}^{\scriptscriptstyle 0},
\tilde{M}_{\scriptscriptstyle 4}^{\scriptscriptstyle 0})
 = Z^{*}YZ^{-1}\,. 
\end{equation}
Here the diagonal elements can be either positive or negative. As in the chargino case, 
we choose the Majorana masses  $M_{\scriptscriptstyle 1}, M_{\scriptscriptstyle 2}$ to be
positive  and allow $\mu$ to be positive or negative. The physical neutralino masses
$|\tilde{M}_{\scriptscriptstyle 1}^{\scriptscriptstyle 0}|$, $|\tilde{M}_{\scriptscriptstyle 2}^{\scriptscriptstyle 0}|$,
$|\tilde{M}_{\scriptscriptstyle 3}^{\scriptscriptstyle 0}|$, $|\tilde{M}_{\scriptscriptstyle 4}^{\scriptscriptstyle 0}|$ 
are defined to be positive. In case of negative eigenvalues, one defines a new matrix $N$ such that,\\
\begin{equation} 
N^{*}YN^{-1} = \diag(|\tilde{M}_{\scriptscriptstyle 1}^{\scriptscriptstyle 0}|
,|\tilde{M}_{\scriptscriptstyle 2}^{\scriptscriptstyle 0}|,
|\tilde{M}_{\scriptscriptstyle 3}^{\scriptscriptstyle 0}|,
|\tilde{M}_{\scriptscriptstyle 4}^{\scriptscriptstyle 0}|)\,,
\end{equation}
and express all the interactions in terms of $N$. $N$ can be trivially obtained from $Z$ by 
multiplying all the entries in the file corresponding to the negative eigenvalue by the imaginary unity;
i.e, $N_{\scriptscriptstyle kl}=\eta^{'}_{k}Z_{\scriptscriptstyle kl}$ (no sum over k), where 
$\eta^{'}_{k}=1$ if the eigenvalue $\tilde{M}_{\scriptscriptstyle k}^{\scriptscriptstyle 0}$ is positive and
$\eta^{'}_{k}=i$ if the eigenvalue $\tilde{M}_{\scriptscriptstyle k}^{\scriptscriptstyle 0}$ is negative 
$( k,l=1,2,3,4)$. 

In the large SUSY masses limit important simplifications do occur. In order to get the four 
neutralino masses larger than the electroweak scale it is necessary to consider the mass 
parameters in the range $M_{\scriptscriptstyle 1},M_{\scriptscriptstyle 2},\mu \gg
m_{\scriptscriptstyle Z}$. Therefore, to a very good approximation, the off-diagonal terms
$\varepsilon'_{\scriptscriptstyle i} (i=1,2,3,4)$ are negligible as compared to $M_{\scriptscriptstyle 1},
M_{\scriptscriptstyle 2}$ and $\mu$. The physical mass eigenstates 
${\tilde {\chi}}_{i}^{\scriptscriptstyle 0}, 
(i=1,\ldots, 4)$ are predominantly $\tilde B, {\tilde W}_{\scriptscriptstyle 3}, ({\tilde H}_{1}^{\scriptscriptstyle 0}+
{\tilde H}_{2}^{\scriptscriptstyle 0})/\sqrt{2}$ and $({\tilde H}_{1}^{\scriptscriptstyle 0}-
{\tilde H}_{2}^{\scriptscriptstyle 0})/\sqrt{2}$, and their corresponding masses are close 
to $M_{\scriptscriptstyle 1}, M_{\scriptscriptstyle 2}, |\mu|$ and $|\mu|$ respectively.

In this case,
\begin{equation} 
Z =\pmatrix{ 1 & 0 & 0 & 0 \cr  0 & 1 & 0 & 0 \cr  
0 & 0 & \frac{1}{\sqrt{2}} & {\frac{1}{\sqrt{2}}} \cr
0 & 0 & {\frac{1}{\sqrt{2}}} & {-\frac{1}{\sqrt{2}}} } \,, 
\end{equation}
and if $\mu\geq 0$ or $\mu<0$, N will be respectively: \\
\begin{equation}  
\label{eq:n}
N =\pmatrix{1 & 0 & 0 & 0 \cr  0 & 1 & 0 & 0 \cr  
0 & 0 & \frac{i}{\sqrt{2}} & \frac{i}{\sqrt{2}} \cr
0 & 0 & \frac{1}{\sqrt{2}} & -\frac{1}{\sqrt{2}} }\,,\ \ \ \ \ 
N =\pmatrix{1 & 0 & 0 & 0 \cr  0 & 1 & 0 & 0 \cr  
0 & 0 & {\frac{1}{\sqrt{2}}} & {\frac{1}{\sqrt{2}}} \cr
0 & 0 & \frac{i}{\sqrt{2}} & -\frac{i}{\sqrt{2}} }\,.
\end{equation}

\subsection{The relevant Lagrangian and notation}
\hspace{0.5cm}Once the mass eigenstates that we plan to the integrate out have been
specified, we need to specify in addition the relevant interaction terms of the 
supersymmetric Lagrangian ${\cal L}_{\rm SUSY}$. Since, in the present paper, we aim to 
calculate the one-loop effective action for external $W^{\pm}, Z$ and $\gamma$ gauge 
bosons, it is easy to see that the relevant interaction terms are those that connect gauge
bosons with sfermions on one hand, and gauge bosons with neutralinos and charginos on the
other hand. The terms connecting inos with sfermions are not relevant at one loop level. We write these 
interaction terms as well as the free Lagrangian in the 
mass-eigenstate basis which is the one that will be used in the rest of the paper. All 
together the relevant terms for the two points functions are the following:\\
\begin{equation}
  {\cal L}_{\rm MSSM}(V, \tilde{f}, \scp, \sne)
  ={\cal L}^{\scriptscriptstyle (0)} (V)+{\cal L}_{\tilde{f}} (V, \tilde{f})+
  {\cal L}_{\rm {\tilde{\chi}}} (V, \tilde{\chi})\,,
\end{equation}
where ${\cal L}^{\scriptscriptstyle (0)} (V)$ is the standard quadratic Lagrangian in
the $R_\xi$ gauge for the electroweak gauge bosons $V=W^{\pm}, Z, \gamma$, 
and ${\cal L}_{\tilde{f}} (V, \tilde{f})$ and ${\cal L}_{\rm {\tilde{\chi}}} (V, \tilde{\chi})$ are 
the Lagrangians for the sfermions and the neutralinos and charginos respectively. We use 
here a compact notation that is convenient for the integration in the path integral 
formalism. ${\cal L}_{\tilde{f}}$ is defined as follows:\\
\begin{equation}
\begin{array}{l}
\label{eq:lagferm}
\displaystyle 
{\cal L}_{\tilde{f}} (V,\tilde{f}) = {\cal L}^{(0)}_{\tilde{f}} ({\tilde{f}}) + 
{\cal L}^{(1)}_{\tilde{f}} (V,{\tilde{f}}) + {\cal L}^{(2)}_{\tilde{f}} (V,{\tilde{f}}),
\end{array}
\end{equation}
where ${\cal L}^{(0)}_{\tilde{f}} (\tilde{f})$ is the sfermions free Lagrangian: \\
\begin{equation}
\begin{array}{l}
\displaystyle 
{\cal L}^{(0)}_{\tilde{f}} ({\tilde{f}}) = 
\sum_{\tilde{f}} [({\partial}_{\mu} \tilde{f}^{+}{\partial}^{\mu} \tilde{f}) 
-(\tilde{f}^{+} \tilde{M}_{f}^{2} \tilde{f})],
\end{array}
\end{equation}
and $\tilde{f}$ is the shorthand notation for  sfermions of all types. It must be
understood as a column matrix with four entries, containing either the four mass 
eigenstates if it refers to squarks or the three mass eigenstates if it refers to sleptons.
If we use the third generation notation, as before, it reads:
\begin{equation}
\label{eq:matf}
\begin{array}{l}
\displaystyle 
\tilde{f} \equiv \left(
\begin{array}{l}
\tilde{t}_{1}\\  \tilde{t}_{2} \\
\tilde{b}_{1}\\  \tilde{b}_{2}
\end{array}
\right)
\hspace*{0.5cm} if \hspace*{0.5cm} \tilde{f} = \tilde{q} \hspace*{0.5cm};
\hspace*{0.5cm} 
\displaystyle 
\tilde{f} \equiv \left(
\begin{array}{l}
\tilde{\nu}\\ 0 \\
\tilde{\tau}_{1}\\ \tilde{\tau}_{2}
\end{array}
\right) 
\hspace*{0.5cm} if \hspace*{0.5cm} \tilde{f} = \tilde{l}
\end{array}
\end{equation}
the sum $\sum_{\tilde{f}}$ is over the three generations and, in the case of squarks, it 
runs also over the $N_{c}$ color indexes. The corresponding squared mass matrices are:
\begin{equation}
\begin{array}{l}
\displaystyle
\tilde{M}_{f}^{2} = \diag(\tilde{m}_{t_{1}}^{2},\tilde{m}_{t_{2}}^{2},
\tilde{m}_{b_{1}}^{2},\tilde{m}_{b_{2}}^{2})
\hspace*{0.5cm} if \hspace*{0.5cm} \tilde{f} = \tilde{q}; \\
\\
\tilde{M}_{f}^{2} = \diag(\tilde{m}_{\nu}^{2},0,\tilde{m}_{\tau_{1}}^{2},
\tilde{m}_{\tau_{2}}^{2})
\hspace*{0.5cm} if \hspace*{1.0cm} \tilde{f} = \tilde{l}.
\end{array}
\end{equation}
The interaction Lagrangian of sfermions and gauge bosons consists of two parts.
${\cal L}^{(1)}_{\tilde{f}}$ gives the interactions of two sfermions and one gauge boson and
 ${\cal L}^{(2)}_{\tilde{f}}$ gives the interactions of two sfermions and two gauge bosons: \\
\begin{eqnarray}
\label{eq:intdfug}
\displaystyle {\cal L}^{(1)}_{\tilde{f}} (V,{\tilde{f}}) &=& 
\sum_{\tilde{f}} \left\{-i e A_{\mu} \tilde{f}^{+} \hat{Q}_{f} 
\stackrel{\leftrightarrow}{{\partial}^{\mu}}
 \tilde{f} - \frac{i g}{c_{\scriptscriptstyle W}} Z_{\mu} \tilde{f}^{+}
\hat{G}_{f} \stackrel{\leftrightarrow}{{\partial}^{\mu}} \tilde{f}
-\frac{i g}{\sqrt{2}} [W^{+}_{\mu} \tilde{f}^{+} \Sigma_{f}^{tb} 
\stackrel{\leftrightarrow}{{\partial}^{\mu}} \tilde{f} + 
W^{-}_{\mu} \tilde{f}^{+} \Sigma_{f}^{bt} \stackrel{\leftrightarrow}{{\partial}^{\mu}} 
\tilde{f}] \right\},\nonumber\\
\\
\label{eq:intdfdg}
\displaystyle {\cal L}^{(2)}_{\tilde{f}} (V,{\tilde{f}}) &=& 
\sum_{\tilde{f}} \left\{\frac{g^{2}}{c_{\scriptscriptstyle W}^{2}} 
Z_{\mu} Z^{\mu} \tilde{f}^{+} \hat{G}_{f} \hat{G}_{f} \tilde{f} + 
\frac{2 g e}{c_{\scriptscriptstyle W}} A_{\mu} Z^{\mu} \tilde{f}^{+}
\hat{Q}_{f} \hat{G}_{f} \tilde{f}
+e^{2} A_{\mu} A^{\mu} \tilde{f}^{+} \hat{Q}_{f} \hat{Q}_{f} \tilde{f} \right.\nonumber\\
&+& \frac{1}{2} g^{2} W_{\mu}^{+} W^{\mu_{-}} \tilde{f}^{+} \Sigma_{f} \tilde{f}+
\frac{eg}{\sqrt{2}} y_{f}A_{\mu}W^{\mu_{+}} \tilde{f}^{+} \Sigma_{f}^{tb}\tilde{f}+
\frac{eg}{\sqrt{2}} y_{f}A_{\mu}W^{\mu_{-}} \tilde{f}^{+} \Sigma_{f}^{bt}\tilde{f}\nonumber\\
&-& \left.\frac{g^{2}}{\sqrt{2}}y_{f} \frac{s_{\scriptscriptstyle W}^{2}}{c_{\scriptscriptstyle W}}
Z_{\mu}W^{\mu_{+}} \tilde{f}^{+} \Sigma_{f}^{tb}\tilde{f}-
\frac{g^{2}}{\sqrt{2}}y_{f} \frac{s_{\scriptscriptstyle W}^{2}}{c_{\scriptscriptstyle W}}
Z_{\mu}W^{\mu_{-}} \tilde{f}^{+} \Sigma_{f}^{bt}\tilde{f}\right\},
\end{eqnarray}
where:
\begin{eqnarray}
\label{eq:rgfs}
 \hat{Q_{f}} = R_{f}^{-1} Q_{f} R_{f} &,& \hat{G_{f}} = R_{f}^{-1} G_{f} R_{f}\,\,, 
 \hspace*{0.3cm}
R_{f} = \left(
      \begin{array}{l} 
        R_{t}    \hspace*{0.5cm} 0 \\
        0    \hspace*{0.5cm} R_{b}
      \end{array}
    \right) \,,\nonumber\\
 Q_{f} = \diag(Q_{t}, Q_{t}, Q_{b}, Q_{b})&,&
    G_{f} = \diag(\frac{1}{2} - Q_{t} s_{\scriptscriptstyle W}^{2}, 
    - Q_{t} s_{\scriptscriptstyle W}^{2}, - \frac{1}{2} Q_{b} s_{\scriptscriptstyle W}^{2}, 
    -Q_{b} s_{\scriptscriptstyle W}^{2})\,,\nonumber  \\
 y_{f} = \frac{1}{3} \hspace*{0.3cm} if \hspace*{0.3cm} \tilde{f} = \tilde{q} &;& 
 y_{f} = -1 \hspace*{0.3cm} if \hspace*{0.3cm} \tilde{f} = \tilde{l}\,,\nonumber  \\   
   \Sigma_{f}^{tb} = \left(
      \begin{array}{l}
        0 \hspace*{0.5cm}
        \hat{\sigma}_{tb} \\
        0 \hspace*{0.5cm} 0
      \end{array}
    \right) &=& \left(\Sigma_{f}^{bt} \right)^{\scriptscriptstyle T} , \hspace*{0.3cm}
    \hat{\sigma}_{tb} = R_{t}^{-1} \sigma R_{b}, \nonumber\\
   \Sigma_{f} = \left(
      \begin{array}{l}
        \hat{\sigma_{t}} \hspace*{.5cm} 0 \\
        0 \hspace*{.5cm} \hat{\sigma_{b}} 
      \end{array}
    \right) &,& \hat{\sigma}_{t,b} = R_{t,b}^{-1} \sigma R_{t,b} \, \,, \hspace{0.6cm}
    \sigma = \left(
      \begin{array}{l}
        1 \hspace*{0.5cm} 0 \\
        0 \hspace*{0.5cm} 0
      \end{array}
    \right),
\end{eqnarray}
and $R_{t}, R_{b}$ have been defined in (\ref{eq:rotatmass}).

The above expressions have been written for $\sfe=\tilde q$. For $\sfe=\tilde l$
analogous expressions are obtained by changing $t\rightarrow \nu,
b \rightarrow \tau$, and  the corresponding changes in the couplings and rotation matrices. For
brevity, we will use in the following the squarks notation. 

The Lagrangian for neutralinos and charginos is:   
 \begin{equation}
\begin{array}{l}
\label{eq:lagcn}
\displaystyle {\cal L}_{\tilde{\chi}} (V, \tilde{\chi}) = 
{\cal L}^{(o)}_{\tilde{\chi}} ({\tilde{\chi}}) + {\cal L}^{(1)}_{\tilde{\chi}} (V, {\tilde{\chi}}),
\end{array}
\end{equation}
where ${\cal L}^{(o)}_{\tilde{\chi}}$ is the free Lagrangian: \\
\begin{equation}
\begin{array}{l}
\displaystyle 
{\cal L}^{(0)}_{\tilde{\chi}} (\tilde{\chi}) = 
\frac{1}{2} \bar{\tilde{\chi}}^{o} (i \partial{\hspace{-6pt}\slash} - \tilde{M}^{o}) 
\tilde{\chi}^{o} + 
\bar{\tilde{\chi}}^{+} (i \partial{\hspace{-6pt}\slash} - \tilde{M}^{+}) 
\tilde{\chi}^{+},
\end{array}
\end{equation}
with,
\begin{equation}
\begin{array}{l}
\displaystyle 
\tilde{\chi}^{o} = \left(
\begin{array}{l}
\tilde{\chi}^{o}_{1} \\  \tilde{\chi}^{o}_{2} \\
\tilde{\chi}^{o}_{3} \\  \tilde{\chi}^{o}_{4} 
\end{array}
\right)
\hspace*{0.5cm}; \hspace*{0.5cm}  
\displaystyle 
\tilde{\chi}^{+} = \left(
\begin{array}{l}
\tilde{\chi}^{+}_{1} \\  \tilde{\chi}^{+}_{2} 
\end{array}
\right)\,\,, 
\end{array}
\end{equation}
and the mass matrices $\tilde{M}^{+}$ and $\tilde{M}^{o}$ have been defined in (\ref{eq:chargediag}) and (\ref{eq:neutdiag})
respectively.

The interaction Lagrangian ${\cal L}^{(1)}_{\tilde{\chi}}$ consists of three parts, \\
\begin{equation}
\begin{array}{l}
\displaystyle 
{\cal L}^{(1)}_{\tilde{\chi}} (V,\tilde{\chi}) = 
{\cal L}^{(1)}_{o} (V,\tilde{\chi}^{o}) + 
{\cal L}^{(1)}_{+} (V,\tilde{\chi}^{+}) + 
{\cal L}^{(1)}_{+o} (V,\tilde{\chi}^{+},\tilde{\chi}^{o}) ,
\end{array}
\end{equation}
${\cal L}^{(1)}_{o}$ gives the interactions between neutralinos and gauge bosons, ${\cal L}^{(1)}_{+}$
the interactions between charginos and gauge bosons and ${\cal L}^{(1)}_{+o}$ the 
interactions that connect charginos, neutralinos and gauge bosons. More explicitly, 
\begin{eqnarray}
\label{eq:lagcnb}
\displaystyle 
{\cal L}^{(1)}_{o} (V,\tilde{\chi}^{o}) &=& \frac{g}{2 \cw} Z_{\mu} \bar{\tilde{\chi}}^{o} 
\gamma^{\mu} ({O''}_{\scriptscriptstyle L} P_{\scriptscriptstyle L} + 
{O''}_{\scriptscriptstyle R} P_{\scriptscriptstyle R} ) \tilde{\chi}^{o} \nonumber
\\
\displaystyle {\cal L}^{(1)}_{+} (V,\tilde{\chi}^{+}) &=& -e A_{\mu} \bar{\tilde{\chi}}^{+} 
\gamma^{\mu} \tilde{\chi}^{+} +  
\frac{g}{c_{\scriptscriptstyle W}} Z_{\mu} \bar{\tilde{\chi}}^{+} 
\gamma^{\mu} ({O'}_{\scriptscriptstyle L} P_{\scriptscriptstyle L} + 
{O'}_{\scriptscriptstyle R} P_{\scriptscriptstyle R} ) \tilde{\chi}^{+} \nonumber
\\
\displaystyle {\cal L}^{(1)}_{+ o} (V,\scp,\sne) &=& g W_{\mu}^{-} \bar{\tilde{\chi}}^{o} 
\gamma^{\mu} (O_{\scriptscriptstyle L} P_{\scriptscriptstyle L} + 
O_{\scriptscriptstyle R} P_{\scriptscriptstyle R} ) \tilde{\chi}^{+} + 
g W_{\mu}^{+} \bar{\tilde{\chi}}^{+} 
\gamma^{\mu} 
(O_{\scriptscriptstyle L}^{+} P_{\scriptscriptstyle L} + 
O_{\scriptscriptstyle R}^{+} P_{\scriptscriptstyle R} ) \tilde{\chi}^{o},
\end{eqnarray}
where $O_{\scriptscriptstyle L,R}$, ${O'}_{\scriptscriptstyle L,R}$ and 
${O''}_{\scriptscriptstyle L,R}$ are the following $4\times2$, $2\times2$ and $4\times4$
matrices, respectively: 
\begin{eqnarray}
\displaystyle 
(O_{\scriptscriptstyle L})_{ij} &=& -\frac{1}{\sqrt{2}} N_{i4} V_{j2}^{*} + 
N_{i2} V_{j1}^{*} \,\,;\hspace*{0.5cm} 
(O_{\scriptscriptstyle R})_{ij} = \frac{1}{\sqrt{2}} N_{i3}^{*} U_{j2} + 
N_{i2}^{*} U_{j1}; \hspace*{0.1cm} i=1, 2, 3, 4;
\hspace*{0.1cm} j=1, 2 \,, \nonumber \\
\displaystyle ({O'}_{\scriptscriptstyle L})_{ij} &=& -V_{i1} V_{j1}^{*} - 
\frac{1}{2} V_{i2} V_{j2}^{*} + \delta_{ij} s_{\scriptscriptstyle W}^{2} \,\,;
\hspace*{0.2cm}  
({O'}_{\scriptscriptstyle R})_{ij} = -U_{i1}^{*} U_{j1} - 
\frac{1}{2} U_{i2}^{*} U_{j2} + \delta_{ij} s_{\scriptscriptstyle W}^{2}; 
\hspace*{0.3cm} i,j = 1, 2  \,, \nonumber \\
\displaystyle 
({O''}_{\scriptscriptstyle L})_{ij} &=& -\frac{1}{2} N_{i3} N_{j3}^{*} +  
\frac{1}{2} N_{i4} N_{j4}^{*} \,\,; \hspace*{0.5cm} 
({O''}_{\scriptscriptstyle R})_{ij} = - ({O''}_{\scriptscriptstyle L})_{ij}^{*}
\hspace*{0.4cm} ; \hspace*{0.4cm} i,j = 1, 2, 3, 4 \,.
\end{eqnarray}

In particular, in the limit of large neutralino and chargino masses and by using the limiting expressions 
of $U, V$ and $N$ given in eqs.~(\ref{eq:uw}), (\ref{eq:v}) and (\ref{eq:n}) respectively, we get the 
following values for the coupling matrices, which are valid for $\mu \geq 0$  ($\mu <0$):
\\
\begin{equation}  
\label{eq:os}
\begin{array}{l}
O_{\scriptscriptstyle L} = O_{\scriptscriptstyle R}  = \pmatrix{ 0 & 0 \cr 
   1 & 0 \cr   0 & \frac{-i}{2} (\frac{1}{2}) \cr
0 & \frac{1}{2} (\frac{-i}{2}) } \,\,\,;\hspace*{0.3cm}
{O'}_{\scriptscriptstyle L} = {O'}_{\scriptscriptstyle R} = \pmatrix{ {\sw}^{2}-1 & 0 \cr 
         0 &  {\sw}^{2}-\frac{1}{2} } \,\,; \\
 \\
{O''}_{\scriptscriptstyle L} = {O''}_{\scriptscriptstyle R} = \pmatrix{  0 & 0 & 0 & 0 \cr        
0 & 0 & 0 & 0 \cr
0 & 0 & 0 & \frac{-i}{2} (\frac{i}{2}) \cr
0 & 0 & \frac{i}{2} (\frac{-i}{2}) & 0 }.
\end{array}
\end{equation}


\section{The large sparticle masses limit.}
\label{sec:limit}

\hspace*{0.5cm} In this section we describe the large sparticle masses limit. We consider 
the situation where {\it all} the sparticle masses are much larger than the electroweak scale 
and therefore much heavier than their corresponding standard partners. In particular
this could be the case if the sparticle masses are well above 
$m_{\scriptscriptstyle Z}, m_{\scriptscriptstyle W}$ and $m_{\scriptscriptstyle t}$ 
but still below the few TeV upper bound that is imposed by the standard solution to the 
hierarchy problem. The SUSY masses are also considered much larger than any of the external 
momenta in the Green functions that are studied in this work. The reason for this choice is because 
we are interested in the low energy limit of the MSSM and, in particular, in looking for any
possible non-decoupling effect of heavy SUSY particles in the low energy observables as for instance
the high precision observables at LEP.

Generically we writte $\tilde{m}_{\scriptscriptstyle i}^{2} \gg 
M_{\scriptscriptstyle EW}^{2}, q^{2}$, where $\tilde{m}_{\scriptscriptstyle i}$ denotes any of 
the physical sparticle masses, $M _{\scriptscriptstyle EW}$ any of the electroweak masses  
$(m_{\scriptscriptstyle Z}, m_{\scriptscriptstyle W}, m_{\scriptscriptstyle t},
\ldots)$ and $q$ denotes any of the external momenta. As for the analytical computation, whenever we refer 
to the large sparticle masses limit of a given one-loop
Feynman integral, we mean the asymptotic limit $\tilde{m}_{\scriptscriptstyle i}
\rightarrow \infty$ for all sparticle masses that are involved in that integral. However, we would like
to emphasize that this asymptotic limit is not fully defined unless one specifies in addition
the relative sizes of the involved masses.
In other words, the result may depend, in general, on the particular way this asymptotic limit is taken.
This can be seen in many examples of one-loop integrals. In order to illustrate  this, let us 
consider, for example, the well known scalar two-point function $B_{\scriptscriptstyle 0}
(q^{2}, m_{\scriptscriptstyle 1}, m_{\scriptscriptstyle 2})$, as defined for instance in
~\cite{BINT}. The result of this function in dimensional regularization is:
\begin{eqnarray}
\displaystyle B_{\scriptscriptstyle 0}
(q^{2}, m_{\scriptscriptstyle 1}, m_{\scriptscriptstyle 2})&=&
{\Delta}_\epsilon- \frac{1}{2}\left(\log\frac{m_{\scriptscriptstyle 1}^{2}}{\mu_{o}^{2}}
+\log\frac{m_{\scriptscriptstyle 2}^{2}}{\mu_{o}^{2}}\right)+1\nonumber \\
&&-\left(\frac{m_{\scriptscriptstyle 1}^{2}+m_{\scriptscriptstyle 2}^{2}}
{m_{\scriptscriptstyle 1}^{2}-m_{\scriptscriptstyle 2}^{2}}\right)
\log\frac{m_{\scriptscriptstyle 1}}{m_{\scriptscriptstyle 2}}+F(q^{2}, 
m_{\scriptscriptstyle 1}, m_{\scriptscriptstyle 2})\,,
\end{eqnarray} 
where $q^{2}$ is the external momentum, $m_{\scriptscriptstyle 1}$ and 
$m_{\scriptscriptstyle 2}$ are the masses of the two internal propagators, 
${\Delta}_\epsilon$ and $\mu_{o}$ are defined in Appendix B and the explicit expression for the 
function $F(q^{2}, m_{\scriptscriptstyle 1}, m_{\scriptscriptstyle 2})$ can be found in~\cite{BINT}.

Let us consider the large masses limit, $m_{\scriptscriptstyle 1}, m_{\scriptscriptstyle 2}
\gg q^{2}$, of $B_{\scriptscriptstyle 0}$ in the two following situations:\\
\hspace*{0.2cm}$m_{\scriptscriptstyle 1}^{2}=m_{\scriptscriptstyle 2}^{2} \equiv m^{2}, 
 m^{2} \gg q^{2}$\, ({\em case A})\hspace*{0.2cm} and \hspace*{0.2cm}
$m_{\scriptscriptstyle 1}^{2}=2m_{\scriptscriptstyle 2}^{2}\equiv 2m^{2}, 
 m^{2} \gg q^{2}$\, ({\em case B}\,).
 
By explicit computation of $B_{\scriptscriptstyle 0}$ in this two different limits we get:\\
\\
\hspace*{0.7cm}{\em case A}
\begin{equation}
\displaystyle B_{\scriptscriptstyle 0}
(q^{2}, m_{\scriptscriptstyle 1}, m_{\scriptscriptstyle 2})=
{\Delta}_\epsilon-\log\frac{m^{2}}{\mu_{o}^{2}}+\left[ \frac{2}{3}
\left(\frac{q^{2}}{4m^{2}}\right)+O\left(\frac{q^{4}}{m^{4}}\right)\right]\,,
\end{equation}  
\hspace*{0.7cm}{\em case B}
\begin{equation}
\displaystyle B_{\scriptscriptstyle 0}
(q^{2}, m_{\scriptscriptstyle 1}, m_{\scriptscriptstyle 2})=
{\Delta}_\epsilon-\log\frac{m^{2}}{\mu_{o}^{2}}+1-2\log2+\left[
O\left(\frac{q^{4}}{m^{4}}\right)\right]\,.
\end{equation}  
\hspace*{0.5cm}The quantities in square brackets represent the decoupling effects since 
they vanish in the asymptotic limit $m^{2} \rightarrow \infty$. The remaining terms contain
all the non-decoupling effects of particles 1 and 2 and they are the object of our interest 
here since they do not vanish in the asymptotic limit $m^{2} \rightarrow \infty$. It is clear
from the above results that the two cases lead to different non-decoupling effects.

For the present study of large sparticle masses limit, $\tilde{m}_{\scriptscriptstyle i}^{2} \gg 
M_{\scriptscriptstyle EW}^{2}, q^{2}$ we consider the two following different possibilities
for the physical masses $(i\neq j)$:\\
\\
\hspace*{0.7cm}{\em case A}
\begin{equation}
\displaystyle |\tilde{m}_{\scriptscriptstyle i}^{2}-
\tilde{m}_{\scriptscriptstyle j}^{2}| \ll
|\tilde{m}_{\scriptscriptstyle i}^{2}+
\tilde{m}_{\scriptscriptstyle j}^{2}| \,\,,
\end{equation}
\hspace*{0.7cm}{\em case B}
\begin{equation}
\displaystyle O\left(|\tilde{m}_{\scriptscriptstyle i}^{2}-
\tilde{m}_{\scriptscriptstyle j}^{2}|\right) \approx
O\left(|\tilde{m}_{\scriptscriptstyle i}^{2}+
\tilde{m}_{\scriptscriptstyle j}^{2}|\right) \,.
\end{equation} 
\hspace*{0.5cm}In the first case, the asymptotic limit $\tilde{m}_{\scriptscriptstyle i,j}^{2}
\rightarrow \infty$ is taken such that,
$|\frac{\tilde{m}_{\scriptscriptstyle i}^{2}-\tilde{m}_{\scriptscriptstyle j}^{2}}
{\tilde{m}_{\scriptscriptstyle i}^{2}+\tilde{m}_{\scriptscriptstyle j}^{2}}| \ll 1$
and therefore this mass ratio is a good parameter for the large mass expansion. The 
extreme situation of total degeneracy $\tilde{m}_{\scriptscriptstyle i}^{2}=
\tilde{m}_{\scriptscriptstyle j}^{2}$ can be considered as a particular example
belonging to this {\em case A}.

In {\em case B} the asymptotic limit $\tilde{m}_{\scriptscriptstyle i,j}^{2}
\rightarrow \infty$ is taken such that,
$|\frac{\tilde{m}_{\scriptscriptstyle i}^{2}-\tilde{m}_{\scriptscriptstyle j}^{2}}
{\tilde{m}_{\scriptscriptstyle i}^{2}+\tilde{m}_{\scriptscriptstyle j}^{2}}| \approx O(1)$
and therefore it is not the proper parameter for the large mass expansion.
Physically this situation corresponds to consider both masses $\tilde{m}_{\scriptscriptstyle i}$
and $\tilde{m}_{\scriptscriptstyle j}$ different and large with their difference being also large.
The previous considered {\em case B} for $B_{\scriptscriptstyle 0}$ is one example
belonging to this situation.

In the following we study how these two cases can be accomplished in the
MSSM for each one of the sparticle sectors that are considered in this work.

\subsection{Large mass limit in the inos sector.}
\hspace*{0.5cm}Given the particular form of the mass matrices in the inos sector,
eqs.(\ref{eq:chargemass}) and (\ref{eq:neumass}), the large chargino masses limit, 
${\tilde{M}^{+^{2}}_{\scriptscriptstyle 1,2}} \gg M_{\scriptscriptstyle EW}^{2}$, and the 
large neutralino masses limit, ${\tilde M}_{\scriptscriptstyle 1,2,3,4}
^{\scriptscriptstyle 0^{2}} \gg M_{\scriptscriptstyle EW}^{2}$, can only be accomplished if
the three involved SUSY mass parameters are taken large, namely, if $M_{\scriptscriptstyle 1}^{2},
 M_{\scriptscriptstyle 2}^{2}, \mu^{2} \gg M_{\scriptscriptstyle EW}^{2}$. As we have 
already mentioned in the previous section, the physical masses are, in this limit:
\begin{equation}
\hspace*{1.2cm}{\tilde{M}^{+}_{\scriptscriptstyle 1}} \approx M_{\scriptscriptstyle 2}\,\,,
{\tilde{M}^{+}_{\scriptscriptstyle 2}} \approx |\mu|\,\,,
{\tilde{M}^{\scriptscriptstyle 0}_{\scriptscriptstyle 1}} \approx M_{\scriptscriptstyle 1}\,\,,
{\tilde{M}^{\scriptscriptstyle 0}_{\scriptscriptstyle 2}} \approx M_{\scriptscriptstyle 2}\,\,
{\tilde{M}^{\scriptscriptstyle 0}_{\scriptscriptstyle 3}} \approx |\mu|\,\,,
{\tilde{M}^{\scriptscriptstyle 0}_{\scriptscriptstyle 4}} \approx |\mu|\,\,.
\end{equation}

In consequence, the two situations {\em A} and {\em B} above can be accomplished in the inos sector by 
choosing the corresponding possibilities for the parameters, $M_{\scriptscriptstyle 1}, 
M_{\scriptscriptstyle 2}$ and $\mu$. For instance, 
$|\tilde{M}^{+^{2}}_{\scriptscriptstyle 1}-\tilde{M}^{+^{2}}_{\scriptscriptstyle 2}| \ll 
|\tilde{M}^{+^{2}}_{\scriptscriptstyle 1}+\tilde{M}^{+^{2}}_{\scriptscriptstyle 2}|$
can be accomplished if one chooses 
$|M^{2}_{\scriptscriptstyle 2}-\mu^{2}| \ll |M^{2}_{\scriptscriptstyle 2}+\mu^{2}|$.
However, by inspection of the coupling matrices in the inos-sector, eq.(\ref{eq:os}), we see
that only the following pairings will occur in the one-loop integrals:
$\,\,(\chiu, {\sne}_{\scriptscriptstyle 2})\,,\,(\chid, {\sne}_{\scriptscriptstyle 3})\,,\,
(\chid, {\sne}_{\scriptscriptstyle 4})\,,\,(\chiu, \chiu)\,,\,(\chid, \chid)\,$ and
$({\sne}_{\scriptscriptstyle 3}, {\sne}_{\scriptscriptstyle 4})$. Therefore, in this sector
the masses that need to be compared when performing the large mass limit always belong
to {\em case A}.

\subsection{Large mass limit in the sfermions sector.}
\hspace*{0.5cm}Here the situation is very different since the mixing may not be negligible
and hence it may play a relevant role in taking the large mass limit. For definitess and for
illustrative purposes, let us consider here the stop-sbottom sector. First of all, the 
requirement that all sparticles be heavier that their corresponding partners reads in this 
case $\tilde{m}_{t_{1,2}}^{2} > m_{t}^{2}$ and $\tilde{m}_{b_{1,2}}^{2} > m_{b}^{2}$ and
imply the following conditions on the mass matrix parameters of eq.(\ref{eq:massmatrix})\,:
$\,{\varepsilon}^{2} < LR+m_{t}^{4}-m_{t}^{2}(L+R)$\hspace*{0.1cm} and
\hspace*{0.1cm}${\varepsilon'}^{2} < L'R'+m_{b}^{4}-m_{b}^{2}(L'+R')\,$.

On the other hand, the wanted large stop and sbottom masses limit, 
$\tilde{m}_{t_{1,2}}^{2}, \tilde{m}_{b_{1,2}}^{2} \gg M_{\scriptscriptstyle EW}^{2}$ can
only be accomplished if $L\,, R\,, L'\,, R' \gg M_{\scriptscriptstyle EW}^{2}$, and 
therefore the above conditions translate into the simpler ones:
${\varepsilon}^{2} < LR$ and ${\varepsilon'}^{2} < L'R'$ respectively.
This means that in taking the large masses limit the mixing can be large but not arbitrarily
large since it is bounded from above by these conditions.

By using the definitions in the MSSM of eq.(\ref{eq:inputssq}), these inequalities can be expressed
in terms of the soft SUSY breaking masses and the $\mu$ parameter as follows:
\begin{eqnarray}
\label{eq:rest1}
m_{\scriptscriptstyle t}^{2} (\at-\mu \cot{\beta})^{2} < 
M_{{\scriptscriptstyle {\tilde Q}}}^{2} M_{{\scriptscriptstyle {\tilde U}}}^{2}
\hspace*{0.1cm} &,& \hspace*{0.1cm}
m_{\scriptscriptstyle b}^{2} (\ab-\mu \tan{\beta})^{2} < 
M_{{\scriptscriptstyle {\tilde Q}}}^{2} M_{{\scriptscriptstyle {\tilde D}}}^{2}\,,\nonumber\\
M_{{\scriptscriptstyle {\tilde Q}}}^{2}, M_{{\scriptscriptstyle {\tilde D}}}^{2},
M_{{\scriptscriptstyle {\tilde U}}}^{2} & \gg & m_{t}^{2}, m_{{\scriptscriptstyle Z}}^{2}\,,
\end{eqnarray}
where the later condition implies, in turn, the limiting values:
\begin{equation} 
\hspace*{1.2cm} \tilde{m}_{t_{1}}^{2}\approx M_{{\scriptscriptstyle {\tilde Q}}}^{2}\hspace*{0.1cm},\,
\tilde{m}_{t_{2}}^{2}\approx M_{{\scriptscriptstyle {\tilde U}}}^{2} \hspace*{0.1cm},\,
\tilde{m}_{b_{1}}^{2}\approx M_{{\scriptscriptstyle {\tilde Q}}}^{2}\hspace*{0.1cm},\,  
\tilde{m}_{b_{2}}^{2}\approx M_{{\scriptscriptstyle {\tilde D}}}^{2}\,\,.
\end{equation}

In summary, in other to get large stop and sbottom masses one needs large 
values of the SUSY breaking masses $M_{{\scriptscriptstyle {\tilde Q}}},
M_{{\scriptscriptstyle {\tilde U}}}$ and $M_{{\scriptscriptstyle {\tilde D}}}$ and in 
order not to get a too large mixing, the  trilinear couplings
$\at, \ab$ and the $\mu$ parameter must be constrained from above by the previous 
inequalities. In particular, $\mu$ cannot be arbitrarily
large when the rest of the parameters are kept fixed, since the lightest mass eigenvalues 
of eq.(\ref{eq:autov}) can be driven to non-physical negative values.

Notice that the $\mu$ parameter enters in both the sfermions and the inos sector and therefore
the large sparticle masses limit in both sectors are {\it not} independent. More specifically, the 
large masses limit in the ino sector which requires a large value of $\mu$, must respect in addition
the above restriction on $\mu$. In practice, this can be implemented by several different choices 
of the SUSY parameters.

Let us study now how the the previous cases {\em A} and {\em B} can be reached in the sfermions 
sector. We present the discussion in terms of the six generic mass matrix parameters 
$L\,, R\,, \varepsilon\,, L'\,, R'\,, \varepsilon'$. Notice that it is just once these are specified 
in terms of the MSSM parameters that there are some correlations among them which are relevant in studing 
the large sparticle masses limit.

Let us consider first the simplest case, corresponding to vanishing mixing:\\
\\
\hspace*{3.5cm}$\varepsilon\, = \,\varepsilon'\,=\,0 \,\,\,\Rightarrow\,\,s_{t}=s_{b}=0\,,\,\,\,$ and\\
\hspace*{3.2cm}$\tilde{m}^2_{t_{\scriptscriptstyle 1}}=L\,\,,\,
\tilde{m}^2_{t_{\scriptscriptstyle 2}}=R\,\,,\,\tilde{m}^2_{b_{\scriptscriptstyle 1}}=L'\,\,,\,
\tilde{m}^2_{b_{\scriptscriptstyle 2}}=R'$.

The only possible pairings in this case are:
$(\tu,\,\tu)\,,(\td,\,\td)\,,(\bu,\,\bu)\,,(\bd,\,\bd)\,\,$ and $(\tu,\,\bu)\,.$
Therefore there are just two masses to be compared, namely, $\tilde{m}_{t_{\scriptscriptstyle 1}}$
and $\tilde{m}_{b_{\scriptscriptstyle 1}}$.
Generically, both possibilities {\em A} and {\em B} could occur. However, in restricting us to the MSSM
parameters of eq.~(\ref{eq:inputssq}) it is clear that only possibility {\em A} does remain, since 
$|L-L'| \ll |L+L'|$ is always true in the large sparticle masses limit of the MSSM. Notice that this 
situation applies to all squarks of the first and second generation and to all sleptons.

Let us consider next the case of non-vanishing mixings $\varepsilon \neq 0\,,
\varepsilon'\neq 0$\,.Notice that it is in fact the most realistic situation
in the stop-sbottom sector of the MSSM since, as we have already said, 
$|\varepsilon|$ and $|\varepsilon'|$ grow linearly with $\mu$ and this must be taken large
$(\mu^{2} \gg M_{\scriptscriptstyle EW}^{2})$ to get heavy inos. Generically, therefore, in taking the large 
masses limit the mixings $\varepsilon$ and $\varepsilon'$ should {\it not} be held fixed.

For non-vanishing mixings, all possible pairings do occur: 
$\,(\tu,\,\tu)\,,(\td,\,\td)\,,(\bu,\,\bu)\,,(\bd,\,\bd)\,\,$ and 
$\,(\tu,\,\td)\,,(\tu,\,\bu)\,,(\tu,\,\bd)\,,(\bu,\,\bd)\,,(\bu,\,\td)\,,(\td,\,\bd)\,.$ 
Therefore, all the mass pairs need to be compared. For 
illustrative purposes let us analize here the two cases {\em A} and {\em B} in particular for
the $(\tu,\,\td)$ pair.\\
\\
\hspace*{0.7cm}{\em case A}
\begin{equation}
\label{eq:casoa}
\displaystyle \left|\frac{\tilde{m}^2_{t_{\scriptscriptstyle 1}}-
\tilde{m}^2_{t_{\scriptscriptstyle 2}}}{\tilde{m}^2_{t_{\scriptscriptstyle 1}}+
\tilde{m}^2_{t_{\scriptscriptstyle 2}}}\right| \ll 1 \,.
\end{equation} 

By using eq.~(\ref{eq:autov}) this condition can be written as,
\begin{equation}
\label{eq:exppar}
{\left[{\left(\frac{L-R}{L+R}\right)}^{2}+{\left(\frac{2\varepsilon}{L+R}\right)}^{2}
\right]}^{\frac{1}{2}} \ll 1 \,
\Longleftrightarrow \,\left(\frac{L-R}{L+R}\right) \ll 1 \,\, {\mbox{and}}\,\,
\left(\frac{2\varepsilon}{L+R}\right) \ll 1
\end{equation}
The proper parameter for the large mass expansion in terms of the physical masses of 
eq.~(\ref{eq:casoa}) is  translated into the two small parameters of eq.~(\ref{eq:exppar}). Furthermore, 
within this case there are still two different possibilities:\\
\\
\hspace*{0.7cm}{\em A1}\,) $\,\,\, L=R$\\
\hspace*{0.5cm}It can be solved exactly for all values of $\varepsilon\,\neq 0\,$ and
$\varepsilon'\,\neq 0\,$ and gives the limiting values,\\
\hspace*{3.0cm}$s_{t}\,=\,c_{t}\,=\,\frac{1}{\sqrt{2}}\,\,\,\Longrightarrow \,\,\, \tan{2\phi_{t}}=
\frac{2\varepsilon}{L-R}\rightarrow \infty$\,.\\
The proper parameter for the expansion in this case is $\left(\frac{2\varepsilon}{L+R}\right)$.

{\em A2}\,) $\,\,\, L\neq R$\\
\hspace*{0.5cm}This is the most plausible situation in the MSSM. In this case the two above conditions in
eq.~(\ref{eq:exppar}) can be written in terms of the MSSM parameters respectively as follows,
\begin{equation}
\label{eq:condm1}
\left|\frac{M_{{\scriptscriptstyle {\tilde Q}}}^{2}-M_{{\scriptscriptstyle {\tilde U}}}^{2}
+O(M_{\scriptscriptstyle EW}^{2})}
{M_{{\scriptscriptstyle {\tilde Q}}}^{2}+M_{{\scriptscriptstyle {\tilde U}}}^{2}}\right| \ll 1
\end{equation}
and,
\begin{equation}
\label{eq:condm2}
\left|\frac{2m_{t}(\at-\mu \cot{\beta})}
{M_{{\scriptscriptstyle {\tilde Q}}}^{2}+M_{{\scriptscriptstyle {\tilde U}}}^{2}}\right| \ll 1
\end{equation}

Notice that if one assumes, as usual, that all the SUSY mass parameters are of the same order, 
namely, if $M_{\scriptscriptstyle {\tilde Q}},\, M_{\scriptscriptstyle {\tilde U}},\, \at,\, \mu \sim
O(M_{\scriptscriptstyle SUSY})$ with $M_{\scriptscriptstyle SUSY} \gg M_{\scriptscriptstyle EW}$
being the effective SUSY breaking mass scale, then eq.~(\ref{eq:condm2}) is automatically fulfiled, since 
this small parameter goes as $O\left(\frac{M_{\scriptscriptstyle EW}}{M_{\scriptscriptstyle SUSY}}
\right)$. In order to get the first condition of eq.~(\ref{eq:condm1}) also automatically accomplished in
the MSSM, one needs in addition to impose the equality of the two soft SUSY breaking parameters 
$M_{\scriptscriptstyle {\tilde Q}}= M_{\scriptscriptstyle {\tilde U}}$. In this case the small
parameter of eq.~(\ref{eq:condm1}) goes as $O\left(\frac{M_{\scriptscriptstyle EW}^{2}}
{M_{\scriptscriptstyle SUSY}^{2}}\right)$. Any departure from this exact equality would lead us 
to a different situation which is considered next.\\
\\
\hspace*{0.7cm}{\em case B}
\begin{equation}
\label{eq:casob}
\displaystyle \left|\frac{\tilde{m}^2_{t_{\scriptscriptstyle 1}}-
\tilde{m}^2_{t_{\scriptscriptstyle 2}}}{\tilde{m}^2_{t_{\scriptscriptstyle 1}}+
\tilde{m}^2_{t_{\scriptscriptstyle 2}}}\right| \approx O(1) \,.
\end{equation} 

This can be written as:
\begin{equation}
\label{eq:exporder}
{\left[{\left(\frac{L-R}{L+R}\right)}^{2}+{\left(\frac{2\varepsilon}{L+R}\right)}^{2}
\right]}^{\frac{1}{2}} \approx O(1) \,.
\end{equation}
\hspace*{0.5cm}If, as before, one assumes $M_{\scriptscriptstyle {\tilde Q}},\, 
M_{\scriptscriptstyle {\tilde U}},\, \at,\, \mu \sim
O(M_{\scriptscriptstyle SUSY})$ then eq.~(\ref{eq:exporder}) can be traslated in terms of the MSSM parameters as:
\begin{equation}
\label{eq:condorder}
\left|\frac{M_{{\scriptscriptstyle {\tilde Q}}}^{2}-M_{{\scriptscriptstyle {\tilde U}}}^{2}}
{M_{{\scriptscriptstyle {\tilde Q}}}^{2}+M_{{\scriptscriptstyle {\tilde U}}}^{2}}\right| \approx O(1)\,,
\end{equation}
since eq.~(\ref{eq:condm2}) still holds.

This condition is naturally reached if one chooses, 
$(M_{\scriptscriptstyle {\tilde Q}}^{2}-M_{\scriptscriptstyle {\tilde U}}^{2})\sim
O(M_{\scriptscriptstyle SUSY}^{2})$. The parameters in eqs.~(\ref{eq:casob}) and (\ref{eq:condorder})
are obviously not the proper parameters for the large mass expansion in this case, but the parameter 
$\left(\frac{2\varepsilon}{L+R}\right)$ given in the MSSM by eq.~(\ref{eq:condm2}) is still a good one.
Other proper choices are also available. In particular, both quantities, $\tan{2\phi_{t}}$ and
$s_{t}$ turn out to be good parameters for the expansion in this case. In fact,
\begin{equation}
\tan{2 \phi_{t}}=\frac{2{\varepsilon}}{{L}-{R}} \simeq \frac{2m_{t}(\at-\mu \cot{\beta})}
{M_{{\scriptscriptstyle {\tilde Q}}}^{2}-M_{{\scriptscriptstyle {\tilde U}}}^{2}} \approx
O\left(\frac{M_{\scriptscriptstyle EW}}{M_{\scriptscriptstyle SUSY}}\right)\,,
\end{equation}
and similarly, $\,s_{t} \approx
O\left(\frac{M_{\scriptscriptstyle EW}}{M_{\scriptscriptstyle SUSY}}\right)\,.$

One can make a parallel discussion for the sbottom sector and conclude similarly that,\\
$\,s_{b} \approx O\left(\frac{M_{\scriptscriptstyle EW}}{M_{\scriptscriptstyle SUSY}}\right)\,.$

In consequence, {\em case B} implies that only the following pairings do remain in the 
large squark masses limit: $\,(\tu,\,\tu)\,,(\td,\,\td)\,,(\bu,\,\bu)\,,(\bd,\,\bd)\,\,$ and 
$(\tu,\,\bu)$ and therefore, once again, the masses to be compared are $\tilde{m}_{t_{\scriptscriptstyle 1}}$
and $\tilde{m}_{b_{\scriptscriptstyle 1}}$, which fulfil automatically the condition 
$|\tilde{m}^2_{t_{\scriptscriptstyle 1}}-\tilde{m}^2_{b_{\scriptscriptstyle 1}}| \ll 
|\tilde{m}^2_{t_{\scriptscriptstyle 1}}+\tilde{m}^2_{b_{\scriptscriptstyle 1}}|$
in the MSSM.

In summary, the sfermion sector of the MSSM behaves in the large masses limit 
$(M_{\scriptscriptstyle SUSY} \gg M_{\scriptscriptstyle EW})$ as follows,
\begin{enumerate}
\item Sfermions ${\tilde{f}_{{\scriptscriptstyle 1}}}, {\tilde{f}_{{\scriptscriptstyle 2}}}$ other
than stop and sbottom. The masses of their standard fermionic partners are neglected
$(\varepsilon=\varepsilon'=0)$ and,
$$\left|\frac{\tilde{m}^2_{f_{\scriptscriptstyle 1}}-
\tilde{m}^2_{f_{\scriptscriptstyle 2}}}{\tilde{m}^2_{f_{\scriptscriptstyle 1}}+
\tilde{m}^2_{f_{\scriptscriptstyle 2}}}\right| \sim 
O\left(\frac{M_{\scriptscriptstyle EW}^{2}}{M_{\scriptscriptstyle SUSY}^{2}}\right) \,\,\ll \,1\,.$$
It is the proper parameter for the large mass expansion of a one loop integral involving the pairing 
$(\tilde{f}_{\scriptscriptstyle 1}, \tilde{f}_{\scriptscriptstyle 2})$. The corresponding SUSY mass parameters
$M_{\scriptscriptstyle {\tilde L}}, M_{\scriptscriptstyle {\tilde E}},\ldots$ are just required to be 
of order $M_{\scriptscriptstyle SUSY}$.
\item Stops and sbottoms $(\varepsilon \neq 0,\,\varepsilon'\neq 0).$\\
\\
{\em 2a}\,) If $M_{\scriptscriptstyle {\tilde Q}}=M_{\scriptscriptstyle {\tilde U}}=
M_{\scriptscriptstyle {\tilde D}} \sim O(M_{\scriptscriptstyle SUSY})$ and 
$\mu,\, \at,\, \ab \sim O(M_{\scriptscriptstyle SUSY})$ then,
$$\left|\frac{\tilde{m}^2_{\scriptscriptstyle i}-
\tilde{m}^2_{\scriptscriptstyle j}}{\tilde{m}^2_{\scriptscriptstyle i}+
\tilde{m}^2_{\scriptscriptstyle j}}\right| \sim 
O\left(\frac{M_{\scriptscriptstyle EW}^{2}}{M_{\scriptscriptstyle SUSY}^{2}}\right) \,\,\ll \,1\,,
\hspace*{0.3cm} i,j=\tu,\td,\bu,\bd \hspace*{0.2cm} (i\neq j)$$
It is the proper parameter for the large mass expansion of a one-loop integral involving the pair $(i,j)$ 
with $i\neq j$. In the asymptotic limit $M_{\scriptscriptstyle SUSY} \rightarrow \infty$ the mixing is
maximal and $s_{t}, s_{b}$ tend to their limiting values, $s_{t}=s_{b}=\frac{1}{\sqrt{2}}$.\\
\\
{\em 2b}\,) If $\,M_{\scriptscriptstyle {\tilde Q}},\, M_{\scriptscriptstyle {\tilde U}},\,
M_{\scriptscriptstyle {\tilde D}},\, \mu,\, \at,\, \ab \sim O(M_{\scriptscriptstyle SUSY})\,\,$ with
$\,(M_{\scriptscriptstyle i}^{2}-M_{\scriptscriptstyle j}^{2}) \sim O(M_{\scriptscriptstyle SUSY}^{2}),
\,\,\,i,j = {\tilde Q}, {\tilde U}, {\tilde D} \\
(i\neq j)\,$ then,
$$s_{t},\, s_{b} \sim O\left(\frac{M_{\scriptscriptstyle EW}}{M_{\scriptscriptstyle SUSY}}\right)\,.$$
In the asymptotic limit $M_{\scriptscriptstyle SUSY} \rightarrow \infty$ then
$s_{t}, s_{b} \rightarrow 0$ and all the one-loop integrals involving pairings $(i,j)$ with $i\neq j$ 
decouple except $(\tu,\,\bu)$ for which:
$$\left|\frac{\tilde{m}^2_{t_{\scriptscriptstyle 1}}-
\tilde{m}^2_{b_{\scriptscriptstyle 1}}}{\tilde{m}^2_{t_{\scriptscriptstyle 1}}+
\tilde{m}^2_{b_{\scriptscriptstyle 1}}}\right| \sim 
O\left(\frac{M_{\scriptscriptstyle EW}^{2}}{M_{\scriptscriptstyle SUSY}^{2}}\right) \,\,\ll \,1\,.$$
\end{enumerate}

In conclusion, the large sparticle masses limit in the MSSM can generically be studied in terms of the 
physical masses by considering the corresponding one-loop Feynman integrals in the limit 
$\tilde{m}^{2}_{\scriptscriptstyle i}, \tilde{m}^{2}_{\scriptscriptstyle j} \gg 
M_{\scriptscriptstyle EW}^{2}, q^{2}$
with either possibility {\em A}\,: $\,\left| \frac{\tilde{m}^{2}_{\scriptscriptstyle i}-
\tilde{m}^{2}_{\scriptscriptstyle j}}{\tilde{m}^{2}_{\scriptscriptstyle i}+
\tilde{m}^{2}_{\scriptscriptstyle j}} \right| \ll 1$ or possibility {\em B}\,:
$\,\left|\frac{\tilde{m}^2_{t_{\scriptscriptstyle 1}}-
\tilde{m}^2_{t_{\scriptscriptstyle 2}}}{\tilde{m}^2_{t_{\scriptscriptstyle 1}}+
\tilde{m}^2_{t_{\scriptscriptstyle 2}}}\right| \approx O(1)\,$ if $(i\neq j)$. 
In this paper we have analized in full detail the possibility {\em A} and we will demonstrate the
decoupling of SUSY particles under this assumption. The different possibility {\em B} is not analized 
generically in this work, but it will be studied for the particular case of $S, T$ and $U$ 
in section 5, where we will demonstrate that decoupling of sparticles does occur in both cases 
{\em A} and {\em B}.


\section{Effective action for the electroweak gauge bosons in the MSSM}
\label{sec:effact}

\hspace*{0.5cm} Our aim is to compute the effective action for the standard particles,
$\Gamma_{eff}[\phi]$, that is defined through functional integration of all the
sparticles of the MSSM. In short notation it is defined by,
\begin{equation}
\label{eq:gammaeff}
{\rm e}^{i\Gamma_{eff}[\phi]}=\int [{\rm d}\tilde\phi]\,{\rm e}^{i \Gamma_{\rm MSSM}
  [\phi,\tilde\phi]}
\end{equation}
where $\phi=l,q,A,W^\pm,Z,g,H$ are the SM particles; $\tilde \phi=\tilde l,\tilde q,\tilde A,
\w,\tilde Z,\tilde g,\tilde H$ their supersymmetric partners, and 
\begin{equation}
\label{eq:gammaeffmssm}
\Gamma_{\rm MSSM}[\phi,\tilde\phi] \equiv \int \dx{\cal L}_{\rm
  MSSM}(\phi,\tilde \phi)\,\,;\,\,{\rm d}x\equiv{\rm d}^4x \,\,.
\end{equation}

In the present paper we are interested in the part of the effective action that
contains the two point Green functions with external gauge bosons
$\Gamma_{\mu\nu}^{V_1 V_2} (k)$, $V_1,V_2=A,Z,W^\pm$. This will allow us to study the 
decoupling properties of the gauge boson self-energies and from them we will deduce the 
corresponding analytical expressions for the well known parameters $S, T$ and $U$. The 
computation of the effective action will be performed at the one loop level by using 
dimensional regularization and will include the integration of all the
sfermions $\sfe$, the neutralinos $\sne$ and the charginos $\scp$. We leave the 
integration of the Higgs sector as well as the computation of other relevant parts of the 
effective action, as for instance the higher point Green functions or the Green 
functions with external fermions, for forthcoming works~\cite{DMS},~\cite{YO}.

We start by defining the piece of the effective action that we want to compute,
\begin{equation}
\label{eq:gammaeffV}
{\rm e}^{i\Gamma_{eff}[V]}=\int [{\rm d}\sfe]\, [{\rm d}\sfe^*]\, 
[{\rm d}\scp]\, [{\rm d}\bar{\sg}^+]\, [{\rm d}\sne]\,
{\rm e}^{i \Gamma_{\rm MSSM}[V,\sfe, \scp,\sne]}
\end{equation}
where,
\begin{eqnarray}
\label{eq:gammaMSSM}
 \Gamma_{\rm MSSM}[V,\sfe, \scp,\sne]&\equiv&
 \int\dx {\cal L}_{\rm MSSM}(V,\sfe, \scp,\sne) \nonumber\\ 
&=& \int\dx {\cal L}^{(0)} (V)+ \int\dx{\cal
 L}_{\sfe}(V,\sfe)+\int\dx{\cal L}_{\sg}(V,\sg) \nonumber\\
&\equiv&\Gamma_0[V]+\Gamma_{\sfe}[V,\sfe]+\Gamma_{\sg}[V,\sg]
\end{eqnarray}
and ${\cal L}_{\sfe}$, ${\cal L}_{\sg}$ have been defined in~(\ref{eq:lagferm}) 
and~(\ref{eq:lagcn}) respectively.

In order to perform the functional integration, it is convenient to write the
classical action in terms of operators. Thus by using eqs.~(\ref{eq:lagferm})
through~(\ref{eq:rgfs}) we get,
\begin{equation}
 \label{eq:gammasf}
\Gamma_{\sfe}[V,\sfe]=\langle \sfe^{+} A_{\sfe}\sfe\rangle 
\end{equation}
where,
\begin{eqnarray}
  \label{eq:gammasfdef1}
  A_{\sfe}&\equiv&A_{\sfe}^{(0)}+A_{\sfe}^{(1)}+A_{\sfe}^{(2)}\nonumber \\
  \label{eq:gammasfdef2}
  \langle \sfe^{+} A_{\sfe}^{(i)}\sfe\rangle &\equiv&\sum_{\sfe}\int\dx\dy \sfe_x^{+}
  A_{\sfe x y}^{(i)} \sfe_y\,\,,\,\,i=0,1,2
\end{eqnarray}
and the operators are:
\begin{eqnarray}
  \label{eq:opersfdef}
  A_{\sfe xy}^{(0)}&\equiv&(-\Box-\tilde M_{f}^2)_x\delta_{xy}\nonumber\\
  A_{\sfe xy}^{(1)}&\equiv&-i\,e\left(\partial_\mu A^\mu \hat{Q}_f+2\,\hat{Q}_f
    A_\mu\partial^\mu\right)_x\delta_{xy}-\frac{i\,g}{c_w}\left(\partial_\mu Z^\mu \hat{G}_f+2\,\hat{G}_f
    Z_\mu\partial^\mu\right)_x\delta_{xy}\nonumber\\
  &-&\frac{i\,g}{\sqrt{2}}\left(\partial_\mu W^{+\mu} \Sigma_f^{tb}+2\,
    \Sigma_f^{tb} W_\mu^+\partial^\mu\right)_x\delta_{xy}+ {\rm h.c.}\nonumber\\
  A_{\sfe xy}^{(2)}&\equiv&\left(e^2 \hat{Q}_f^2 A_\mu
  A^\mu+\frac{2\,g\,e}{c_w}A_\mu Z^\mu
  \hat{Q}_f\hat{G}_f+\frac{g^2}{c_w^2}\hat{G}_f^2 Z_\mu Z^\mu+\frac{1}{2}g^2
  \Sigma_f W_\mu^+ W^{\mu-}\right.\nonumber\\
  &+&\left. \frac{eg}{\sqrt{2}} y_{f}A_{\mu}W^{\mu_{+}} \Sigma_{f}^{tb}+
\frac{eg}{\sqrt{2}} y_{f}A_{\mu}W^{\mu_{-}}\Sigma_{f}^{bt}
-\frac{g^{2}}{\sqrt{2}}y_{f} \frac{s_{\scriptscriptstyle W}^{2}}{c_{\scriptscriptstyle W}}
Z_{\mu}W^{\mu_{+}}\Sigma_{f}^{tb}-
\frac{g^{2}}{\sqrt{2}}y_{f} \frac{s_{\scriptscriptstyle W}^{2}}{c_{\scriptscriptstyle W}}
Z_{\mu}W^{\mu_{-}} \Sigma_{f}^{bt}\right)_x \delta_{xy}\,\,.\nonumber\\
\end{eqnarray}
\hspace*{0.5cm}In these formulae and from now on we will use the compact notation,
\begin{eqnarray}
\label{eq:compact}
&&\phi(x)\equiv\phi_x\,,\,\delta(x-y)\equiv\delta_{xy}\,,\,A(x,y)\equiv A_{xy}\nonumber\\
&&\Tr A={\rm tr} \int\dx A_{xx}=\sum_a \int \dx A_{xx}^{aa}\,.
\end{eqnarray}
\hspace*{0.5cm}Analogously, by using eqs.~(\ref{eq:lagcn}) through~(\ref{eq:lagcnb}) we get,
\begin{equation}
  \label{eq:gammasg}
  \Gamma_{\sg}[V,\sg]=\frac{1}{2}\langle\bar{\sg}^0(A_0^{(0)}+A_0^{(1)})
  \sne\rangle + \langle\bar{\sg}^+(A_+^{(0)}+A_+^{(1)})\scp\rangle +
   \langle\bar{\sg}^0 A_{0+}^{(1)}\scp\rangle + \langle\bar{\sg}^+ A_{+0}^{(1)}\sne\rangle
\end{equation}
where,
\begin{eqnarray}
  \label{eq:gammsgdef}
  \langle\bar{\sg}^0 A_0^{(i)}\sne\rangle&\equiv&\int \dx\dy \bar{\sg}^0_x
  A_{0\,xy}^{(i)} \sne_y\,\,, \hspace*{0.4cm}
  \langle\bar{\sg}^+ A_+^{(i)}\scp\rangle \equiv \int \dx\dy \bar{\sg}^+_x
  A_{+\,xy}^{(i)} \scp_y \,\,;\,\,i=0,1 \nonumber\\
  \langle\bar{\sg}^0 A_{0+}^{(1)}\scp\rangle&\equiv&\int \dx\dy \bar{\sg}^0_x
  A_{0+\,xy}^{(1)} \scp_y\,\,, \hspace*{0.2cm}
  \langle\bar{\sg}^+ A_{+0}^{(1)}\sne\rangle \equiv \int \dx\dy \bar{\sg}^+_x
  A_{+0\,xy}^{(1)} \sne_y 
\end{eqnarray}
and the operators are:
\begin{eqnarray}
  \label{eq:opersgdef}
  A_{0\,xy}^{(0)}&\equiv&\left(i \slas{\partial}-\tilde M^0\right)_x \delta_{xy}\,\,, \hspace*{0.5cm}
  A_{+\,xy}^{(0)} \equiv \left(i \slas{\partial}-\tilde M^+\right)_x \delta_{xy}\nonumber\\
  A_{0\,xy}^{(1)}&\equiv&\left[\frac{g}{c_w}Z_\mu \gamma^\mu\left(O_L^{\prime\prime}
  P_L+O_R^{\prime\prime} P_R\right)\right]_x \delta_{xy}\,\,, \hspace*{0.2cm} 
  A_{+\,xy}^{(1)} \equiv \left[\frac{g}{c_w}Z_\mu \gamma^\mu\left(O_L^{\prime}
  P_L+O_R^{\prime} P_R\right)-e\,A_\mu \gamma^\mu\right]_x \delta_{xy}\nonumber\\
  A_{0+\,xy}^{(1)}&\equiv&\left[g\, W_\mu^- \gamma^\mu\left(O_L
      P_L+O_R P_R\right)\right]_x \delta_{xy}\,\,, \hspace*{0.3cm} 
  A_{+0\,xy}^{(1)} \equiv \left[g\, W_\mu^+ \gamma^\mu\left(O_L^{+}
      P_L+O_R^{+} P_R\right)\right]_x \delta_{xy}
\end{eqnarray}
being $P_L=1/2(1-{\gamma}_5)$ and $P_R=1/2(1+{\gamma}_5)$.

Given all the ingredients, we now proceed with the integration. The formula of the effective 
action (\ref{eq:gammaeffV}) can be factorized into three pieces: \\
\begin{equation}
\begin{array}{l}
\displaystyle 
e^{i \Gamma_{eff} [{\scriptscriptstyle V}]} = e^{i \Gamma_{o} [{\scriptscriptstyle V}]} 
e^{i \Gamma_{eff}^{\tilde{f}} [{\scriptscriptstyle V}]}
e^{i \Gamma_{eff}^{\tilde{\chi}} [{\scriptscriptstyle V}]} 
\end{array}
\end{equation}
where,
\begin{equation}
\begin{array}{l}
\label{eq:gammaeffF}
\displaystyle 
e^{i \Gamma_{eff}^{\tilde{f}} [{\scriptscriptstyle V}]} = \int [d\tilde{f}] [d\tilde{f}^{*}] 
e^{i \Gamma_{\tilde{f}} [{\scriptscriptstyle V},\tilde{f}]} 
\end{array}
\end{equation}
\begin{equation}
\begin{array}{l}
\label{eq:gammaeffCN}
\displaystyle e^{i \Gamma_{eff}^{\tilde{\chi}} [{\scriptscriptstyle V}]} = 
\int [d\tilde{\chi}^{+}] [d\bar{\tilde{\chi}}^{+}] [d\tilde{\chi}^{o}]  
e^{i \Gamma_{\tilde{\chi}} [{\scriptscriptstyle V},\tilde{\chi}]} 
\end{array}
\end{equation}
\hspace*{0.5cm} We next compute $\Gamma_{eff}^{\tilde{f}} [V]$ and $\Gamma_{eff}^{\tilde{\chi}} [V]$ 
separately. By substituting (\ref{eq:gammasf}) into (\ref{eq:gammaeffF}), and by performing 
a standard Gaussian integration we get,
$$\Gamma_{eff}^{\tilde{f}} [V] = i \Tr \log A_{\tilde{f}} = 
i \Tr \log [A_{\tilde{f}}^{(o)} (1 + A_{\tilde{f}}^{(o)-1}
(A_{\tilde{f}}^{(1)} + A_{\tilde{f}}^{(2)}) )]$$
The logarithm can be expanded and the $V$ independent terms can be left apart since 
they are irrevelant for the present computation. We get,\\
\begin{equation}
\begin{array}{l}
\displaystyle 
\Gamma_{eff}^{\tilde{f}} [V] = i \sum_{k=1}^{\infty} \frac{(-1)^{k+1}}{k} \Tr 
[G_{\tilde{f}} (A_{\tilde{f}}^{(1)} + A_{\tilde{f}}^{(2)})]^{k} 
\end{array}
\end{equation}
where,
$\displaystyle G_{\tilde{f}} \equiv A_{\tilde{f}}^{(o)^{-1}}$
and represents the free sfermion propagator matrix in the position space,
\begin{equation}
\label{eq:propG}
\displaystyle G_{\tilde{f} x y}^{ij} \equiv 
\int \frac{d^{{\scriptscriptstyle D}} q}{(2 \pi)^{{\scriptscriptstyle D}}} 
{\mu}_{o}^{4 - {\scriptscriptstyle D}} e^{-i q (x-y)} (q^{2} - \tilde{M}_{f}^{2})_{ij}^{-1} 
\end{equation}
with,\\
\hspace*{1cm} $\displaystyle (q^{2} - \tilde{M}_{f}^{2})^{-1} = \diag(\frac{1}{q^{2} - \tilde{m}_{t_{1}}^{2}},
\frac{1}{q^{2} - \tilde{m}_{t_{2}}^{2}}, \frac{1}{q^{2} - \tilde{m}_{b_{1}}^{2}},
\frac{1}{q^{2} - \tilde{m}_{b_{2}}^{2}})$ \hspace*{0.4cm} if \hspace*{0.4cm} 
$\tilde{f}=\tilde{q}$ \hspace*{0.2cm} ; or\\
\\
\hspace*{1cm} $\displaystyle (q^{2} - \tilde{M}_{f}^{2})^{-1} = 
\diag(\frac{1}{q^{2} - \tilde{m}_{\nu}^{2}}, \frac{1}{q^{2}},
\frac{1}{q^{2} - \tilde{m}_{\tau_{1}}^{2}},
\frac{1}{q^{2} - \tilde{m}_{\tau_{2}}^{2}})$ \hspace*{0.5cm} if \hspace*{0.5cm} 
$\tilde{f}=\tilde{l}$\\
\\
and, as always, a sum over the three generations and over the $N_{c}$ colors of 
the squarks are understood.

Finally, if we keep just the terms that contribute to the two-point functions, the 
effective action generated from sfermions integration can be written as, \\
\begin{equation}
\begin{array}{l}
\displaystyle 
\Gamma_{eff}^{\tilde{f}} [V] = i \Tr (G_{\tilde{f}} A_{\tilde{f}}^{(2)}) - 
\frac{i}{2} \Tr (G_{\tilde{f}} A_{\tilde{f}}^{(1)})^{2} + O(V^{3}) 
\end{array}
\end{equation}

Notice that in the case of sleptons we have formally integrated over the four 
components of $\tilde{f}=\tilde{l}$. This integration, in principle, would include
a rigth-handed sneutrino. However, due to the fact that this sneutrino doesn't
couple to any of the gauge bosons (see the definitions of ${\hat Q}_{f}, {\hat G}_{f}, \Sigma_{f}, 
\Sigma_{f}^{tb}$ in~(\ref{eq:rgfs})), it does not finally contribute to the effective action, as it must be.

We next compute the effective action generated from neutralinos and charginos 
integration. By substituting (\ref{eq:gammasg}) into (\ref{eq:gammaeffCN}) we get, 
\begin{equation}
\begin{array}{l}
\displaystyle 
e^{i \Gamma_{eff}^{\tilde{\chi}} [{\scriptscriptstyle V}]} = 
\int [d\tilde{\chi}^{+}] [d\bar{\tilde{\chi}}^{+}] [d\tilde{\chi}^{o}]
e^{i \{
\frac{1}{2} 
<\bar{\tilde{\chi}}^{o} (A_{o}^{(o)} + A_{o}^{(1)}) \tilde{\chi}^{o}> + 
<\bar{\tilde{\chi}}^{+} (A_{+}^{(o)} + A_{+}^{(1)}) \tilde{\chi}^{+}> + 
<\bar{\tilde{\chi}}^{o} A_{o+}^{(1)} \tilde{\chi}^{+}> + 
<\bar{\tilde{\chi}}^{+} A_{o+}^{(1)} \tilde{\chi}^{o}> \}}
\end{array}
\end{equation}
By performing first the integration over the charginos we find,
$$\displaystyle 
e^{i \Gamma_{eff}^{\tilde{\chi}} [{\scriptscriptstyle V}]} = 
det(A_{+}^{(o)} + A_{+}^{(1)}) \int [d\tilde{\chi}^{o}] e^{i \frac{1}{2} 
<\bar{\tilde{\chi}}^{o} [A_{o}^{(o)} + A_{o}^{(1)}-
2A_{o+}^{(1)} (A_{+}^{(o)} + A_{+}^{(1)})^{-1} A_{+o}^{(1)}] \tilde{\chi}^{o}>}$$
and next, by integrating over neutralinos we get,
$$\displaystyle 
e^{i \Gamma_{eff}^{\tilde{\chi}} [{\scriptscriptstyle V}]} = 
det(A_{+}^{(o)} + A_{+}^{(1)}) det[A_{o}^{(o)} + A_{o}^{(1)} - 
2A_{o+}^{(1)} (A_{+}^{(o)} + A_{+}^{(1)})^{-1} A_{+o}^{(1)}]^{\frac{1}{2}}.$$
The effective action can therefore be written as,
\begin{equation}
\begin{array}{l}
\displaystyle 
\Gamma_{eff}^{\tilde{\chi}} [V] = -i \Tr \log(A_{+}^{(o)} + A_{+}^{(1)}) 
- \frac{i}{2} \Tr  \log[A_{o}^{(o)} + A_{o}^{(1)} - 
2A_{o+}^{(1)} (A_{+}^{(o)} + A_{+}^{(1)})^{-1} A_{+o}^{(1)}]
\end{array}
\end{equation}
These logarithms can be worked out as before. By ignoring the $V$-independent 
terms and by expanding the logarithm we find, 
\begin{eqnarray}
\label{eq:trlog1}
\displaystyle 
-i \Tr \log(A_{+}^{(o)} + A_{+}^{(1)}) &=& 
-i \Tr \log[A_{+}^{(o)} (1 + A_{+}^{(o)^{-1}} A_{+}^{(1)})] \nonumber\\
&=& -i \Tr \log(1 + k_{+} A_{+}^{(1)}) 
= -i \sum_{k=1}^{\infty} \frac{(-1)^{k+1}}{k} \Tr (k_{+} A_{+}^{(1)})^{k}\,\,, 
\end{eqnarray}
where $k_{+} \equiv A_{+}^{(o)^{-1}}$ and represents the free chargino propagator 
matrix in the position space,\\
\begin{equation}
\label{eq:propcK}
\begin{array}{l}
\displaystyle 
k_{+ x y}^{ij} \equiv \int \frac{d^{{\scriptscriptstyle D}} q}
{(2 \pi)^{{\scriptscriptstyle D}}} {\mu}_{o}^{4 - {\scriptscriptstyle D}} 
e^{-i q (x - y)} (q{\hspace{-6pt}\slash} -\tilde{M}_{+})_{ij}^{-1}, \hspace*{0.5cm} i,j=1,2.
\end{array}
\end{equation}
Similarly, we find, 
\begin{eqnarray}
\label{eq:trlog2}
\displaystyle  &-&\frac{i}{2} \Tr \log[A_{o}^{(o)} + A_{o}^{(1)} - 
2A_{o+}^{(1)} (A_{+}^{(o)} + A_{+}^{(1)})^{-1} A_{+o}^{(1)}]=\nonumber\\
& & =-\frac{i}{2} {\Tr} \log[A_{o}^{(o)} (1 + A_{o}^{(o)^{-1}} A_{o}^{(1)} - 
2A_{o}^{(o)^{-1}} A_{o+}^{(1)} (A_{+}^{(o)} + A_{+}^{(1)})^{-1}
A_{+o}^{(1)})]\nonumber \\
& & =-\frac{i}{2} {\Tr} \log[1 + k_{o} A_{o}^{(1)} - 
2k_{o} A_{o+}^{(1)} (1 + k_{+} A_{+}^{(1)})^{-1} k_{+} A_{+o}^{(1)}] \nonumber \\
& & =-\frac{i}{2} \sum_{k=1}^{\infty} \frac{(-1)^{k+1}}{k}
{\Tr} [k_{o} A_{o}^{(1)} - 2k_{o} A_{o+}^{(1)} 
\sum_{r=0}^{\infty} (-1)^{r} (k_{+} A_{+}^{(1)})^{r} k_{+} A_{+o}^{(1)} ]^{k} 
\end{eqnarray}
where $k_{o} \equiv A_{o}^{(o)^{-1}}$ and represents the free neutralino propagator
matrix in the position space,\\
\begin{equation}
\label{eq:propnK}
\begin{array}{l}
\displaystyle 
k_{o x y}^{ij} \equiv \int \frac{d^{{\scriptscriptstyle D}} q}
{(2 \pi)^{{\scriptscriptstyle D}}} {\mu}_{o}^{4 - {\scriptscriptstyle D}} 
e^{-i q (x - y)} (q{\hspace{-5pt}\slash} -\tilde{M}_{o})_{ij}^{-1}, \hspace*{0.5cm}i, j=1,2,3,4 
\end{array}
\end{equation}
The sum of (\ref{eq:trlog1}) and (\ref{eq:trlog2}) gives the total contribution to the one-loop effective 
action generated from neutralino and chargino integration 
$\Gamma_{eff}^{\tilde{\chi}} [V]$.

Finally, if we keep just the terms that contribute to the two points functions we obtain,\\
\begin{equation}
\begin{array}{l}
\displaystyle 
\Gamma_{eff}^{\tilde{\chi}} [V] = \frac{i}{2} \Tr (k_{+} A_{+}^{(1)})^{2} + 
\frac{i}{4} \Tr (k_{o} A_{o}^{(1)})^{2} + i \Tr (k_{o} A_{o+}^{(1)} k_{+} A_{+o}^{(1)})+O(V^{3}). 
\end{array}
\end{equation}
\hspace*{0.5cm} The total resulting effective action for the two-point Green functions can be 
finally summarized by the following expression,
\begin{eqnarray}
\label{eq:gammaop}
\displaystyle \Gamma_{eff} [V] &=& \Gamma_{o} [V]+\Gamma_{eff}^{\tilde{f}} [V]+
\Gamma_{eff}^{\tilde{\chi}} [V] \nonumber \\
\displaystyle &=& \Gamma_{o} [V] +i \Tr (G_{\tilde{f}} A_{\tilde{f}}^{(2)}) 
-\frac{i}{2} \Tr (G_{\tilde{f}} A_{\tilde{f}}^{(1)})^{2}+ \frac{i}{2} \Tr (k_{+} A_{+}^{(1)})^{2}  \nonumber \\
&& \displaystyle+ \frac{i}{4} \Tr (k_{o} A_{o}^{(1)})^{2} + 
i \Tr (k_{o} A_{o+}^{(1)} k_{+} A_{+o}^{(1)}) + O( V^{3})
\end{eqnarray}
\hspace*{0.5cm} Diagrammatically, $\Gamma_{o}$ gives the contribution to the free two-point 
functions, the second and third terms give the two types of one loop contributions
with all kind of sfermions in the loop, as in Fig.{\bf 1a}, the fourth term gives the 
one-loop contributions with charginos in the loop, the fifth term is the corresponding 
contribution with neutralinos in the loop and the last term gives the mixed one-loop 
contributions with both charginos and neutralinos in the loop. The corresponding diagrams to 
the last three terms are shown in Fig.{\bf 1b}.


\section{Two-point functions for the electroweak gauge bosons in the large SUSY
 masses limit}
\label{sec:selflm}

\hspace*{0.5cm} In this section we present the two point functions for the electroweak gauge bosons 
to one loop and analize the limit of large masses of the SUSY particles.

In order to get the explicit expressions for the two-point functions one must work 
out the traces in the formula~(\ref{eq:gammaop}). This involves lengthy 
algebraic standard manipulations that we do not present here for brevity. Basically one must 
substitute into~(\ref{eq:gammaop}) the operators $A$ and propagators $G$ and $K$ of 
eqs.~(\ref{eq:opersfdef}, \ref{eq:opersgdef}, \ref{eq:propG}, \ref{eq:propcK}, \ref{eq:propnK}) , 
express the one-loop integrals in momentum space of D dimensions, compute all the appearing 
Dirac traces and Fourier transform the result back to the position space. The traces also 
involve to perform the sum in the corresponding matrix indexes, the sum over the various 
types of sfermions and the sum in color indexes in the case of squarks. The result for the 
effective action is the following:
\begin{eqnarray}
\displaystyle \Gamma_{eff} [V] &=& 
\frac{1}{2} \int dx dy A_{x}^{\mu} \Gamma_{\mu \nu}^{AA} (x,y) A_{y}^{\nu} + 
\frac{1}{2} \int dx dy Z_{x}^{\mu} \Gamma_{\mu \nu}^{ZZ} (x,y) Z_{y}^{\nu} +
\int dx dy A_{x}^{\mu} \Gamma_{\mu \nu}^{AZ} (x,y) Z_{y}^{\nu} \nonumber \\ 
&+& \int dx dy W_{x}^{+ \mu} \Gamma_{\mu \nu}^{+-} (x,y) W_{y}^{- \nu} + O( V ^{3})
\end{eqnarray}
where $\Gamma_{\mu \nu} (x,y)$ are the two-point functions in position space. Their 
relation with the correspon\-ding functions in momentum space is defined  by,
$$\displaystyle (2 \pi)^{4} \delta(k_{1} + k_{2}) \Gamma_{\mu \nu}(k_{1}) \equiv 
\int dx dy e^{-i k_{1} x - i k_{2} y} \Gamma_{\mu \nu}(x,y)$$
Finally, for the two-point functions in momentum space we find,
\begin{eqnarray}
\label{eq:gammaAA} 
\displaystyle \Gamma^{A\, A}_{\mu\, \nu}(k) &=& 
-k^{2} g_{\mu \nu} + \left(1 - \frac{1}{\xi_{{\scriptscriptstyle A}}}\right) 
k_{\mu} k_{\nu} \nonumber \\
\displaystyle \hspace*{1.0cm} &+&
i e^{2} \sum_{\tilde{f}} \left\{ 2 \sum_{a} I_{o}(\tilde{m}_{f_{a}}^{2})
(\hat{Q}_{f}^{2})_{aa} g_{\mu \nu} - \sum_{ab} (\hat{Q}_{f})_{ab} 
(\hat{Q}_{f})_{ba} I_{f_{\mu \nu}}^{ab}(k, \tilde{m}_{f_{a}}, \tilde{m}_{f_{b}})
\right\} \nonumber \\
\displaystyle \hspace*{1.0cm} &+& 2 i e^{2} \sum_{i=1}^{2}  
\left\{ T_{\mu \nu}^{ii}(k, \tilde{M}_{i}^{+}, \tilde{M}_{i}^{+}) +
 2 {\hat{I}}^{ii}(k, \tilde{M}_{i}^{+}, \tilde{M}_{i}^{+}) g_{\mu \nu} \right\} \\ 
\nonumber \\
\nonumber \\
\label{eq:gammaZZ} 
\displaystyle \Gamma^{Z\, Z}_{\mu\, \nu} (k) &=& 
(m_{{\scriptscriptstyle Z}}^{2}-k^{2}) g_{\mu \nu} + 
\left(1 - \frac{1}{\xi_{{\scriptscriptstyle Z}}}\right) k_{\mu} k_{\nu} \nonumber \\
\hspace*{1.0cm} &+& \displaystyle i \frac{g^{2}}{c_{{\scriptscriptstyle W}}^{2}} 
\sum_{\tilde{f}} \left\{ 2 \sum_{a} I_{o}(\tilde{m}_{f_{a}}^{2}) 
(\hat{G}_{f}^{2})_{aa} g_{\mu \nu} - \sum_{ab} (\hat{G}_{f})_{ab} 
(\hat{G}_{f})_{ba} I_{f_{\mu \nu}}^{ab}(k , \tilde{m}_{f_{a}}, \tilde{m}_{f_{b}})
\right\} \nonumber \\
\hspace*{1.0cm} &+& \displaystyle  \frac{i}{2} 
\frac{g^{2}}{c_{{\scriptscriptstyle W}}^{2}} \sum_{i,j=1}^{4} 
\left\{ \,({O''}_{{\scriptscriptstyle L}}^{ij} {O''}_{{\scriptscriptstyle L}}^{ji} 
+ {O''}_{{\scriptscriptstyle R}}^{ij} {O''}_{{\scriptscriptstyle R}}^{ji})\,
T_{\mu \nu}^{ij}(k, \tilde{M}_{i}^{0}, \tilde{M}_{j}^{0}) \right. \nonumber \\
\hspace*{1.0cm} && + \left.
2 \,({O''}_{{\scriptscriptstyle L}}^{ij} {O''}_{{\scriptscriptstyle R}}^{ji} +
{O''}_{{\scriptscriptstyle R}}^{ij} {O''}_{{\scriptscriptstyle L}}^{ji})\, 
{\hat{I}}^{ij}(k, \tilde{M}_{i}^{0}, \tilde{M}_{j}^{0})\,  g_{\mu \nu} \right\} \nonumber \\
\hspace{1.0cm} \displaystyle &+& i \frac{g^{2}}{c_{{\scriptscriptstyle W}}^{2}} \sum_{i,j=1}^{2} 
\left\{ \,({O'}_{{\scriptscriptstyle L}}^{ij} {O'}_{{\scriptscriptstyle L}}^{ji} + 
{O'}_{{\scriptscriptstyle R}}^{ij} {O'}_{{\scriptscriptstyle R}}^{ji})\,
T_{\mu \nu}^{ij}(k,  \tilde{M}_{i}^{+}, \tilde{M}_{j}^{+}) \right. \nonumber \\
\hspace*{1.0cm} && + \left. 
2 \,({O'}_{{\scriptscriptstyle L}}^{ij} {O'}_{{\scriptscriptstyle R}}^{ji} +
{O'}_{{\scriptscriptstyle R}}^{ij} {O'}_{{\scriptscriptstyle L}}^{ji})\, 
{\hat{I}}^{ij}(k,  \tilde{M}_{i}^{+}, \tilde{M}_{j}^{+})\, g_{\mu \nu}  \right\} \\
\nonumber \\
\label{eq:gammaAZ} 
\displaystyle \Gamma^{A\, Z}_{\mu\, \nu}(k) &=&\Gamma^{Z\, A}_{\mu\, \nu}(k)= 
\frac{i g e}{\cw} \sum_{\widetilde{f}} \left\{ 
2 \sum_{a} I_0(\tilde{m}^2_{f_a}) (\widehat{Q}_{f} \widehat{G}_{f})_{a\, a}  g_{\mu\, \nu}
- \sum_{ab}  (\widehat{Q}_{f})_{a\, b} (\widehat{G}_{f})_{b\, a} 
I^{a\,b}_{f_{\mu\, \nu}}(k , \tilde{m}_{f_{a}}, \tilde{m}_{f_{b}}) \right\} \nonumber \\
\hspace*{1cm} \displaystyle &-& \frac{i g e}{\cw}  \sum_{i=1}^2  
\left(  {O'}^{i\, i}_L + {O'}^{i\, i}_R \right)\left(
T_{\mu\, \nu}^{i\, i}(k, \tilde{M}_{i}^{+}, \tilde{M}_{i}^{+}) 
+ 2 {\hat{I}}^{ii}(k, \tilde{M}_{i}^{+}, \tilde{M}_{i}^{+}) g_{\mu\, \nu}\right)\\
\nonumber \\
\label{eq:gammaWW} 
\displaystyle \Gamma^{+\, -}_{\mu\, \nu}(k) &=&\Gamma^{-\, +}_{\mu\, \nu}(k)= 
(m_{{\scriptscriptstyle W}}^2-k^2)g_{\mu\, \nu} + \left(1-\frac{1}{\xi_{\scriptscriptstyle W}}
\right) k_{\mu}k_{\nu} \nonumber \\
\hspace{1.0cm} \displaystyle
&+& \frac{i g^2}{2} \sum_{\widetilde{f}}  \left\{ \sum_{a} (\Sigma_{f})_{a\, a}
I_0(\tilde{m}^2_{f_a}) g_{\mu\, \nu}-\sum_{a,b} (\Sigma_{f}^{t\,b})_{a\, b} 
(\Sigma_{f}^{t\,b})_{a\, b} I^{a\,b}_{f_{\mu\, \nu}}(k, \tilde{m}_{f_{a}}, \tilde{m}_{f_{b}}) 
\right\} \nonumber \\
\hspace*{1cm} \displaystyle &+& i g^{2} \sum_{i=1}^4 \sum_{j=1}^2 \left\{
\left({O}^{i\, j }_L {O}^{+\, j\, i}_L + {O}^{i\, j}_R  {O}^{+\, j\, i}_R \right)
T^{i\,j}_{\mu\, \nu}(k, \tilde{M}_{i}^{0}, \tilde{M}_{j}^{+}) \right. \nonumber \\
\hspace{1.0cm} \displaystyle 
&+& \left. 2 \left({O}^{i\,j}_L {O}^{+\,j\,i}_R + {O}^{i\, j}_R  {O}^{+\, j\, i}_L \right)
{\hat{I}}^{ij}(k, \tilde{M}_{i}^{0}, \tilde{M}_{j}^{+})  g_{\mu\, \nu} \right\} 
\end{eqnarray}
Here the indexes $a$ and $b$ run over the four entries of the sfermions column matrix 
in~(\ref{eq:matf}), and the sum in $\tilde{f}$ refers to the sum over squarks and 
sleptons of each generation as well as to the sum in color indexes for the squarks case. 
The indexes $i, j$ vary as $i,j=1,2,3,4$ if they refer to neutralinos and as $i,j=1,2$ if 
they refer to charginos.

The one-loop integrals in eqs.~(\ref{eq:gammaAA}) through~(\ref{eq:gammaWW}) are defined in
dimensional regularization by,\\
\begin{equation}
\label{eq:int0}
I_{0}(\mfa) =  \int d\widehat{q} \frac{1}{\left[q^2 - \mfa \right]}
\end{equation}
\\
\begin{equation}
I^{a\,b}_{f_{\mu\, \nu}}(k, \tilde{m}_{f_{a}}, \tilde{m}_{f_{b}}) =  \int d\widehat{q} 
\frac{(2q+k)_{\mu} (2q+k)_{\nu}}
{\left[(k+q)^2 - \tilde{m}^2_{f_a}\right] \left[q^2 - \tilde{m}^2_{f_b}\right]}
\end{equation}
\\
\begin{equation}
I^{i\,j}(k, \tilde{M}_{i}, \tilde{M}_{j}) =  \int d\widehat{q} 
\frac{1}{\left[q^2 - \tilde{M}^2_{i}\right] \left[(q+k)^2 - \tilde{M}^2_{j}\right]}
\end{equation}
\\
\begin{equation}
\begin{array}{l}
T^{i\,j}_{\mu\, \nu} = 4 I^{i\,j}_{\mu\, \nu}- 
2 g_{\mu\, \nu} g^{\alpha\, \beta} I^{i\,j}_{\alpha\, \beta}+2 ({I'}^{i\,j}_{\mu\, \nu}
+ {I'}^{i\,j}_{\nu\, \mu})-2  g_{\mu\, \nu} g^{\alpha\, \beta} {I'}^{i\,j}_{\alpha\, \beta},
 \quad T^{j\,i}_{\nu\, \mu}= T^{i\,j}_{\mu\, \nu}
\end{array}
\end{equation}
\\
\begin{equation}
I^{i\,j}_{\mu\, \nu}(k, \tilde{M}_{i}, \tilde{M}_{j}) = \int d\widehat{q} 
\frac{q_{\mu} q_{\nu} }{ \left[q^2 - \tilde{M}^2_{i}\right]\left[(k+q)^2 - 
\tilde{M}^2_{j}\right]}
\end{equation}
\\
\begin{equation}
{I'}^{i\,j}_{\mu\, \nu}(k, \tilde{M}_{i}, \tilde{M}_{j}) =\int d\widehat{q} 
\frac{q_{\mu} k_{\nu} }
{ \left[q^2 - \tilde{M}^2_{i}\right]\left[(k+q)^2 - \tilde{M}^2_{j}\right]}
\end{equation}
\\
\begin{equation}
\label{eq:intM}
{\hat I}^{i\,j}(k, \tilde{M}_{i}, \tilde{M}_{j}) = \int d\widehat{q} 
\frac{\tilde{M}_{i} \tilde{M}_{j} }
{ \left[q^2 - \tilde{M}^2_{i}\right]\left[(k+q)^2 - \tilde{M}^2_{j}\right]}
\end{equation}
Here one extra integral $I^{i\,j}$ has been included for completeness. In the above integrals
$\tilde{M}_{i}$ should be understood as $\tilde{M}^{0}_{i}$ if the index i (i=1,...,4)
refers to neutralinos or as $\tilde{M}^{+}_{i}$ if the index i (i=1,2) refers to charginos, and:
$$\int d\widehat{q} \equiv \int \frac{d^{{\scriptscriptstyle D}} q}
{(2 \pi)^{{\scriptscriptstyle D}}} {\mu}_{o}^{4 - {\scriptscriptstyle D}}.$$
Notice, that for the kind of loop integrals that we are considering there should not be relevant
difference in the results with respect to other regularization methods, as for instance, 
dimensional reduction. We have not done, however, this check explicitly.

The self-energies $\Sigma^{X\, Y}$ are defined from the two-point functions as 
usual,\\
\begin{equation}
\label{eq:gasi}
\Gamma^{X\, Y}_{\mu\, \nu} (k) = {\Gamma_0}^{X\, Y}_{\mu\, \nu} (k) +
\Sigma^{X\, Y} (k) \, g_{\mu\, \nu} + R^{X\, Y} (k) \, k_{\mu} k_{\nu}
\end{equation}
where ${\Gamma_0}^{X\, Y}_{\mu\, \nu}$ are the two-point functions at tree level 
and $X, Y =  A, Z, W^{\pm}$.

The transverse and longitudinal parts of the two-point functions, 
$\Sigma_{\scriptscriptstyle T}^{X\, Y}$ and $\Sigma_{\scriptscriptstyle L}^{X\, Y}$,
are defined by,
\begin{equation}
\label{eq:tl}
\Gamma^{X\, Y}_{\mu\, \nu} (k) = {\Gamma_0}^{X\, Y}_{\mu\, \nu} (k) +
\Sigma_{\scriptscriptstyle T}^{X\, Y} (k) \left( g_{\mu\, \nu}-
\frac{k_{\mu} k_{\nu}}{k^2}\right)+ \Sigma_{\scriptscriptstyle L}^{X\, Y} (k)\,
\frac{k_{\mu} k_{\nu}}{k^2}
\end{equation}

The expressions for the self energies $\Sigma^{X\, Y} (k)$ and the $R^{X\, Y} (k)$ functions 
can be easily read from eqs.~(\ref{eq:gammaAA}) through~(\ref{eq:gammaWW}). It is convenient
to express them in terms of new functions $A's, B's, C's$ and $D's$ being defined as follows,\\
\begin{equation}
\begin{array}{l}
\label{eq:ABCD}
\displaystyle
I^{a\,b}_{f_{\mu\, \nu}}(k, \tilde{m}_{f_{a}}, \tilde{m}_{f_{b}}) = 
A^{a\,b}_{f}(k, \tilde{m}_{f_{a}}, \tilde{m}_{f_{b}}) g_{\mu\, \nu} +
B^{a\,b}_{f}(k, \tilde{m}_{f_{a}}, \tilde{m}_{f_{b}}) k_{\mu} k_{\nu} \\
\\
T^{i\,j}_{\mu\, \nu}(k, \tilde{M}_{i}, \tilde{M}_{j}) = C^{i\, j}
(k, \tilde{M}_{i}, \tilde{M}_{j}) g_{\mu\, \nu} +
D^{i\,j}(k, \tilde{M}_{i}, \tilde{M}_{j}) k_{\mu} k_{\nu}.
\end{array}
\end{equation}

The results for $\Sigma^{X\, Y} (k)$ and $R^{X\, Y} (k)$ are collected in Appendix A.
This complete our computation of the effective action and the two-point functions for 
gauge bosons which should be noticed are exact to one-loop. 

Since we are interested in the large mass limit of the SUSY particles we need to have at 
hand not just the exact results of the above integrals but their asymptotic expressions to
be valid in that limit. We will present in the following the asymptotic results of the two-point 
functions and the gauge bosons self-energies. 

We have analized the integrals by means of the so-called m-Theorem \cite{GMR}. This 
theorem provides a powerful technique to study the asymptotic behaviour of Feynman 
integrals in the limit where some of the masses are large. Notice that this is not
trivial since some of these integrals are divergent and the interchange of the 
integral with the limit is not allowed. Thus, one should first compute the integrals in 
dimensional regularization and at the end take the large mass limit. Instead of this direct
way it is also possible to proceed as follows: First, one rearranges the integrand through
algebraic manipulations in order to decrease the ultraviolet divergence degree of some parts 
of the integral up to separate the MS (or DR) regularized integral into two parts, one of which
is a divergent contribution in 4 dimensions that 
can be evaluated exactly using the standard techniques, and the other one is a convergent part 
which satisfies
the requirements demanded by the m-Theorem and therefore, goes to zero in the infinite mass
limit. By means of this procedure we guarantee rigurously the correct asymptotic behaviour
of the integrals.

We give the details of the computation of the Feynman integrals by means
of the m-Theorem in  Appendix B. These integrals have been used
to obtain the final results for the transverse and longitudinal parts,
$\Sigma_{\scriptscriptstyle T}^{X\, Y} (k)$ and $\Sigma_{\scriptscriptstyle L}^{X\, Y} (k)$,
given in Appendix C.

Some comments on these results are in order:
\begin{itemize}
\item These asymptotic expressions are completely general and depend just on the
physical masses of the SUSY particles and on the generic
coefficients $c_{q}, s_{q}, c_{l}, s_{l}, {O}^{j\, i}_{L,R}, {O'}^{j\, i}_{L,R},
{O''}^{j\, i}_{L,R}$. Notice that they do not depend on the particular
mechanism that generates the SUSY masses.
\item We have done this computation, in addition, by diagrammatical methods
and we have found the same results. It involves the evaluation of Feynman diagrams with 
all kind of sparticles in the loops. We have shown in Fig.1 the various diagrams 
contributing to the two-point functions $\Gamma^{A\, A}, \Gamma^{A\, Z}, 
\Gamma^{Z\, Z}$ and $\Gamma^{W\, W}$. This diagrammatic computation provides a good check
of our previous results from functional methods, and at the same time helps to illustrate
which sparticle masses must be compared to which in the large mass expansion. Notice
that the longitudinal components of the two-point functions involving the photon field 
fulfil the expected Slavnov-Taylor identities.
\item In all these asymptotic expressions, the physical sparticle masses are assumed to
be much larger than the external momenta. As we can see, from Fig.1, the masses 
that must be compared to each other are the ones appering in the same one-loop 
diagram. Thus, for instance, the self-energies $\Sigma^{A\, A}$ and $\Sigma^{A\, Z}$, where no 
mixed diagrams with different sfermions contribute, do not need of any reference on 
the relative size of the sfermion masses. $\Sigma^{Z\, Z}$ and $\Sigma^{W\, W}$, 
on the contrary, do require this comparison. In the case of $\Sigma^{Z\, Z}$ one 
needs to compare squarks of the same charge, sleptons of the same charge, charginos of 
the same charge and neutralinos among them. No comparison among sfermions of
different generations is required since we have not considered intergenerational
mixing in this paper. In the case of $\Sigma^{W\, W}$ one needs to compare, in each
generation, the squarks of different charge, the sleptons of different charge and the 
netralinos with the charginos.The realistic and more interesting situation will be when all the
sparticles masses must be compared at the same time and, obviously, the final result 
will depend on the kind of SUSY hierarchy masses that had been previously established.
This will happen in the observables where all the four self-energies do contribute.
\item As discussed in the introduction, decoupling of heavy SUSY particles in the 
Appelquist-Carazzone Theorem sense will occur if the virtual effects due to these
particles on the effective low-energy SM action can be absorbed into a redefinition
of the SM parameters and wave functions renormalization, or else they are suppressed by 
inverse powers of the heavy SUSY particle masses. From our results in Appendix C it is
clear that we get indeed decoupling in the two point electroweak gauge boson functions.
This can be easily understood due to the specific analytical form of our formulae given 
generically by $\Sigma^{X\, Y} (k)=\Sigma^{X\, Y}_{(0)}+\Sigma^{X\, Y}_{(1)} k^{2}$ 
and $R^{X\, Y} (k)=R^{X\, Y}_{(0)}$, where $\Sigma^{X\, Y}_{(0)}, \Sigma^{X\, Y}_{(1)}$
and $R^{X\, Y}_{(0)}$ are functions of the SUSY large masses but are $k$ independent.
Equivalently, the transverse and longitudinal two-point functions of Appendix C fulfil 
$\Sigma_{\scriptscriptstyle T}^{X\, Y}=\Sigma_{\scriptscriptstyle T\, (0)}^{X\, Y}+
\Sigma_{\scriptscriptstyle T\, (1)}^{X\, Y} \,k^{2}$ and $(\Sigma_{\scriptscriptstyle L}^{X\, Y}-
\Sigma_{\scriptscriptstyle T}^{X\, Y}) \propto k^{2}$, which together are sufficient conditions
to get decoupling.
\end{itemize} 


\section{Decoupling of sparticles in $S, T$ and $U$}
\label{sec:stu}
\renewcommand\baselinestretch{1.3}
\hspace*{0.5cm} The radiative corrections from SUSY particles to the observables $S, T$ and
$U$ have been analyzed exhaustively in the literature~\cite{CHA} \cite{HA2}, but neither their 
complete analytical expressions in the large sparticle masses limit  nor a general and 
systematic study of sparticles decoupling have been provided so far. We present in this 
section our results to one loop for these analytical expressions of $S, T$ and $U$ in a
complete general form. Next we analyze under which particular conditions the sparticles 
decoupling takes place and finally we discuss how and why does it occur in
the very special case of the MSSM with soft SUSY breaking terms.

The definition that we use for $S, T$ and $U$ are the usual ones~\cite{PT}:
\begin{equation}
\begin{array}{l}
\displaystyle S = -\frac{16 \pi}{e^{2}} \sw \cw
[s_{{\scriptscriptstyle W}} 
c_{{\scriptscriptstyle W}} \Sigma_{{\scriptscriptstyle AA}}^{'} (0) - 
s_{{\scriptscriptstyle W}} s_{{\scriptscriptstyle W}} 
\Sigma_{{\scriptscriptstyle ZZ}}^{'}(0) + (c_{{\scriptscriptstyle W}}^{2} - 
s_{{\scriptscriptstyle W}}^{2}) \Sigma_{{\scriptscriptstyle AZ}}^{'} (0)],
\end{array}
\end{equation}
\begin{equation}
\begin{array}{l}
\displaystyle T = \frac{4 \pi}{e^{2}} 
\left[ \frac{\Sigma_{\scriptscriptstyle WW} (0)}{m_{\scriptscriptstyle W}^{2}} - 
 \frac{\Sigma_{{\scriptscriptstyle ZZ}} (0)}{m_{\scriptscriptstyle Z}^{2}} -2 \frac{\sw}{\cw} 
\frac{\Sigma_{{\scriptscriptstyle AZ}} (0)}{m_{\scriptscriptstyle Z}^{2}}\right]\,,  
\end{array}
\end{equation}
\begin{equation}
\begin{array}{l}
\displaystyle U = \frac{16 \pi}{e^{2}} [\Sigma_{{\scriptscriptstyle WW}}^{'} (0) - {\cw}^{2}
\Sigma_{{\scriptscriptstyle ZZ}}^{'}(0)-{\sw}^{2} \Sigma_{{\scriptscriptstyle AA}}^{'} (0)-
2 \sw \cw  \Sigma_{{\scriptscriptstyle AZ}}^{'} (0)]\,.
\end{array}
\end{equation}
\hspace*{0.5cm}The contribution to $S, T$ and $U$ are known to be finite and well defined
separately for each sparticle sector, so that we can analyze them separately as
well. As we have already said we consider in this paper all the sparticle
contributions except that of the Higgs sector. Consequently, we define: 
\begin{equation}
\begin{array}{l}
\displaystyle 
S_{\rm SUSY} = S_{\tilde{q}} + S_{\tilde{l}} + S_{\tilde{\chi}}+ S_{{\scriptscriptstyle H}}\\
\\
T_{\rm SUSY}= T_{\tilde{q}} + T_{\tilde{l}} + T_{\tilde{\chi}}+ T_{{\scriptscriptstyle H}}\\
\\
U_{\rm SUSY} = U_{\tilde{q}} + U_{\tilde{l}} + U_{\tilde{\chi}}+ U_{{\scriptscriptstyle H}}
\end{array}
\end{equation}
By using the corresponding expressions for the self-energies given in Appendix C we 
obtain the following results in the large masses limit, $m_{\tilde{q}_{i}}^{2}$, 
$m_{\tilde{l}_{j}}^{2}$, $\tilde{M}_{l}^{+^{2}}$, $\tilde{M}_{m}^{o^{2}} \gg k^{2}$, $\forall i,j,l,m$
and $k$ being the external momentum.

\subsection{Squarks:}

Under the conditions: \hspace{0.1cm} 
$\displaystyle |\tilde{m}_{t_{1}}^{2} - \tilde{m}_{t_{2}}^{2}| \ll 
|\tilde{m}_{t_{1}}^{2} + \tilde{m}_{t_{2}}^{2}|$,\hspace{0.1cm} 
$|\tilde{m}_{b_{1}}^{2} - \tilde{m}_{b_{2}}^{2}| \ll
|\tilde{m}_{b_{1}}^{2} + \tilde{m}_{b_{2}}^{2}|$ \hspace{0.1cm} we get,
\begin{equation}
\begin{array}{l}
\displaystyle 
S_{\tilde{q}} = -\sum_{\tilde{q}} \frac{N_{c}}{36 \pi} \left\{
\log \frac{\tilde{m}_{t_{1}}^{2}}{\tilde{m}_{b_{1}}^{2}} +
s_{b}^{2} \log \frac{\tilde{m}_{b_{1}}^{2}}{\tilde{m}_{b_{2}}^{2}} -
s_{t}^{2} \log \frac{\tilde{m}_{t_{1}}^{2}}{\tilde{m}_{t_{2}}^{2}} 
\displaystyle -3 c_{t}^{2} s_{t}^{2} 
\log \frac{(\tilde{m}_{t_{1}}^{2} + \tilde{m}_{t_{2}}^{2})^{2}}
{4 \tilde{m}_{t_{1}}^{2} \tilde{m}_{t_{2}}^{2}} 
-3 c_{b}^{2} s_{b}^{2} 
\log \frac{(\tilde{m}_{b_{1}}^{2} + \tilde{m}_{b_{2}}^{2})^{2}}
{4 \tilde{m}_{b_{1}}^{2} \tilde{m}_{b_{2}}^{2}} \right\}.
\end{array}
\end{equation}
For the observables $T_{\tilde{q}}$ and $U_{\tilde{q}}$ we must
consider together:\\
\vskip 0.1cm
$\displaystyle |\tilde{m}_{t_{1}}^{2} - \tilde{m}_{t_{2}}^{2}| \ll  
|\tilde{m}_{t_{1}}^{2} + \tilde{m}_{t_{2}}^{2}|, \hspace*{0.2cm} 
|\tilde{m}_{b_{1}}^{2} - \tilde{m}_{b_{2}}^{2}| \ll  
|\tilde{m}_{b_{1}}^{2} + \tilde{m}_{b_{2}}^{2}|, \hspace*{0.2cm} 
|\tilde{m}_{t_{i}}^{2} - \tilde{m}_{b_{j}}^{2}| \ll 
|\tilde{m}_{t_{i}}^{2} + \tilde{m}_{b_{j}}^{2}|$ \hspace*{0.3cm} $i,j=1,2$\\
\\
obtaining,
\begin{eqnarray}
\displaystyle T_{\tilde{q}} &=& \sum_{\tilde{q}} \frac{N_{c}}{8 \pi} 
\frac{1}{s_{{\scriptscriptstyle W}}^{2} 
m_{{\scriptscriptstyle W}}^{2}} \left\{
c_{t}^{2} c_{b}^{2} h(\tilde{m}_{t_{1}}^{2}, \tilde{m}_{b_{1}}^{2}) +  
c_{t}^{2} s_{b}^{2} h(\tilde{m}_{t_{1}}^{2}, \tilde{m}_{b_{2}}^{2}) - 
c_{t}^{2} s_{t}^{2} h(\tilde{m}_{t_{1}}^{2}, \tilde{m}_{t_{2}}^{2}) \right.\nonumber
\\
\displaystyle 
&+& \left.s_{t}^{2} c_{b}^{2} h(\tilde{m}_{t_{2}}^{2}, \tilde{m}_{b_{1}}^{2}) - 
s_{t}^{2} s_{b}^{2} h(\tilde{m}_{t_{2}}^{2}, \tilde{m}_{b_{2}}^{2}) + 
s_{b}^{2} c_{b}^{2} h(\tilde{m}_{b_{1}}^{2}, \tilde{m}_{b_{2}}^{2}) \right\}, \\
\nonumber \\   \nonumber \\
\displaystyle U_{\tilde{q}} &=& \sum_{\tilde{q}} \frac{N_{c}}{12 \pi} \left\{
c_{t}^{2} c_{b}^{2} \log \frac{4 \tilde{m}_{t_{1}}^{2} \tilde{m}_{b_{1}}^{2}}
{(\tilde{m}_{t_{1}}^{2} + \tilde{m}_{b_{1}}^{2})^{2}} + 
c_{t}^{2} s_{b}^{2} \log \frac{4 \tilde{m}_{t_{1}}^{2} \tilde{m}_{b_{2}}^{2}}
{(\tilde{m}_{t_{1}}^{2} + \tilde{m}_{b_{2}}^{2})^{2}} + 
s_{t}^{2} s_{b}^{2} \log \frac{4 \tilde{m}_{t_{2}}^{2} \tilde{m}_{b_{2}}^{2}}
{(\tilde{m}_{t_{2}}^{2} + \tilde{m}_{b_{2}}^{2})^{2}} \right. \nonumber
\\
\displaystyle &+& \left. s_{t}^{2} c_{b}^{2} 
\log \frac{4 \tilde{m}_{t_{2}}^{2} \tilde{m}_{b_{1}}^{2}}
{(\tilde{m}_{t_{2}}^{2} + \tilde{m}_{b_{1}}^{2})^{2}} - 
c_{t}^{2} s_{t}^{2} \log \frac{4 \tilde{m}_{t_{1}}^{2} \tilde{m}_{t_{2}}^{2}}
{(\tilde{m}_{t_{1}}^{2} + \tilde{m}_{t_{2}}^{2})^{2}} -  
s_{b}^{2} c_{b}^{2} \log \frac{4 \tilde{m}_{b_{1}}^{2} \tilde{m}_{b_{2}}^{2}}
{(\tilde{m}_{b_{1}}^{2} + \tilde{m}_{b_{2}}^{2})^{2}} \right\},
\end{eqnarray}
where:
\begin{equation}
\label{eq:h}
h({m}_{1}^{2}, {m}_{2}^{2}) \equiv {m}_{1}^{2} \log \frac{2{m}_{1}^{2}}{{m}_{1}^{2} + {m}_{2}^{2}} +
{m}_{2}^{2} \log \frac{2{m}_{2}^{2}}{{m}_{1}^{2} + {m}_{2}^{2}}.
\end{equation}

\subsection{Sleptons:}

If $\displaystyle |\tilde{m}_{\tau_{1}}^{2} - \tilde{m}_{\tau_{2}}^{2}| \ll 
|\tilde{m}_{\tau_{1}}^{2} + \tilde{m}_{\tau_{2}}^{2}|$:
\begin{eqnarray}
\displaystyle S_{\tilde{l}} &=& -\sum_{\tilde{l}}  \frac{1}{36 \pi} \left\{
\log \frac{\tilde{m}_{\nu}^{2}}{\tilde{m}_{\tau_{1}}^{2}} + 
s_{\tau}^{2} \log \frac{\tilde{m}_{\tau_{1}}^{2}}{\tilde{m}_{\tau_{2}}^{2}} - 
3 c_{\tau}^{2} s_{\tau}^{2} \log \frac{(\tilde{m}_{\tau_{1}}^{2}+\tilde{m}_{\tau_{2}}^{2})^{2}}
{4 \tilde{m}_{\tau_{1}}^{2} \tilde{m}_{\tau_{2}}^{2}} \right\}, \\
\nonumber \\
\displaystyle T_{\tilde{l}} &=& \sum_{\tilde{l}}  \frac{1}{8 \pi} 
\frac{1}{s_{{\scriptscriptstyle W}}^{2} m_{{\scriptscriptstyle W}}^{2}} \left\{
c_{\tau}^{2}  h(\tilde{m}_{\nu}^{2}, \tilde{m}_{\tau_{1}}^{2}) + 
s_{\tau}^{2}  h(\tilde{m}_{\nu}^{2}, \tilde{m}_{\tau_{2}}^{2}) + 
s_{\tau}^{2} c_{\tau}^{2}  h(\tilde{m}_{\tau_{1}}^{2}, \tilde{m}_{\tau_{2}}^{2})\right\},\\
\nonumber \\
\displaystyle U_{\tilde{l}} &=& \sum_{\tilde{l}}  \frac{1}{12 \pi} \left\{
c_{\tau}^{2} \log \frac{4 \tilde{m}_{\nu}^{2} \tilde{m}_{\tau_{1}}^{2}}
{(\tilde{m}_{\nu}^{2} + \tilde{m}_{\tau_{1}}^{2})^{2}} + 
s_{\tau}^{2} \log \frac{4 \tilde{m}_{\nu}^{2} \tilde{m}_{\tau_{2}}^{2}}
{(\tilde{m}_{\nu}^{2} + \tilde{m}_{\tau_{2}}^{2})^{2}} - s_{\tau}^{2} c_{\tau}^{2} 
\log \frac{4 \tilde{m}_{\tau_{1}}^{2} \tilde{m}_{\tau_{2}}^{2}}
{(\tilde{m}_{\tau_{1}}^{2} + \tilde{m}_{\tau_{2}}^{2})^{2}} \right\},
\end{eqnarray}
\\
where the last two equations have been obtained considering in addition:\\
\\
\hspace*{3cm} $\displaystyle |\tilde{m}_{\nu}^{2} - \tilde{m}_{\tau_{i}}^{2}| \ll 
|\tilde{m}_{\nu}^{2} + \tilde{m}_{\tau_{i}}^{2}|$ \hspace*{0.5cm} $i=1,2$.

\subsection{Neutralinos and Charginos:}

If $\displaystyle |\tilde{M}_{1}^{+^{2}} - \tilde{M}_{2}^{+^{2}}| \ll 
|\tilde{M}_{1}^{+^{2}} + \tilde{M}_{2}^{+^{2}}|, \hspace*{0.2cm}
|\tilde{M}_{i}^{o^{2}} - \tilde{M}_{j}^{o^{2}}| \ll |\tilde{M}_{i}^{o^{2}} + \tilde{M}_{j}^{o^{2}}|$ 
\hspace*{0.4cm} $i,j=1,2,3,4$:
\begin{eqnarray}
\label{eq:Sneuchar}
\displaystyle 
S_{\tilde{\chi}} &=& \frac{-1}{3 \pi} \log \frac{2 \tilde{M}_{2}^{+^{2}}}
{\tilde{M}_{3}^{o^{2}} + \tilde{M}_{4}^{o^{2}}}, \\
\nonumber \\
\label{eq:Tneuchar}
\displaystyle 
T_{\tilde{\chi}} &=& \frac{1}{4 \pi m^{2}_{\scriptscriptstyle W} {\sw}^{2}} 
\left\{ -2{(\tilde{M}_{1}^{+} - \tilde{M}_{2}^{o})}^{2}
\log \frac{\tilde{M}_{1}^{+^{2}} + \tilde{M}_{2}^{o^{2}}}{2{\mu}_{o}^{2}} -
\frac{1}{2} {(\tilde{M}_{2}^{+} - \tilde{M}_{3}^{o})}^{2}
\log \frac{\tilde{M}_{2}^{+^{2}} + \tilde{M}_{3}^{o^{2}}}{2{\mu}_{o}^{2}}\right.\nonumber \\
&-& \displaystyle \left.\frac{1}{2} {(\tilde{M}_{2}^{+} - \tilde{M}_{4}^{o})}^{2}
\log \frac{\tilde{M}_{2}^{+^{2}} + \tilde{M}_{4}^{o^{2}}}{2{\mu}_{o}^{2}} +
\frac{1}{2} {(\tilde{M}_{3}^{o} - \tilde{M}_{4}^{o})}^{2}
\log \frac{\tilde{M}_{3}^{o^{2}} + \tilde{M}_{4}^{o^{2}}}{2{\mu}_{o}^{2}} \right\},\\
\nonumber \\
\label{eq:Uneuchar}
\displaystyle 
U_{\tilde{\chi}} &=& \frac{4}{3{\sw}^{2}} \log \frac{\tilde{M}_{2}^{o^{2}} + \tilde{M}_{1}^{+^{2}}}{2\tilde{M}_{1}^{+^{2}}} +
\frac{1}{3{\sw}^{2}} \log \left[ 
\frac{(\tilde{M}_{3}^{o^{2}} + \tilde{M}_{2}^{+^{2}})(\tilde{M}_{4}^{o^{2}} + \tilde{M}_{2}^{+^{2}})}
{2\tilde{M}_{2}^{+^{2}}(\tilde{M}_{3}^{o^{2}} + \tilde{M}_{4}^{o^{2}})}\right],
\end{eqnarray}
where the expressions (\ref{eq:Tneuchar}) and (\ref{eq:Uneuchar}) are valid considering together
the above two conditions and also: \\
\hspace*{2.0cm} $\displaystyle  |M_{i}^{+^{2}} - M_{j}^{o^{2}}| \ll |M_{i}^{+^{2}} + M_{j}^{o^{2}}|,$ 
\hspace*{0.5cm} $i=1,2$; \hspace*{0.3cm} $j=1,2,3,4$. 

Here we have used the values of the coupling matrices $O_{{\scriptscriptstyle L,R}}, {O'}_{{\scriptscriptstyle L,R}}, 
{O''}_{{\scriptscriptstyle L,R}}$ corresponding to the large neutralinos and 
charginos masses limit that are given in eqs~(\ref{eq:os}). Notice that the above
expressions are valid for both $\mu \ge 0$ and $\mu < 0$.

\subsection{Discussion and comments:}
\begin{itemize}
\item The above expressions for $S, T$ and $U$ are general and depend just on the
  physical sparticle masses and the generic coefficients $c_{f}, s_{f}$
  ($f=t,b,\tau ,... $ and $0\leq c_{f}, s_{f}\leq 1$). The results for these parameters
of the various sectors are finite as they must
 be. The cancellation  of divergences occur between the $\tilde{t}$ and
 $\tilde{b}$ contributions of each generation of squarks, between the $\tilde{\nu}$ and $\tilde{\tau}$
contributions of each generation of sleptons and between the charginos and neutralinos.
\item The corrections to these formulae are always suppressed  by extra factors of the type 
$${\left[\frac{\tilde{m}_{1}^{2}-\tilde{m}_{2}^{2}}{\tilde{m}_{1}^{2}+\tilde{m}_{2}^{2}} 
\right]}^{n}$$
which are forced to be small under our assumption $|\tilde{m}_{1}^{2}-\tilde{m}_{2}^{2}| \ll
|\tilde{m}_{1}^{2}+\tilde{m}_{2}^{2}|$ for the various types of sparticles, and vanish in the infinite 
masses limit.
\item Although the three parameters do not require the same set of conditions on the sparticle 
masses, the physical and realistic situation corresponds to have fixed all the SUSY spectra 
at once, and therefore all these conditions must hold together. Thus, by considering: \\
$|\tilde{m}_{t_{1}}^{2}-\tilde{m}_{t_{2}}^{2}| \ll 
|\tilde{m}_{t_{1}}^{2}+\tilde{m}_{t_{2}}^{2}|$, \hspace*{0.1cm}
$|\tilde{m}_{b_{1}}^{2}-\tilde{m}_{b_{2}}^{2}| \ll 
|\tilde{m}_{b_{1}}^{2}+\tilde{m}_{b_{2}}^{2}|$ and
$|\tilde{m}_{t_{{\scriptscriptstyle i}}}^{2}-\tilde{m}_{b_{{\scriptscriptstyle j}}}^{2}| \ll 
|\tilde{m}_{t_{{\scriptscriptstyle i}}}^{2}+\tilde{m}_{b_{{\scriptscriptstyle j}}}^{2}|,$ \\
$(i,j=1,2)$, is equivalent to say that all the squarks of the same generation have masses of 
similar large size. Similarly, in the sleptons sector the conditions are: \\
\hspace*{1cm} $|\mnu -{\tilde{m}^2}_{{\scriptscriptstyle \tau}_{\scriptscriptstyle i}}|
\ll |\mnu +{\tilde{m}^2}_{{\scriptscriptstyle \tau}_{\scriptscriptstyle i}}|
\hspace*{0.1cm} (i=1,2) \hspace*{0.1cm}$ and $|\mtau-\mtad| \ll |\mtau-\mtad|$ \\
and imply that the sleptons of the same generation have also large masses of 
similar size. The conditions on the charginos and the neutralinos sector are, \\
\hspace*{2cm} $|M_{1}^{+^{2}} - M_{2}^{+^{2}}| \ll |M_{1}^{+^{2}} + M_{2}^{+^{2}}|,$ \\
\hspace*{2cm} $|M_{i}^{0^{2}} - M_{j}^{0^{2}}| \ll |M_{i}^{0^{2}} + M_{j}^{0^{2}}| 
\hspace*{0.35cm}; \hspace*{0.4cm}(i,j=1,2,3,4),$\\
\hspace*{2.0cm} $\displaystyle  |M_{i}^{+^{2}} - M_{j}^{o^{2}}| \ll |M_{i}^{+^{2}} + M_{j}^{o^{2}}| \hspace*{0.2cm},$ 
\hspace*{0.2cm} $(i=1,2; \hspace*{0.2cm} j=1,2,3,4)$,\\
and imply analogously that all the large masses $M_{i}^{+}$ and $M_{j}^{0}$ are comparable.
\end{itemize}
\hspace{0.5cm}Let us now comment on how the decoupling occurs in the various sectors.

Interestingly, in the squarks sector and by looking just at the $S_{\tilde{q}}$ parameter, there is 
apparently no decoupling since the dominant contribution goes as:
\begin{equation}
\displaystyle S_{\tilde{q}} \rightarrow -\sum_{\tilde{q}} \frac{N_{c}}{36 \pi} 
\log \frac{\tilde{m}_{t_{1}}^{2}}{\tilde{m}_{b_{1}}^{2}},
\hspace{2.0cm} (\tilde{m}_{q_{i}}^{2} \gg k^{2})
\end{equation}
which under the corresponding conditions $|\tilde{m}_{t_{1}}^{2}-\tilde{m}_{t_{2}}^{2}| \ll 
|\tilde{m}_{t_{1}}^{2}+\tilde{m}_{t_{2}}^{2}|$ and 
$|\tilde{m}_{b_{1}}^{2}-\tilde{m}_{b_{2}}^{2}| \ll 
|\tilde{m}_{b_{1}}^{2}+\tilde{m}_{b_{2}}^{2}|$ does not vanish in the infinite 
$\tilde{m}_{t_{1}}$ and $\tilde{m}_{b_{1}}$ limit. However, when the 
three parameters $S_{\tilde{q}}, T_{\tilde{q}}$  and $U_{\tilde{q}}$ are analized
together and all the conditions, including the third one,
$|\tilde{m}_{t_{{\scriptscriptstyle i}}}^{2}-\tilde{m}_{b_{{\scriptscriptstyle j}}}^{2}| \ll 
|\tilde{m}_{t_{{\scriptscriptstyle i}}}^{2}+\tilde{m}_{b_{{\scriptscriptstyle j}}}^{2}|$, 
$(i,j=1,2)$, are required together, then the above dominant term in $S_{\tilde{q}}$ 
also vanishes in the infinite squark masses limit as it was expected.

In order to show the decoupling  explicitly one can go a step further and make 
an expansion of $S_{\tilde{q}}, T_{\tilde{q}}$  and $U_{\tilde{q}}$ in powers
 of the proper dimensionless quantities which accordingly must vanish in the
considered infinite mass limit. More explicitly, we define the proper 
expansion parameters in the (third generation) squarks sector as:
$$ \frac{\tilde{m}_{t_{1}}^{2} - \tilde{m}_{b_{1}}^{2}}
{\tilde{m}_{t_{1}}^{2} + \tilde{m}_{b_{1}}^{2}},
\hspace{0.5cm} \frac{\tilde{m}_{b_{1}}^{2} - \tilde{m}_{b_{2}}^{2}}
{\tilde{m}_{b_{1}}^{2} + \tilde{m}_{b_{2}}^{2}},
\hspace{0.5cm} \frac{\tilde{m}_{t_{i}}^{2} - \tilde{m}_{b_{j}}^{2}}
{\tilde{m}_{t_{i}}^{2} + \tilde{m}_{b_{j}}^{2}},
\hspace{0.8cm} (i,j=1,2).$$
\hspace*{0.5cm} In terms of these parameters we get the following dominant terms in
the power expansions of  $S_{\tilde{q}}, T_{\tilde{q}}$  and $U_{\tilde{q}}$:
\begin{eqnarray}
\displaystyle 
S_{\tilde{q}} &\rightarrow& -\sum_{\tilde{q}} \frac{N_{c}}{18 \pi} \left\{ \left(
\frac{\tilde{m}_{t_{1}}^{2} - \tilde{m}_{b_{1}}^{2}}
{\tilde{m}_{t_{1}}^{2} + \tilde{m}_{b_{1}}^{2}} \right) + 
s_{b}^{2} \left( \frac{\tilde{m}_{b_{1}}^{2} - \tilde{m}_{b_{2}}^{2}}
{\tilde{m}_{b_{1}}^{2} + \tilde{m}_{b_{2}}^{2}} \right) -  
s_{t}^{2} \left( \frac{\tilde{m}_{t_{1}}^{2} - \tilde{m}_{t_{2}}^{2}}
{\tilde{m}_{t_{1}}^{2} + \tilde{m}_{t_{2}}^{2}} \right) \right.\nonumber \\ 
 && \displaystyle \left. -3 c_{t}^{2} s_{t}^{2} \left(
\frac{\tilde{m}_{t_{1}}^{2} - \tilde{m}_{t_{2}}^{2}}
{\tilde{m}_{t_{1}}^{2} + \tilde{m}_{t_{2}}^{2}} \right)^{2} - 
3 c_{b}^{2} s_{b}^{2} \left( \frac{\tilde{m}_{b_{1}}^{2} - \tilde{m}_{b_{2}}^{2}}
{\tilde{m}_{b_{1}}^{2} + \tilde{m}_{b_{2}}^{2}} \right)^{2} \right\}+ 
O\left(\frac{\tilde{m}_{i}^{2} - \tilde{m}_{j}^{2}}{\tilde{m}_{i}^{2} 
+ \tilde{m}_{j}^{2}}\right)^{3} , \\
\nonumber \\ 
\displaystyle T_{\tilde{q}} &\rightarrow& \sum_{\tilde{q}} \frac{N_{c}}{16 \pi}
\frac{1}{s_{{\scriptscriptstyle W}}^{2} m_{{\scriptscriptstyle W}}^{2}}
\left\{ c_{t}^{2} c_{b}^{2} (\tilde{m}_{t_{1}}^{2} - \tilde{m}_{b_{1}}^{2}) \left(
\frac{\tilde{m}_{t_{1}}^{2} - \tilde{m}_{b_{1}}^{2}}{\tilde{m}_{t_{1}}^{2} + \tilde{m}_{b_{1}}^{2}} \right) + 
c_{t}^{2} s_{b}^{2} (\tilde{m}_{t_{1}}^{2} - \tilde{m}_{b_{2}}^{2}) \left(
\frac{\tilde{m}_{t_{1}}^{2} - \tilde{m}_{b_{2}}^{2}}{\tilde{m}_{t_{1}}^{2} 
+ \tilde{m}_{b_{2}}^{2}} \right) \right. \nonumber \\ 
&& \displaystyle 
-c_{t}^{2} s_{t}^{2} (\tilde{m}_{t_{1}}^{2} - \tilde{m}_{t_{2}}^{2}) \left(
\frac{\tilde{m}_{t_{1}}^{2} - \tilde{m}_{t_{2}}^{2}}{\tilde{m}_{t_{1}}^{2} + \tilde{m}_{t_{2}}^{2}} \right) +  
s_{t}^{2} c_{b}^{2} (\tilde{m}_{t_{2}}^{2} - \tilde{m}_{b_{1}}^{2}) \left(
\frac{\tilde{m}_{t_{2}}^{2} - \tilde{m}_{b_{1}}^{2}}{\tilde{m}_{t_{2}}^{2} 
+ \tilde{m}_{b_{1}}^{2}} \right) \nonumber \\
&& \displaystyle \left. 
+s_{t}^{2} s_{b}^{2} (\tilde{m}_{t_{2}}^{2} - \tilde{m}_{b_{2}}^{2}) \left(
\frac{\tilde{m}_{t_{2}}^{2} - \tilde{m}_{b_{2}}^{2}}{\tilde{m}_{t_{2}}^{2} + \tilde{m}_{b_{2}}^{2}} \right) -
s_{b}^{2} c_{b}^{2} (\tilde{m}_{b_{1}}^{2} - \tilde{m}_{b_{2}}^{2}) \left(
\frac{\tilde{m}_{b_{1}}^{2} - \tilde{m}_{b_{2}}^{2}}{\tilde{m}_{b_{1}}^{2} 
+ \tilde{m}_{b_{2}}^{2}} \right) \right\} \nonumber \\
&& \displaystyle + O \left[ \left( \frac{\tilde{m}_{i}^{2} - \tilde{m}_{j}^{2}}
{M_{{\scriptscriptstyle W}}^{2}} \right) 
\left( \frac{\tilde{m}_{i}^{2} - \tilde{m}_{j}^{2}}{\tilde{m}_{i}^{2} 
+ \tilde{m}_{j}^{2}} \right)^{2} \right],\\
\nonumber \\
\displaystyle U_{\tilde{q}} &\rightarrow& \sum_{\tilde{q}} \frac{N_{c}}{12 \pi} 
\left\{ -c_{t}^{2} c_{b}^{2}  \left( \frac{\tilde{m}_{t_{1}}^{2} - 
\tilde{m}_{b_{1}}^{2}}{\tilde{m}_{t_{1}}^{2} + \tilde{m}_{b_{1}}^{2}}
\right)^{2} - c_{t}^{2} s_{b}^{2} \left( \frac{\tilde{m}_{t_{1}}^{2} - 
\tilde{m}_{b_{2}}^{2}}{\tilde{m}_{t_{1}}^{2} + \tilde{m}_{b_{2}}^{2}}
\right)^{2} - s_{t}^{2} s_{b}^{2} \left( \frac{\tilde{m}_{t_{2}}^{2} - \tilde{m}_{b_{2}}^{2}}
{\tilde{m}_{t_{2}}^{2} + \tilde{m}_{b_{2}}^{2}} \right)^{2}\right.\nonumber \\
&&\displaystyle \left. -s_{t}^{2} c_{b}^{2} \left(
\frac{\tilde{m}_{t_{2}}^{2} - \tilde{m}_{b_{1}}^{2}}{\tilde{m}_{t_{2}}^{2} + 
\tilde{m}_{b_{1}}^{2}} \right)^{2} +   c_{t}^{2} s_{t}^{2} \left(
\frac{\tilde{m}_{t_{1}}^{2} - \tilde{m}_{t_{2}}^{2}}{\tilde{m}_{t_{1}}^{2} + 
\tilde{m}_{t_{2}}^{2}} \right)^{2} + s_{b}^{2} c_{b}^{2} \left(
\frac{\tilde{m}_{b_{1}}^{2} - \tilde{m}_{b_{2}}^{2}}{\tilde{m}_{b_{1}}^{2} + 
\tilde{m}_{b_{2}}^{2}} \right)^{2} \right\} + O \left(
\frac{\tilde{m}_{i}^{2} - \tilde{m}_{j}^{2}}{\tilde{m}_{i}^{2} + \tilde{m}_{j}^{2}} 
\right)^{4}. \nonumber \\
\end{eqnarray}
\hspace*{0.5cm}First, we see that in the limit of exact custodial ${SU(2)}_{\scriptscriptstyle V}$ 
symmetry, which corresponds to $\tilde{m}_{t_{1}}=\tilde{m}_{b_{1}} \equiv \tilde{m}_{1}$, $\tilde{m}_{t_{2}}=\tilde{m}_{b_{2}} \equiv 
\tilde{m}_{2}$ and $c_{t}=c_{b} \equiv c$,  $s_{t}=s_{b} \equiv s$, both $T_{\tilde{q}}$ and $U_{\tilde{q}}$
vanish as it is expected, whereas $S_{\tilde{q}}$ goes as,
\begin{equation}
\begin{array}{l}
\displaystyle
S_{\tilde{q}} \rightarrow  \sum_{\tilde{q}} \frac{N_{c}}{3 \pi} c^{2} s^{2} 
\left( \frac{\tilde{m}_{1}^{2} - \tilde{m}_{2}^{2}}{\tilde{m}_{1}^{2} + \tilde{m}_{2}^{2}} \right)^{2} + 
O \left( \frac{\tilde{m}_{1}^{2} - \tilde{m}_{2}^{2}}{\tilde{m}_{1}^{2} + \tilde{m}_{2}^{2}} \right)^{4}
\end{array}
\end{equation}
\hspace*{0.5cm}Second, the above formulae show that the decoupling indeed occurs in the three parameters 
since they go to zero as some power of the parameters 
$\left[ ({\tilde{m}_{i}^{2} - \tilde{m}_{j}^{2}})/({\tilde{m}_{i}^{2} + \tilde{m}_{j}^{2}}) \right]$
which vanish in the infinite masses limit, $\tilde{m}_{i}^{2}, \tilde{m}_{j}^{2} \rightarrow \infty$,
with $|\tilde{m}_{i}^{2} - \tilde{m}_{j}^{2}| \ll |\tilde{m}_{i}^{2} + 
\tilde{m}_{j}^{2}|$. Besides, the decoupling is much faster in $U_{\tilde{q}}$
than in $S_{\tilde{q}}$ and $T_{\tilde{q}}$. These results confirm the numerical analyses performed 
in the literature and agree with the qualitative behaviour discussed in \cite{CHA}, \cite{HA2},
\cite{SOLA}.

However, we would like to emphasize once more that, contrary to most of the 
studies in the literature (with the exception of those on $\Delta \rho$), our results
for $S, T$ and $U$ in this section
are model independent and do not make any reference on whether there is or not a 
common effective scale of supersymmetry breaking. We have neither assumed here
the common assumption for the MSSM masses (see section 2) that 
$(\tilde{m}_{i}^{2} - \tilde{m}_{j}^{2}) \sim O(m_{\scriptscriptstyle Z}^{2})$
or $\sim O(m_{\scriptscriptstyle f}^{2})$. Generically speaking, these mass differences could 
well be larger than the gauge boson or fermion masses and therefore  the rapidity  of decoupling 
can vary from one SUSY breaking model to another. In particular by comparing 
$S_{\tilde{q}}$ and $T_{\tilde{q}}$ in the non-custodial symmetric case, we see that their 
dominant contributions go respectively as 
\hspace*{5cm} $$\left( \frac{\tilde{m}_{i}^{2} - \tilde{m}_{j}^{2}}
{\tilde{m}_{i}^{2} + \tilde{m}_{j}^{2}} \right)  \hspace*{0.4cm} {\rm and}
\hspace*{0.4cm} \left( \frac{\tilde{m}_{i}^{2} - \tilde{m}_{j}^{2}}{m_{\scriptscriptstyle W}^{2}} \right)
\left( \frac{\tilde{m}_{i}^{2} - \tilde{m}_{j}^{2}}{\tilde{m}_{i}^{2} + \tilde{m}_{j}^{2}} \right).$$\\
Therefore the later could be enhanced (suppressed) respect 
to the first one if \\
$\hspace*{4.5cm} (\tilde{m}_{i}^{2} - \tilde{m}_{j}^{2}) \gg 
{m_{\scriptscriptstyle W}}^{2} \hspace*{0.5cm} 
((\tilde{m}_{i}^{2} - \tilde{m}_{j}^{2}) \ll {m_{\scriptscriptstyle W}}^{2})$.

In summary, from our previous analysis we can infer with complete generality that
the largest contributions from the squarks sector to $S_{\tilde{q}}, T_{\tilde{q}}$
and $U_{\tilde{q}}$ come from the squarks pairs with the largest 
$\left[ ({\tilde{m}_{i}^{2} - \tilde{m}_{j}^{2}})/
({\tilde{m}_{i}^{2} + \tilde{m}_{j}^{2}}) \right]$ values.

Parallel results for the sleptons sector can be obtained by making the following 
replacements in the above formulae: $\tilde{q} \rightarrow \tilde{l},
N_{c} \rightarrow 1, \tilde{m}_{t_{1}} \rightarrow  \tilde{m}_{\nu},
\tilde{m}_{b_{1}} \rightarrow  \tilde{m}_{\tau_{1}}, 
\tilde{m}_{b_{2}} \rightarrow  \tilde{m}_{\tau_{2}}, 
c_{t} \rightarrow 1, s_{t} \rightarrow 0, c_{b} \rightarrow c_{\tau}$ and
$s_{b} \rightarrow s_{\tau}$ .

With regard to the neutralinos and charginos sector and by looking at 
eqs.(\ref{eq:Sneuchar} - \ref{eq:Uneuchar}) we first notice that, in the large masses 
limit, the first chargino $\chiu$ and the two first neutralinos
${\tilde{\chi}^{o}}_{{\scriptscriptstyle 1}}$ and ${\tilde{\chi}^{o}}_{{\scriptscriptstyle 2}}$
decouple completely in the $S$ parameter. These are precisely the chargino and
neutralinos, that in the large masses limit become predominantly gauginos. The
decoupling of the other eigenstates $\chid, {\tilde{\chi}^{o}}_{{\scriptscriptstyle 3}}$
and ${\tilde{\chi}^{o}}_{{\scriptscriptstyle 4}}$ in $S$ is not evident at a first sight, 
since it depends on the relative size of the $\chid$ mass with respect to the masses 
of the neutralinos ${\tilde{\chi}^{o}}_{{\scriptscriptstyle 3}}$
and ${\tilde{\chi}^{o}}_{{\scriptscriptstyle 4}}$. However, we have seen in the
first section, that in the large masses limit, their corresponding squared mass eigenvalues approach
to a common value ${\mu}^{2}$ and, in consequence, the decoupling in $S_{\tilde{\chi}}$ does
finally occur. Notice that this result is not model dependent either, since this common
value ${\mu}^{2}$ is the unique squared mass parameter that is allowed by supersymmetry 
to be present at the  Lagrangian level and do not depend on the particular assumed SUSY 
breaking mechanism. Similarly, in the $T_{\tilde{\chi}}$ and $U_{\tilde{\chi}}$ parameters 
the decoupling occurs exactly if the mass eigenvalues in the large mass limit are considered,
i.e, ${\tilde{M}}_{1}^{+^{2}} \rightarrow M_{2}^{2}, \hspace*{0.1cm}
{\tilde{M}}_{2}^{+^{2}}\rightarrow \mu^{2}, \hspace*{0.1cm}{\tilde{M}}_{1}^{o^{2}}
\rightarrow M_{1}^{2}, \hspace*{0.1cm}{\tilde{M}}_{2}^{o^{2}}\rightarrow M_{2}^{2}$
and ${\tilde{M}}_{3}^{o^{2}}={\tilde{M}}_{4}^{o^{2}} \rightarrow \mu^{2}$. 

Notice that the discussion on how the decoupling occurs in the particular case of the
MSSM with soft SUSY breaking terms is evident from the above analisis together with the 
arguments given in section 2. In conclusion, the decoupling of sparticles in $S, T$ and 
$U$ in our asymptotic limit takes place.

Finally, we discuss in the rest of this section the alternative possibility for the squarks 
sector of the MSSM:
$$\left|\frac{\tilde{m}_{t_{1}}^{2} - \tilde{m}_{t_{2}}^{2}}
{\tilde{m}_{t_{1}}^{2} + \tilde{m}_{t_{2}}^{2}}\right|\sim O(1)\,\,, \hspace*{1.5cm}
\left|\frac{\tilde{m}_{b_{1}}^{2} - \tilde{m}_{b_{2}}^{2}}{\tilde{m}_{b_{1}}^{2} 
+ \tilde{m}_{b_{2}}^{2}}\right|\sim O(1)\,\,,$$
which has been considered in detail in section 2. As we have explained there,
this case implies that $s_{t}$ and $s_{b}$ go to zero in the asymptotic limit of 
$M_{\scriptscriptstyle SUSY} \rightarrow \infty$. From the exact results of 
$S, T$ and $U$ parameters for the squarks sector which can be taken, for instance, from 
eqs. (B.3), (B.6) and (B.9) of ref.~\cite{CHA}, and by taking the limit 
$s_{t},\, s_{b} \rightarrow 0$ we obtain:
\begin{eqnarray}
S &\longrightarrow & -\frac{N_{c}}{36 \pi}\,\log
\frac{\tilde{m}_{t_{1}}^{2}}{\tilde{m}_{b_{1}}^{2}}\,,\nonumber\\
T &\longrightarrow&  \frac{N_{c}}{16 \pi}\,
\frac{1}{s_{{\scriptscriptstyle W}}^{2} m_{{\scriptscriptstyle W}}^{2}} 
\,g(\tilde{m}_{t_{1}}, \tilde{m}_{b_{1}})\,\,,\nonumber\\
U &\longrightarrow& \frac{N_{c}}{12 \pi} 
\,f(\tilde{m}_{t_{1}}, \tilde{m}_{b_{1}})\,,
\end{eqnarray}
where,
\begin{eqnarray}
g({m}_{1}, {m}_{2}) &\equiv& {m}_{1}^{2}+ {m}_{2}^{2} -
2 \frac{{m}_{1}^{2}{m}_{2}^{2}}{{m}_{1}^{2} - {m}_{2}^{2}} 
\log \frac{{m}_{1}^{2}}{{m}_{2}^{2}}\,,\nonumber\\
\nonumber\\
f({m}_{1}, {m}_{2}) &\equiv& -\frac{5}{3}+
\frac{4{m}_{1}^{2}{m}_{2}^{2}}{({m}_{1}^{2}-{m}_{2}^{2})^{2}}
+\frac{({m}_{1}^{2}+{m}_{2}^{2})({m}_{1}^{4}+{m}_{2}^{4}-4{m}_{1}^{2}{m}_{2}^{2})}
{({m}_{1}^{2}-{m}_{2}^{2})^{3}} \log \frac{{m}_{1}^{2}}{{m}_{2}^{2}}\,,
\end{eqnarray}
with $g({m}_{1}, {m}_{2})\approx 0 \,$ and $f({m}_{1}, {m}_{2})\approx 0 \,$
if $\,{m}_{1}\approx {m}_{2}$.

Notice that only squarks $\tu$ and $\bu$ remain in the above expressions and they are precisely 
the squarks whose masses in the large masses limit do always get close, namely 
$\tilde{m}_{t_{1}}^{2} \approx M_{{\scriptscriptstyle {\tilde Q}}}^{2}$
and $\tilde{m}_{b_{1}}^{2} \approx M_{{\scriptscriptstyle {\tilde Q}}}^{2}$.

From the above expressions and by following the discussion presented in section 2 we conclude that 
the three parameters also vanish in this case as it was expected. It completes our proof of 
decoupling of sparticles in these observables.


\section{Conclusions}
\label{sec:con}

\hspace*{0.5cm} Althought there are indications that the common assumption of decoupling of
heavy supersymmetric particles in the MSSM leading to the SM as the remaining
low energy effective theory is correct a formal proof is still lacking.

This formal proof should be performed along the lines stated in the Decoupling
Theorem and by means of the powerful techniques of the Effective Field Theories.
The computation of the effective action for the standard particles which results
by integrating out all the heavy supersymmetric particles will provide the answer
to this question. If the contribution from the heavy sparticles to the effective 
action can be absorbed into redefinitions of the Standard Model parameters or they 
are suppressed by inverse powers of the heavy sparticles masses, then the decoupling 
will be demonstrated.

In this paper we have computed the two-point functions part of this effective 
action for the electroweak gauge bosons, $W^{\pm}, Z$ and $\gamma$, that results by 
integrating out the squarks, sleptons, charginos and neutralinos to one loop level.

We have analized carefully the large SUSY masses limit of these two-point Green
functions and we have presented analytical results for them as well as for the self-
energies and the $R^{X\, Y}$ functions, which are valid in that limit. These formulae are 
given in terms of the sparticle masses and, therefore, they are general. Namely, they do not depend on the 
particular choice for the soft-breaking terms. In our opinion, it is more convenient for 
the analysis of the phenomenon  of decoupling to use the physical sparticle masses
themselves, being the proper parameters, rather than some other possible mass parameters 
of the MSSM as, for instance, the $\mu$-parameter or the soft-SUSY breaking 
parameters.

The results for the two-point functions of the electroweak gauge bosons indicate that there 
is indeed decoupling of squarks, sleptons , gauginos and neutralinos in the limit 
where the sparticle masses are all large as compared to the $W^{\pm}$ and $Z$
masses and the external momentum. In taking the large mass limit we have not
assumed exact universality of the masses but we have always worked under the 
plausible assumption that the differences of their squared masses are much smaller than their sums. As can 
be seen from the formulae of Appendix C, all the remaining contributions to the two
point functions from sparticles can be absorbed into redefinitions  of the Standard 
Model parameters $m_{\scriptscriptstyle Z}, m_{\scriptscriptstyle W}$ and $e$ and the 
gauge bosons wave  functions. The contributions which are not shown in our formulae all 
vanish in the asymptotic limit of very large sparticle masses.
     
We have shown that the decoupling of sparticles also takes place in the 
$S, T$ and $U$ parameters, and we have presented explicit formulae for these parameters, 
which illustrate analytically how this decoupling occurs.
 
Finally, we have explored to what extent the hypothesis of generation of SUSY masses by 
soft-SUSY breaking terms is relevant for decoupling and we have found instead that the 
requirement of $SU(3)_{\rm c} \times \gs$ gauge invariance of the explicit mass terms by itself is 
sufficient to get it.

\vskip 3.0cm
{\bf Acknowledgements}\\
\\
We are indebted to A.Casas for his advising in theoretical aspects of the MSSM and for 
interesting discussions. We thank L.Diaz-Cruz for participating in the very 
early stages of this work. M.J.H wishes to acknowledge D.Espriu, H.Haber, 
W.Hollik and J.Sol{\`a} for interesting discussions on the subject of decoupling.
A.D thanks the CERN-TH division for his kind hospitality during the last part 
of this work. This work has been partially supported by the Spanish Ministerio de Educaci\'on y Ciencia 
under projects CICYT AEN96-1664 and AEN93-0776, and the fellowship AP95 00503301.


\input{apendice.A}
\vspace{0.5cm} 
\begin{large}
{\bf Appendix B.}
\end{large}

\vspace{0.3cm}
\setcounter{equation}{0}
\renewcommand{\theequation}{B.\arabic{equation}}
\hspace{0.5cm}In this appendix, we give the results of the one loop integrals of eqs.~(\ref{eq:int0}-
\ref{eq:intM}) in the large masses limit and include an example of the large mass 
techniques used in the calculation.

We start by giving the results of the integrals. Here and from now on,
\begin{equation}
\hspace*{0.6cm} \displaystyle {\Delta}_\epsilon=\frac{2}{\epsilon }-{\gamma }_{\epsilon} 
+\log (4\pi) \hspace*{0.2cm}, \hspace*{0.2cm} \epsilon = 4-D\,,
\end{equation}
and $\mu_{o}$ is the usual mass scale of dimensional regularization. 

The integrals are given by:
\begin{equation}
\displaystyle I_{0}(\mfa) = \frac{i}{16 \pi^{2}} 
\left( {\Delta}_\epsilon+1-\log \frac{\mfa}{\mu_{o}^{2}} \right) {\mfa}\\
\end{equation}
This is a well known result in dimensional regularization and is exact for all $\mfa$.

$\bullet$ If $\mfa, \mfb \gg k^{2}$ and $|\mfa-\mfb| \ll |\mfa+\mfb|$:
\begin{eqnarray}
\displaystyle A^{ab}_{{\scriptscriptstyle f}} (k, \tilde{m}_{f_{a}}, \tilde{m}_{f_{b}})
&=&  \frac{i}{16 \pi^{2}} \left\{
(\mfa+\mfb) \left({\Delta}_\epsilon+1-\log \frac{(\mfa+\mfb)}{2\mu_{o}^{2}}\right) -
\frac{1}{3} k^{2}  \left({\Delta}_\epsilon-\log 
\frac{(\mfa+\mfb)}{2\mu_{o}^{2}}\right)\right\} \nonumber \\
\nonumber \\
\displaystyle B^{ab}_{{\scriptscriptstyle f}} (k, \tilde{m}_{f_{a}}, 
\tilde{m}_{f_{b}})&=& \frac{i}{16 \pi^{2}} 
\frac{1}{3} \left({\Delta}_\epsilon-\log \frac{(\mfa+\mfb)}{2\mu_{o}^{2}}\right)
\end{eqnarray}

$\bullet$ If $\mi, \mj \gg k^{2}$ and $|\mi-\mj| \ll |\mi+\mj|$:\\
\begin{eqnarray}
\displaystyle I^{ij} (k, \tilde{M}_{i}, \tilde{M}_{j}) &=& \frac{i}{16 \pi^{2}} 
\left( {\Delta}_\epsilon-\log \frac{(\mi+\mj)}{2\mu_{o}^{2}}\right)\,,\nonumber 
\\
\displaystyle I_{\mu \nu}^{ij} (k, \tilde{M}_{i}, \tilde{M}_{j}) &=& \frac{i}{16 \pi^{2}} \left\{
\frac{1}{4} (\mi+\mj) g_{\mu \nu} \left( {\Delta}_\epsilon+1-\log \frac{(\mi+\mj)}{2\mu_{o}^{2}}\right)
-\frac{k^{2}}{12} g_{\mu \nu} \left( {\Delta}_\epsilon-\log \frac{(\mi+\mj)}{2\mu_{o}^{2}}\right)\right.\nonumber\\
&&
\displaystyle +\left.\frac{1}{3} k_{\mu} k_{\nu} \left( {\Delta}_\epsilon-\log 
\frac{(\mi+\mj)}{2\mu_{o}^{2}} \right) \right\}\,,\nonumber 
\\
\displaystyle {I'}^{i\,j}_{\mu\, \nu} (k, \tilde{M}_{i}, \tilde{M}_{j}) &=& 
- \frac{i}{16 \pi^{2}} \frac{1}{2} k_{\mu} k_{\nu}
\left( {\Delta}_\epsilon-\log \frac{(\mi+\mj)}{2\mu_{o}^{2}} \right)\,,\nonumber 
\\
\displaystyle C^{ij} (k, \tilde{M}_{i}, \tilde{M}_{j}) &=& \frac{i}{16 \pi^{2}} 
\left\{-(\mi+\mj) \left( {\Delta}_\epsilon-\log \frac{(\mi+\mj)}{2\mu_{o}^{2}}\right) +
\frac{2}{3} k^{2} \left({\Delta}_\epsilon-\frac{1}{2} -\log 
\frac{(\mi+\mj)}{2\mu_{o}^{2}}\right)\right\}\,,\nonumber 
\\
\displaystyle D^{ij} (k, \tilde{M}_{i}, \tilde{M}_{j}) &=& 
- \frac{i}{16 \pi^{2}} \frac{2}{3} 
\left( {\Delta}_\epsilon-\log \frac{(\mi+\mj)}{2\mu_{o}^{2}} \right)\,,\nonumber 
\\
\displaystyle {\hat I}^{i\,j} (k, \tilde{M}_{i}, \tilde{M}_{j}) &=& 
\frac{i}{16 \pi^{2}} \left\{ \frac{1}{2} (\mi+\mj) \left({\Delta}_\epsilon-
\log \frac{(\mi+\mj)}{2\mu_{o}^{2}}\right)+ \frac{1}{6} k^{2} \right. \nonumber\\
&&
\displaystyle
- \left. \frac{1}{2} (\tilde{M}_{i}-\tilde{M}_{j})^{2} \left({\Delta}_\epsilon-\log \frac{(\mi+\mj)}{2\mu_{o}^{2}}
\right) \right\}\,.
\end{eqnarray}

The corrections to these formulas are suppressed by inverse powers of $(\mfa+\mfb)$
or $(\mi+\mj)$ respectively and vanish in the infinite masses limit. For the 
calculation of these integrals we have used the m-Theorem \cite{GMR}.

{\bf m-Theorem:}
Consider the integral,
$$I(p,m) = m^{\beta} \int d^{4}k 
\frac{M(k)}{\prod_{i} {(l_{i}^{2}+m_{i}^{2})}^{n_{i}}}$$
where $l_{i}=k+\sum_{i=1}^{E}  b_{ij}p_{j}$, being $p_{j}$ the external
momenta which lay in a bounded subdomain of $\Re^{4}$, $m_{i}=0$ or m, 
$\beta$ is an arbitrary real number and $M(k)$ is a monomial in the components
of $k$.\\
If:\\
the integral $I(p,m)$ is both UV and IR convergent by power counting at
non-exceptional external momenta and if $$d-w<0,$$ being $d$ the mass dimension of
$I(p,m)$ and $w$ the minimum of zero and the infrared divergence degree of $I(p,m)$ at zero
external momenta,\\
then:\\
$I(p,m)$ goes to zero when $m$ goes to infinity.

As an example of the applications of this theorem let us consider the following
integral:
\begin{equation}
\label{eq:exint}
I_{\mu \nu}(k, m) =  \int d\widehat{q} 
\frac{q_{\mu} q_{\nu}}{[(k+q)^2 -m^2] [q^2 - m^2]}
\end{equation}

The superficial degree of UV divergence of $I_{\mu\, \nu}$ is 2 at
$D=4$. The first step is to rearrange algebraically the integrand by using:
\begin{equation}
\label{eq:algebra}
 \frac{1}{[(k+q)^2 -m^2]}= \frac{1}{[q^2 -m^2]}-
\frac{2qk+k^{2}}{[(k+q)^2 -m^2] [q^2 - m^2]}
\end{equation}
in order to lower the degree of UV divergence of the integral corresponding to the
second term in (\ref{eq:algebra}). This algebraic rearrangement must be
applied as many times as necessary until obtaining a convergent integral which 
satisfies the requirements of the m-Theorem. 

By substituting (\ref{eq:algebra}) in (\ref{eq:exint}), one gets:
\begin{eqnarray}
I_{\mu\, \nu}(k, m) &=&  \int d\widehat{q} 
\frac{q_{\mu} q_{\nu}}{[q^2 - m^2]^{2}}-
\int d\widehat{q} 
\frac{(2qk)q_{\mu} q_{\nu}}
{[(k+q)^2 -m^2] {[q^2 - m^2]}^{2}}-
\int d\widehat{q} 
\frac{k^2 q_{\mu} q_{\nu}}
{[(k+q)^2 -m^2] {[q^2 - m^2]}^{2}} \nonumber \\
&\equiv& M_{\mu\, \nu}-N_{\mu\, \nu}-k^2 P_{\mu\, \nu}
\end{eqnarray}
where $M_{\mu\, \nu}$ can be evaluated exactly using the standard techniques of
dimensional regularization.

Let us concentrate in the last term $P_{\mu\, \nu}$ that is logarithmically divergent.
By using again (\ref{eq:algebra}) one gets,
\begin{equation}
P_{\mu\, \nu} =  \int d\widehat{q} 
\frac{q_{\mu} q_{\nu}}{[q^2 - m^2]^{3}}-
\int d\widehat{q} 
\frac{(2qk)q_{\mu} q_{\nu}}
{[(k+q)^2 -m^2] {[q^2 - m^2]}^{3}}-
\int d\widehat{q} 
\frac{k^2 q_{\mu} q_{\nu}}
{[(k+q)^2 -m^2] {[q^2 - m^2]}^{3}} 
\end{equation}
\\
Now, the last two integrals are convergent at $D=4$ and satisfy the requirements
of the m-Theorem. Therefore, these two terms go to zero when $D\rightarrow
4$ and $m \rightarrow \infty$. The first term can be evaluated exactly using the
standard techniques obtaining:
\begin{equation}
\displaystyle P_{\mu\, \nu} = (\frac{i}{16 \pi^{2}}) \frac{1}{4}g_{\mu \nu}
\left( {\Delta}_\epsilon-\log \frac{m^{2}}{\mu_{o}^{2}} \right) 
\\
\end{equation}     
Finally, by using the same procedure to compute $N_{\mu\, \nu}$, one gets:
\begin{equation}
\displaystyle I_{\mu\, \nu}(k, m) = \frac{i}{16 \pi^{2}} 
\left\{ \frac{1}{2} g_{\mu \nu} \left({\Delta}_\epsilon+1-\log \frac{m^2}{\mu_{o}^{2}}
\right) m^{2}
+ \left( \frac{1}{3}k_{\mu}k_{\nu}-\frac{k^{2}}{12}g_{\mu \nu}\right)
\left({\Delta}_\epsilon-\log \frac{m^{2}}{\mu_{o}^{2}}\right)\right\}
\end{equation}

Another example that we want to show due to the more intricate manipulations is 
${\hat I}^{i\,j}$ of eq.~(\ref{eq:intM}). In this case we define two new parameters 
that do not appear in the other integrals, ${\mbox{\={m}}}^{2} \equiv
(\tilde{M}_{i}-\tilde{M}_{j})^{2}$ and ${\mbox{\={M}}}^{2} \equiv 
(\tilde{M}_{i}+\tilde{M}_{j})^{2}$. In terms of these and the other more commom
parameters, ${\mbox{\^{M}}}^2$ and ${\mbox{\^{m}}}^2$, ${\hat I}^{i\,j}$ can be 
written as:
\begin{equation}
\label{eq:intex}
{\hat I}^{i\,j} = \frac{1}{4} \int d\widehat{q} 
\frac{{\mbox{\={M}}}^{2}-{\mbox{\={m}}}^{2} }
{ [q^2 - {\mbox{\^{M}}}^2+{\mbox{\^{m}}}^2][(k+q)^2 -{\mbox{\^{M}}}^2-{\mbox{\^{m}}}^2 ]}
\end{equation}
where ${\mbox{\^{M}}}^2 \equiv \frac{1}{2}(\mi+\mj)$ and ${\mbox{\^{m}}}^2
\equiv \frac{1}{2}(\mi-\mj)$.

By using the relations between all parameters, ${\mbox{\={m}}}^{2}, {\mbox{\={M}}}^{2}, 
{\mbox{\^{m}}}^2, {\mbox{\^{M}}}^2$, the above integral can be written as: 
$${\hat I}^{i\,j} =I_{1}-\frac{1}{2}I_{2}$$ being,
\begin{equation} 
\label{eq:intdes}
\begin{array}{l}
\displaystyle I_{1} = \int d\widehat{q} 
\frac{{\mbox{\^{M}}}^{2} }
{ [q^2 - {\mbox{\^{M}}}^2+{\mbox{\^{m}}}^2][(k+q)^2 -{\mbox{\^{M}}}^2-{\mbox{\^{m}}}^2 
]}\,,\nonumber
\\
\displaystyle I_{2} = \int d\widehat{q} 
\frac{{\mbox{\={m}}}^{2} }
{ [q^2 - {\mbox{\^{M}}}^2+{\mbox{\^{m}}}^2][(k+q)^2 -{\mbox{\^{M}}}^2-{\mbox{\^{m}}}^2 ]}\,.
\end{array}
\end{equation}

The first step is again to rearrange algebraically the integrand as many times
as necessary until obtaining a convergent integral. The appropriate algebraic
manipulations  in this case are:
\begin{equation}
\label{eq:algebra1}
\displaystyle \frac{1}{[q^2 - {\mbox{\^{M}}}^2 \pm {\mbox{\^{m}}}^2]}=
 \frac{1}{[q^2-{\mbox{\^{M}}}^2]} \mp \frac{{\mbox{\^{m}}}^{2}}
{[q^2 - {\mbox{\^{M}}}^2 \pm {\mbox{\^{m}}}^2] [q^2 -{\mbox{\^{M}}}^2]}
\end{equation}
\begin{equation}
\label{eq:algebra2}
\displaystyle \frac{1}{[(k+q)^2-{\mbox{\^{M}}}^2-{\mbox{\^{m}}}^2]}= \frac{1}{[q^2 -{\mbox{\^{M}}}^2-{\mbox{\^{m}}}^2]}-
\frac{2qk+k^{2}}
{[(k+q)^2 -{\mbox{\^{M}}}^2-{\mbox{\^{m}}}^2] [q^2 -{\mbox{\^{M}}}^2-{\mbox{\^{m}}}^2]}
\end{equation}
In particular for $I_{2}$, by using~(\ref{eq:algebra1}) twice, (\ref{eq:algebra2})
once and by applying the m-Theorem to the convergent integrals we get:
\begin{equation}
\displaystyle I_{2} = \frac{i}{16 \pi^{2}} 
{\mbox{\={m}}}^{2} \left( {\Delta}_\epsilon-\log \frac{{\mbox{\^{M}}}^2}{{\mu}_{o}^{2}} \right) 
\\
\end{equation}
In order to compute $I_{1}$, notice that it can be written as:
\begin{eqnarray}
\label{eq:desI1} 
\displaystyle I_{1} &=& - \int d\widehat{q} 
\frac{1}{[(k+q)^2 -{\mbox{\^{M}}}^2-{\mbox{\^{m}}}^2 ]}+
 \int d\widehat{q} 
\frac{q^2+{\mbox{\^{m}}}^{2} }
{ [q^2 - {\mbox{\^{M}}}^2+{\mbox{\^{m}}}^2][(k+q)^2
    -{\mbox{\^{M}}}^2-{\mbox{\^{m}}}^2 ]}\,\nonumber \\
&=& -I_{1}'+ I_{1}''.
\end{eqnarray}
\hspace{0.5cm}By doing the change of variables $(k+q) \rightarrow q$ and by considering 
(\ref{eq:algebra1}) twice, the first integral of (\ref{eq:desI1}) can be written as:
\begin{equation}
\displaystyle I_{1}'= \int d\widehat{q} 
\frac{1}{[q^2 -{\mbox{\^{M}}}^{2} ]} + 
 \int d\widehat{q}\frac{{\mbox{\^{m}}}^{2}}{{[q^2 -{\mbox{\^{M}}}^2]}^{2}}+
\int d\widehat{q} \frac{{\mbox{\^{m}}}^{4} }
{ [{q^2 - {\mbox{\^{M}}}^2]}^2 [q^2-{\mbox{\^{M}}}^2-{\mbox{\^{m}}}^2 ]}\,
\end{equation} 
where the first two terms can be again evaluated exactly by using the standard techniques
and the last one goes to zero by virtue of the m-Theorem.

Similarly, to compute the second part, $I_{1}''$, we consider (\ref{eq:algebra1}) and
(\ref{eq:algebra2}) as many times as necessary and apply the m-Theorem. We obtain:
\begin{equation}
\displaystyle I_{1} = \frac{i}{16 \pi^{2}} 
\left\{ {\mbox{\^{M}}}^2 \left({\Delta}_\epsilon-\log 
\frac{{\mbox{\^{M}}}^2}{\mu_{o}^{2}} \right) + \frac{k^2}{6} \right\}.
\end{equation}
\hspace*{0.5cm}Finally, by collecting all the results we get,
\begin{eqnarray}
\displaystyle {\hat I}^{i\,j} (k, \tilde{M}_{i}, \tilde{M}_{j}) &=& 
\frac{i}{16 \pi^{2}} \left\{ \frac{1}{2} (\mi+\mj) \left({\Delta}_\epsilon-
\log \frac{(\mi+\mj)}{2\mu_{o}^{2}}\right)+ \frac{1}{6} k^{2}\right. \nonumber
\\ 
\displaystyle  &-& \left.\frac{1}{2} {(\tilde{M}_{i}-\tilde{M}_{j})}^{2}
\left({\Delta}_\epsilon-\log \frac{(\mi+\mj)}{2\mu_{o}^{2}}\right) \right\}.
\end{eqnarray}


\vspace{0.4cm} 
\begin{large}
{\bf Appendix C.}
\end{large}
\vspace{0.3cm}
\setcounter{equation}{0}
\renewcommand{\theequation}{C.\arabic{equation}}

In this appendix we present the asymptotic results in the large sparticle masses limit for the 
two-point functions $\Sigma^{{\scriptscriptstyle XY}}_{\scriptscriptstyle T}$ and 
$\Sigma^{{\scriptscriptstyle XY}}_{\scriptscriptstyle L}$ of eq.~(\ref{eq:tl}).
\begin{eqnarray}
\label{eq:SumAA}
\displaystyle \Sigma^{{\scriptscriptstyle AA}}_{\scriptscriptstyle T} (k) &=& 
\Sigma_{{\scriptscriptstyle T}}^{{\scriptscriptstyle AA}}
 (k)_{\tilde{q}}+\Sigma_{{\scriptscriptstyle T}}^{{\scriptscriptstyle AA}} 
(k)_{\tilde{\l}}+ \Sigma_{{\scriptscriptstyle T}}^{{\scriptscriptstyle AA}}
(k)_{\tilde{\chi}} +\Sigma_{{\scriptscriptstyle T}}^{{\scriptscriptstyle AA}} 
(k)_{\scriptscriptstyle H} \nonumber \\
\label{eq:RFAA}
\displaystyle \Sigma^{{\scriptscriptstyle AA}}_{\scriptscriptstyle L} (k) &=& 
\Sigma_{{\scriptscriptstyle L}}^{{\scriptscriptstyle AA}}
 (k)_{\tilde{q}}+\Sigma_{{\scriptscriptstyle L}}^{{\scriptscriptstyle AA}} 
(k)_{\tilde{\l}}+ \Sigma_{{\scriptscriptstyle L}}^{{\scriptscriptstyle AA}}
(k)_{\tilde{\chi}} +\Sigma_{{\scriptscriptstyle L}}^{{\scriptscriptstyle AA}} 
(k)_{\scriptscriptstyle H}
\end{eqnarray}

$\bullet$ If \hspace*{0.5cm} $\tilde{m}_{t_{1}}^{2}, \tilde{m}_{t_{2}}^{2}, 
\tilde{m}_{b_{1}}^{2}, \tilde{m}_{b_{2}}^{2} \gg k^{2}$ :
\begin{eqnarray}
\label{eq:SumqAA}
\displaystyle \Sigma_{{\scriptscriptstyle T}}^{{\scriptscriptstyle AA}}(k)_{\tilde{q}} &=& 
-N_{c} \frac{e^{2}}{16 \pi^{2}} k^{2} \frac{1}{3} \sum_{\tilde{q}}
\left\{ \frac{4}{9} \left(2 \Delta_{\epsilon} -\log \frac{\tilde{m}_{t_{1}}^{2}}{\mu_{o}^{2}} 
-\log \frac{\tilde{m}_{t_{2}}^{2}}{\mu_{o}^{2}}\right) +\frac{1}{9} \left(2 \Delta_{\epsilon}  
-\log \frac{\tilde{m}_{b_{1}}^{2}}{\mu_{o}^{2}}-\log \frac{\tilde{m}_{b_{2}}^{2}}{\mu_{o}^{2}}
\right) \right\}, \nonumber \\
\end{eqnarray}
\hspace{0.7cm}$\bullet$ If \hspace*{0.5cm} $\tilde{m}_{\tau_{1}}^{2}, \tilde{m}_{\tau_{2}}^{2} \gg k^{2}$ :
\begin{eqnarray}
\displaystyle \Sigma_{{\scriptscriptstyle T}}^{{\scriptscriptstyle AA}}(k)_{\tilde{\l}}&=& 
-\frac{e^{2}}{16 \pi^{2}} k^{2} \frac{1}{3} 
\sum_{\tilde{l}} \left(2 \Delta_{{\scriptscriptstyle \epsilon}}  
-\log \frac{\tilde{m}_{\tau_{1}}^{2}}{\mu_{o}^{2}} 
-\log \frac{\tilde{m}_{\tau_{2}}^{2}}{\mu_{o}^{2}}\right), 
\end{eqnarray}
\hspace*{0.5cm}$\hspace*{0.3cm} \bullet$ If \hspace*{0.5cm} $\tilde{M}_{1}^+{}^{2}, \tilde{M}_{2}^+{}^{2} \gg k^{2}$ :
\begin{eqnarray}
\label{eq:SumcnAA}
\displaystyle \Sigma_{{\scriptscriptstyle T}}^{{\scriptscriptstyle AA}}(k)_{\tilde{\chi}}
&=& -\frac{e^{2}}{16 \pi^{2}} k^{2} 
\frac{4}{3} \left( 2 \Delta_{\epsilon}-\log \frac{\tilde{M}_{1}^{+^{2}}}{\mu_{o}^{2}}
-\log \frac{\tilde{M}_{2}^{+^{2}}}{\mu_{o}^{2}}\right), 
\end{eqnarray}
and for the longitudinal parts, we get:
\begin{equation}
\label{eq:LongAA}
\Sigma_{{\scriptscriptstyle L}}^{{\scriptscriptstyle AA}}(k)_{\tilde{q}}=
\Sigma_{{\scriptscriptstyle L}}^{{\scriptscriptstyle AA}}(k)_{\tilde{\l}}=
\Sigma_{{\scriptscriptstyle L}}^{{\scriptscriptstyle AA}}(k)_{\tilde{\chi}}=0
\end{equation}
\begin{eqnarray}
\label{eq:SumAZ}
\displaystyle \Sigma^{{\scriptscriptstyle AZ}}_{\scriptscriptstyle T} (k) &=& 
\Sigma_{{\scriptscriptstyle T}}^{{\scriptscriptstyle AZ}}
 (k)_{\tilde{q}}+\Sigma_{{\scriptscriptstyle T}}^{{\scriptscriptstyle AZ}} 
(k)_{\tilde{\l}}+ \Sigma_{{\scriptscriptstyle T}}^{{\scriptscriptstyle AZ}}
(k)_{\tilde{\chi}} +\Sigma_{{\scriptscriptstyle T}}^{{\scriptscriptstyle AZ}} 
(k)_{\scriptscriptstyle H} \nonumber \\
\label{eq:RFAZ}
\displaystyle \Sigma^{{\scriptscriptstyle AZ}}_{\scriptscriptstyle L} (k) &=& 
\Sigma_{{\scriptscriptstyle L}}^{{\scriptscriptstyle AZ}}
 (k)_{\tilde{q}}+\Sigma_{{\scriptscriptstyle L}}^{{\scriptscriptstyle AZ}} 
(k)_{\tilde{\l}}+ \Sigma_{{\scriptscriptstyle L}}^{{\scriptscriptstyle AZ}}
(k)_{\tilde{\chi}} +\Sigma_{{\scriptscriptstyle L}}^{{\scriptscriptstyle AZ}} 
(k)_{\scriptscriptstyle H}
\end{eqnarray}
\hspace*{0.5cm} $\bullet$ If \hspace*{0.5cm} $\tilde{m}_{t_{1}}^{2}, \tilde{m}_{t_{2}}^{2}, 
\tilde{m}_{b_{1}}^{2}, \tilde{m}_{b_{2}}^{2} \gg k^{2}$ :
\begin{eqnarray}
\label{eq:SumqAZ}
\displaystyle \Sigma_{{\scriptscriptstyle T}}^{{\scriptscriptstyle AZ}}
 (k)_{\tilde{q}}&=& -N_{c} \frac{e^{2}}{16 \pi^{2}} k^{2} \frac{1}
{3 s_{{\scriptscriptstyle W}} c_{{\scriptscriptstyle W}}} 
\displaystyle \sum_{\tilde{q}} \left\{
\frac{2}{3} \left(\frac{1}{2} c_{t}^{2} - \frac{2}{3} s_{{\scriptscriptstyle W}}^{2}\right)
\left(\Delta_{\epsilon} - \log \frac{\tilde{m}_{t_{1}}^{2}}{\mu_{o}^{2}}\right) 
\right. \nonumber \\
\displaystyle &+& \frac{2}{3} \left(\frac{1}{2} s_{t}^{2} - 
\frac{2}{3} s_{{\scriptscriptstyle W}}^{2}\right) \left(\Delta_{\epsilon} 
-\log \frac{\tilde{m}_{t_{2}}^{2}}{\mu_{o}^{2}}\right) - 
\frac{1}{3} \left(-\frac{1}{2} c_{b}^{2} + \frac{1}{3} s_{{\scriptscriptstyle W}}^{2}\right) 
\left(\Delta_{\epsilon} -\log \frac{\tilde{m}_{b_{1}}^{2}}{\mu_{o}^{2}}\right)\nonumber \\
\displaystyle &-& \left.\frac{1}{3} 
\left(-\frac{1}{2} s_{b}^{2} + \frac{1}{3} s_{{\scriptscriptstyle W}}^{2}\right) 
\left(\Delta_{\epsilon} -\log \frac{\tilde{m}_{b_{2}}^{2}}{\mu_{o}^{2}}\right) \right\},
\end{eqnarray}
\hspace*{0.5cm} $\bullet$ If \hspace*{0.5cm} $\tilde{m}_{\tau_{1}}^{2}, \tilde{m}_{\tau_{2}}^{2} \gg k^{2}$ :
\begin{eqnarray}
\label{eq:SumlAZ}
\displaystyle \Sigma_{{\scriptscriptstyle T}}^{{\scriptscriptstyle AZ}}
 (k)_{\tilde{\l}} &=& 
-\frac{e^{2}}{16 \pi^{2}} k^{2} \frac{1}{3 s_{{\scriptscriptstyle W}} 
c_{{\scriptscriptstyle W}}} \sum_{\tilde{l}} \left\{
\left(\frac{1}{2} c_{\tau}^{2} - s_{{\scriptscriptstyle W}}^{2}\right) 
\left(\Delta_{\epsilon} - \log \frac{\tilde{m}_{\tau_{1}}^{2}}{\mu_{o}^{2}}\right) + 
\left(\frac{1}{2} s_{\tau}^{2} - s_{{\scriptscriptstyle W}}^{2}\right) 
\left(\Delta_{\epsilon} -\log \frac{\tilde{m}_{\tau_{2}}^{2}}{\mu_{o}^{2}}\right) \right\}, \nonumber \\
\end{eqnarray}
\hspace{0.7cm}$\bullet$ If \hspace*{0.5cm} $\tilde{M}_{1}^+{}^{2}, \tilde{M}_{2}^+{}^{2} \gg k^{2}$ :
\begin{eqnarray}
\label{eq:SumcnAZ}
\displaystyle \Sigma_{{\scriptscriptstyle T}}^{{\scriptscriptstyle AZ}}
 (k)_{\tilde{\chi}} &=& 
\frac{e^{2}}{16 \pi^{2}}\frac{1}{s_{\scriptscriptstyle W} c_{\scriptscriptstyle W}} \frac{4}{3}k^{2}
\left\{ \left({\sw}^{2}-1\right)  \left(\Delta_{\epsilon}  
-\log \frac{\tilde{M}_{1}^+{}^{2}}{\mu_{o}^{2}}\right) 
+ \left({\sw}^{2}-\frac{1}{2}\right)  \left(\Delta_{\epsilon} 
-\log \frac{\tilde{M}_{2}^+{}^{2}}{\mu_{o}^{2}}\right) \right\}\,\,, \nonumber \\
\end{eqnarray}
and for the longitudinal parts, we get:
\begin{equation}
\label{eq:LongAZ}
\Sigma_{{\scriptscriptstyle L}}^{{\scriptscriptstyle AZ}}(k)_{\tilde{q}}=
\Sigma_{{\scriptscriptstyle L}}^{{\scriptscriptstyle AZ}}(k)_{\tilde{\l}}=
\Sigma_{{\scriptscriptstyle L}}^{{\scriptscriptstyle AZ}}(k)_{\tilde{\chi}}=0
\end{equation} 
\begin{eqnarray}
\label{eq:SumZZ}
\displaystyle \Sigma^{{\scriptscriptstyle ZZ}}_{\scriptscriptstyle T} (k) &=& 
\Sigma_{{\scriptscriptstyle T}}^{{\scriptscriptstyle ZZ}}
 (k)_{\tilde{q}}+\Sigma_{{\scriptscriptstyle T}}^{{\scriptscriptstyle ZZ}} 
(k)_{\tilde{\l}}+ \Sigma_{{\scriptscriptstyle T}}^{{\scriptscriptstyle ZZ}}
(k)_{\tilde{\chi}} +\Sigma_{{\scriptscriptstyle T}}^{{\scriptscriptstyle ZZ}} 
(k)_{\scriptscriptstyle H} \nonumber \\
\label{eq:RFZZ}
\displaystyle \Sigma^{{\scriptscriptstyle ZZ}}_{\scriptscriptstyle L} (k) &=& 
\Sigma_{{\scriptscriptstyle L}}^{{\scriptscriptstyle ZZ}}
 (k)_{\tilde{q}}+\Sigma_{{\scriptscriptstyle L}}^{{\scriptscriptstyle ZZ}} 
(k)_{\tilde{\l}}+ \Sigma_{{\scriptscriptstyle L}}^{{\scriptscriptstyle ZZ}}
(k)_{\tilde{\chi}} +\Sigma_{{\scriptscriptstyle L}}^{{\scriptscriptstyle ZZ}} 
(k)_{\scriptscriptstyle H}
\end{eqnarray}

$\bullet$ If \hspace*{0.5cm} $\tilde{m}_{t_{1}}^{2}, \tilde{m}_{t_{2}}^{2}, 
\tilde{m}_{b_{1}}^{2}, \tilde{m}_{b_{2}}^{2} \gg k^{2}; \hspace*{0.2cm} 
|\tilde{m}_{t_{1}}^{2} - \tilde{m}_{t_{2}}^{2}| \ll
|\tilde{m}_{t_{1}}^{2} + \tilde{m}_{t_{2}}^{2}|; \hspace*{0.2cm}
|\tilde{m}_{b_{1}}^{2} - \tilde{m}_{b_{2}}^{2}| \ll 
|\tilde{m}_{b_{1}}^{2} + \tilde{m}_{b_{2}}^{2}|$ :
\begin{eqnarray}
\label{eq:SumqZZ}
\displaystyle \Sigma_{{\scriptscriptstyle T}}^{{\scriptscriptstyle ZZ}}
 (k)_{\tilde{q}} &=& -N_{c} \frac{e^{2}}{16 \pi^{2}} \frac{1}
{s_{{\scriptscriptstyle W}}^{2} c_{{\scriptscriptstyle W}}^{2}} \sum_{\tilde{q}} \left\{ \left[
-\frac{1}{2} c_{t}^{2} s_{t}^{2} h({\tilde{m}}_{t_{1}}^{2}, {\tilde{m}}_{t_{2}}^{2}) 
- \frac{1}{2} c_{b}^{2} s_{b}^{2} h({\tilde{m}}_{b_{1}}^{2}, 
{\tilde{m}}_{b_{2}}^{2}) \right. \right] \nonumber \\
\displaystyle &+& \frac{1}{3} k^{2} \left[
\left(\frac{c_{t}^{2}}{2} - \frac{2 s_{{\scriptscriptstyle W}}^{2}}{3}\right)^{2} 
\left(\Delta_\epsilon - \log \frac{\tilde{m}_{t_{1}}^{2}}{\mu_{o}^{2}}\right) \right. 
+\left(\frac{s_{t}^{2}}{2} - \frac{2 s_{{\scriptscriptstyle W}}^{2}}{3}\right)^{2} 
\left(\Delta_\epsilon - \log \frac{\tilde{m}_{t_{2}}^{2}}{\mu_{o}^{2}}\right) \nonumber \\
 &+& \displaystyle
\left(-\frac{c_{b}^{2}}{2} + \frac{s_{{\scriptscriptstyle W}}^{2}}{3}\right)^{2}
\left(\Delta_\epsilon - \log \frac{\tilde{m}_{b_{1}}^{2}}{\mu_{o}^{2}}\right) + 
\left(-\frac{s_{b}^{2}}{2} + \frac{s_{{\scriptscriptstyle W}}^{2}}{3}\right)^{2}
\left(\Delta_\epsilon - \log \frac{\tilde{m}_{b_{2}}^{2}}{\mu_{o}^{2}}\right) \nonumber \\
 &+& \displaystyle \left. \left.\frac{1}{2} s_{t}^{2} c_{t}^{2} 
\left(\Delta_\epsilon - \log \frac{\tilde{m}_{t_{1}}^{2} + \tilde{m}_{t_{2}}^{2}}
{2 \mu_{o}^{2}}\right) + \frac{1}{2} s_{b}^{2} c_{b}^{2} 
\left(\Delta_\epsilon - \log \frac{\tilde{m}_{b_{1}}^{2} + \tilde{m}_{b_{2}}^{2}}{2 
\mu_{o}^{2}}\right) \right] \right\}, \\
\nonumber \\
\label{eq:RFqZZ}
\nonumber \\
\displaystyle \Sigma_{{\scriptscriptstyle L}}^{{\scriptscriptstyle ZZ}}
 (k)_{\tilde{q}}  &=& N_{c} \frac{e^{2}}{16 \pi^{2}} \frac{1}
{s_{{\scriptscriptstyle W}}^{2} c_{{\scriptscriptstyle W}}^{2}} \frac{1}{2} 
\sum_{\tilde{q}} \left\{ 
c_{t}^{2} s_{t}^{2} h({\tilde{m}}_{t_{1}}^{2}, {\tilde{m}}_{t_{2}}^{2}) 
+  c_{b}^{2} s_{b}^{2} h({\tilde{m}}_{b_{1}}^{2}, {\tilde{m}}_{b_{2}}^{2}) \right\},
\end{eqnarray}

$\bullet$ If \hspace*{0.4cm} $\tilde{m}_{\nu}^{2}, \tilde{m}_{\tau_{1}}^{2}, 
\tilde{m}_{\tau_{2}}^{2} \gg k^{2}; \hspace*{0.4cm}
|\tilde{m}_{\tau_{1}}^{2} - \tilde{m}_{\tau_{2}}^{2}| \ll
|\tilde{m}_{\tau_{1}}^{2} + \tilde{m}_{\tau_{2}}^{2}|$:
\begin{eqnarray}
\label{eq:SumlZZ}
\displaystyle \Sigma_{{\scriptscriptstyle T}}^{{\scriptscriptstyle ZZ}}
 (k)_{\tilde{\l}} &=& -\frac{e^{2}}{16 \pi^{2}} \frac{1}{s_{{\scriptscriptstyle W}}^{2} 
 c_{{\scriptscriptstyle W}}^{2}} \sum_{\tilde{\l}} \left\{ 
-\frac{1}{2} c_{\tau}^{2} s_{\tau}^{2} h({\tilde{m}}_{\tau_{1}}^{2}, 
{\tilde{m}}_{\tau_{2}}^{2}) \right. +\frac{1}{3} k^{2} \left[ \frac{1}{4}
\left(\Delta_\epsilon - \log \frac{\tilde{m}_{\nu}^{2}}{\mu_{o}^{2}}\right) \right. \nonumber \\
\displaystyle &+& \left. \left(\frac{-c_{\tau}^{2}}{2} + s_{{\scriptscriptstyle W}}^{2}\right)^{2} 
\left(\Delta_\epsilon - \log \frac{\tilde{m}_{\tau_{1}}^{2}}{\mu_{o}^{2}}\right) + 
\left(-\frac{s_{\tau}^{2}}{2} + s_{{\scriptscriptstyle W}}^{2}\right)^{2} 
\left(\Delta_\epsilon - \log \frac{\tilde{m}_{\tau_{2}}^{2}}{\mu_{o}^{2}}\right) 
\right. \nonumber \\ 
\displaystyle &+& \left. \left. \frac{1}{2} s_{\tau}^{2} c_{\tau}^{2} 
\left(\Delta_\epsilon - \log \frac{\tilde{m}_{\tau_{1}}^{2} + 
\tilde{m}_{\tau_{2}}^{2}}{2 \mu_{o}^{2}}\right) \right] \right\}, \\
\nonumber \\
\displaystyle \Sigma_{{\scriptscriptstyle L}}^{{\scriptscriptstyle ZZ}}
 (k)_{\tilde{\l}} &=& \frac{e^{2}}{16 \pi^{2}} \frac{1}
{s_{{\scriptscriptstyle W}}^{2} c_{{\scriptscriptstyle W}}^{2}}  \frac{1}{2} 
\sum_{\tilde{\l}} \left\{ c_{\tau}^{2} s_{\tau}^{2} h({\tilde{m}}_{\tau_{1}}^{2}, 
{\tilde{m}}_{\tau_{2}}^{2}) \right\},
\end{eqnarray}

$\bullet$ If \hspace*{0.3cm} $ \tilde{M}_{1,2}^+{}^{2}, 
\tilde{M}_{j}^o{}^{2} \gg k^{2}, \hspace*{0.1cm} 
|\tilde{M}_{1}^+{}^{2} - \tilde{M}_{2}^+{}^{2}| \ll 
|\tilde{M}_{1}^+{}^{2} + \tilde{M}_{2}^+{}^{2}|$  
and $|\tilde{M}_{i}^o{}^{2} - \tilde{M}_{j}^o{}^{2}| \ll 
|\tilde{M}_{i}^o{}^{2} + \tilde{M}_{j}^o{}^{2}|$\\
\hspace*{1cm} $i,j=1,2,3,4$:
\begin{eqnarray}
\label{eq:SumcnZZ}
\displaystyle \Sigma_{{\scriptscriptstyle T}}^{{\scriptscriptstyle ZZ}}(k)_{\tilde{\chi}} &=& 
-\frac{e^{2}}{16 \pi^{2}} \frac{1}
{s_{{\scriptscriptstyle W}}^{2} c_{{\scriptscriptstyle W}}^{2}} \left\{
-\frac{1}{2} {(\tilde{M}_{3}^{o} - \tilde{M}_{4}^{o})}^{2}
\left(\Delta_{\epsilon} - \log \frac{\tilde{M}_{3}^o{}^{2} 
+ \tilde{M}_{4}^o{}^{2}}{2 \mu_{o}^{2}}\right) \right. \nonumber \\
\displaystyle &+& \frac{1}{3} k^{2} \left[ 4 \left(\sw^{2}-1\right)^{2}  
\left(\Delta_{\epsilon} - \log \frac{\tilde{M}_{1}^+{}^{2}}
{\mu_{o}^{2}}\right) + 4\left(\sw^{2}-\frac{1}{2}\right)^{2} 
\left(\Delta_{\epsilon} - \log \frac{\tilde{M}_{2}^+{}^{2}}{\mu_{o}^{2}}\right) 
\right. \nonumber \\
\displaystyle &+& \left. \left.\left({\Delta}_{\epsilon} - 
\log \frac{\tilde{M}_{3}^o{}^{2}+\tilde{M}_{4}^o{}^{2}}{2 \mu_{o}^{2}}\right)
\right] \right\}, \\
\nonumber \\
\label{eq:RFcnZZ}
\displaystyle \Sigma_{{\scriptscriptstyle L}}^{{\scriptscriptstyle ZZ}}(k)_{\tilde{\chi}} 
&=& \frac{e^{2}}{16 \pi^{2}} \frac{1}
{2 s_{{\scriptscriptstyle W}}^{2} c_{{\scriptscriptstyle W}}^{2}} \left\{
{(\tilde{M}_{3}^{o} - \tilde{M}_{4}^{o})}^{2}
\left(\Delta_{\epsilon} - \log \frac{\tilde{M}_{3}^o{}^{2} 
+ \tilde{M}_{4}^o{}^{2}}{2 \mu_{o}^{2}}\right) \right\},
\end{eqnarray}
where $h({m}_{1}^{2}, {m}_{2}^{2})$ of eqs.~(\ref{eq:SumqZZ}, \ref{eq:SumlZZ}) has been
defined in eq.~(\ref{eq:h}).
\begin{eqnarray}
\label{eq:SumWW}
\displaystyle \Sigma^{{\scriptscriptstyle WW}}_{\scriptscriptstyle T} (k) &=& 
\Sigma_{{\scriptscriptstyle T}}^{{\scriptscriptstyle WW}}
 (k)_{\tilde{q}}+\Sigma_{{\scriptscriptstyle T}}^{{\scriptscriptstyle WW}} 
(k)_{\tilde{\l}}+ \Sigma_{{\scriptscriptstyle T}}^{{\scriptscriptstyle WW}}
(k)_{\tilde{\chi}} +\Sigma_{{\scriptscriptstyle T}}^{{\scriptscriptstyle WW}} 
(k)_{\scriptscriptstyle H} \nonumber \\
\label{eq:RFWW}
\displaystyle \Sigma^{{\scriptscriptstyle WW}}_{\scriptscriptstyle L} (k) &=& 
\Sigma_{{\scriptscriptstyle L}}^{{\scriptscriptstyle WW}}
 (k)_{\tilde{q}}+\Sigma_{{\scriptscriptstyle L}}^{{\scriptscriptstyle WW}} 
(k)_{\tilde{\l}}+ \Sigma_{{\scriptscriptstyle L}}^{{\scriptscriptstyle WW}}
(k)_{\tilde{\chi}} +\Sigma_{{\scriptscriptstyle L}}^{{\scriptscriptstyle WW}} 
(k)_{\scriptscriptstyle H}
\end{eqnarray}

$\bullet$ If \hspace*{0.4cm} $\tilde{m}_{t_{1}}^{2}, \tilde{m}_{t_{2}}^{2}, 
\tilde{m}_{b_{1}}^{2}, \tilde{m}_{b_{2}}^{2} \gg k^{2}; \hspace*{0.3cm}
|\tilde{m}_{t_{i}}^{2} - \tilde{m}_{b_{j}}^{2}| \ll
|\tilde{m}_{t_{i}}^{2} + \tilde{m}_{b_{j}}^{2}|$ \hspace*{0.4cm} 
$i,j=1,2$:
\begin{eqnarray}
\label{eq:SumqWW}
\displaystyle \Sigma_{{\scriptscriptstyle T}}^{{\scriptscriptstyle WW}}(k)_{\tilde{q}}&=& 
-N_{c} \frac{e^{2}}{16 \pi^{2}} \frac{1}{s_{{\scriptscriptstyle W}}^{2}} 
\sum_{\tilde{q}} \left\{ \left[ 
-\frac{1}{2} c_{t}^{2} c_{b}^{2} h (\tilde{m}_{t_{1}}^{2}, 
\tilde{m}_{b_{1}}^{2}) -\frac{1}{2} c_{t}^{2} s_{b}^{2} 
h ({\tilde{m}}_{t_{1}}^{2}, {\tilde{m}}_{b_{2}}^{2}) \right.\right. \nonumber \\
&-& \displaystyle \frac{1}{2} s_{t}^{2} c_{b}^{2} h (\tilde{m}_{t_{2}}^{2}, 
\tilde{m}_{b_{1}}^{2}) \left. -\frac{1}{2} s_{t}^{2} s_{b}^{2} 
h (\tilde{m}_{t_{2}}^{2}, \tilde{m}_{b_{2}}^{2}) \right]
+\frac{1}{6} k^{2} \left[ c_{t}^{2} c_{b}^{2} \left(\Delta_{\epsilon} - 
\log \frac{\tilde{m}_{t_{1}}^{2} + \tilde{m}_{b_{1}}^{2}}{2 \mu_{o}^{2}}\right) 
\right. \nonumber \\
\displaystyle &+& c_{t}^{2} s_{b}^{2} 
\left(\Delta_{\epsilon} - \log \frac{\tilde{m}_{t_{1}}^{2} + \tilde{m}_{b_{2}}^{2}}
{2 \mu_{o}^{2}}\right) + s_{t}^{2} c_{b}^{2} 
\left(\Delta_{\epsilon} - \log \frac{\tilde{m}_{t_{2}}^{2} + \tilde{m}_{b_{1}}^{2}}
{2 \mu_{o}^{2}}\right) \nonumber \\
&+& \displaystyle \left. \left.  s_{t}^{2} s_{b}^{2} 
\left(\Delta_{\epsilon} - \log \frac{\tilde{m}_{t_{2}}^{2} + \tilde{m}_{b_{2}}^{2}}
{2 \mu_{o}^{2}}\right)\right] \right\}, \\
\nonumber \\
\label{eq:RFqWW}
\displaystyle \Sigma_{{\scriptscriptstyle L}}^{{\scriptscriptstyle WW}}(k)_{\tilde{q}}&=& 
N_{c} \frac{e^{2}}{16 \pi^{2}} \frac{1}{s_{{\scriptscriptstyle W}}^{2}} \frac{1}{2}
\sum_{\tilde{q}} \left\{ c_{t}^{2} c_{b}^{2} h (\tilde{m}_{t_{1}}^{2}, 
\tilde{m}_{b_{1}}^{2}) +c_{t}^{2} s_{b}^{2} 
h ({\tilde{m}}_{t_{1}}^{2}, {\tilde{m}}_{b_{2}}^{2})\right. \nonumber \\
\displaystyle &+& \left. s_{t}^{2} c_{b}^{2} h (\tilde{m}_{t_{2}}^{2}, 
\tilde{m}_{b_{1}}^{2}) + s_{t}^{2} s_{b}^{2} 
h (\tilde{m}_{t_{2}}^{2}, \tilde{m}_{b_{2}}^{2}) \right\},
\end{eqnarray}

$\bullet$ If \hspace*{0.5cm} $\tilde{m}_{\nu}^{2}, \tilde{m}_{\tau_{1}}^{2}, 
\tilde{m}_{\tau_{2}}^{2} \gg k^{2}, \hspace*{0.4cm}
|\tilde{m}_{\nu}^{2} - \tilde{m}_{\tau_{i}}^{2}| \ll 
|\tilde{m}_{\nu}^{2} + \tilde{m}_{\tau_{i}}^{2}|$ \hspace*{0.5cm} 
$i=1,2$:
\begin{eqnarray}
\label{eq:SumlWW}
\displaystyle \Sigma_{{\scriptscriptstyle T}}^{{\scriptscriptstyle WW}}(k)_{\tilde{\l}} &=& 
-\frac{e^{2}}{16 \pi^{2}} \frac{1}{s_{{\scriptscriptstyle W}}^{2}}  \sum_{\tilde{\l}} \left\{ \left[
-\frac{1}{2} c_{\tau}^{2} h (\tilde{m}_{\nu}^{2}, \tilde{m}_{\tau_{1}}^{2}) 
\right. -\frac{1}{2} s_{\tau}^{2} h (\tilde{m}_{\nu}^{2}, \tilde{m}_{\tau_{2}}^{2}) \right] \nonumber \\
\displaystyle &+& \frac{1}{6} k^{2} \left. \left[ c_{\tau}^{2} \left(\Delta_{\epsilon} - 
\log \frac{\tilde{m}_{\nu}^{2} + \tilde{m}_{\tau_{1}}^{2}}{2 \mu_{o}^{2}}\right) + 
 s_{\tau}^{2} \left(\Delta_{\epsilon} -
\log \frac{\tilde{m}_{\nu}^{2} + \tilde{m}_{\tau_{2}}^{2}}{2 \mu_{o}^{2}}\right) \right] \right\}, \\
\nonumber \\
\label{eq:RFlWW}
\nonumber \\
\displaystyle \Sigma_{{\scriptscriptstyle L}}^{{\scriptscriptstyle WW}}(k)_{\tilde{\l}} &=& 
\frac{e^{2}}{16 \pi^{2}} \frac{1}{s_{{\scriptscriptstyle W}}^{2}} \frac{1}{2} \sum_{\tilde{\l}} 
\left\{ c_{\tau}^{2} h (\tilde{m}_{\nu}^{2}, \tilde{m}_{\tau_{1}}^{2}) 
+s_{\tau}^{2} h (\tilde{m}_{\nu}^{2}, \tilde{m}_{\tau_{2}}^{2})  \right\},
\end{eqnarray}

$\bullet$ If \hspace*{0.5cm} $ \tilde{M}_{i}^o{}^{2}, 
\tilde{M}_{j}^+{}^{2} \gg k^{2}, \hspace*{0.2cm} 
|\tilde{M}_{i}^o{}^{2} - \tilde{M}_{j}^+{}^{2}| \ll 
|\tilde{M}_{i}^o{}^{2} + \tilde{M}_{j}^+{}^{2}|$ ,\hspace*{0.4cm} 
$i=1,2,3,4;$ \hspace*{0.4cm} $j=1,2$:
\begin{eqnarray}
\label{eq:SumcnWW}
\displaystyle \Sigma_{{\scriptscriptstyle T}}^{{\scriptscriptstyle WW}}(k)_{\tilde{\chi}} &=& 
-\frac{e^{2}}{16 \pi^{2}}  \frac{1}{s_{{\scriptscriptstyle W}}^{2}} 
\left\{ - 2 (\tilde{M}_{1}^{+} - \tilde{M}_{2}^{o})^{2} 
\left(\Delta_{\epsilon} - \log \frac{\tilde{M}_{1}^+{}^{2} + 
\tilde{M}_{2}^o{}^{2}}{2 \mu_{o}^{2}}\right) \right. \nonumber \\
\displaystyle &-& \frac{1}{2} (\tilde{M}_{2}^{+} - \tilde{M}_{3}^{o})^{2} 
\left(\Delta_{\epsilon} - \log \frac{\tilde{M}_{2}^+{}^{2} + 
\tilde{M}_{3}^o{}^{2}}{2 \mu_{o}^{2}}\right)-\frac{1}{2} (\tilde{M}_{2}^{+}-\tilde{M}_{4}^{o})^{2} 
\left(\Delta_{\epsilon} - \log \frac{\tilde{M}_{2}^+{}^{2} + 
\tilde{M}_{4}^o{}^{2}}{2 \mu_{o}^{2}}\right) \nonumber \\
\displaystyle &+& k^{2} \left[ \frac{4}{3}\left(\Delta_{\epsilon} - \log \frac{\tilde{M}_{1}^+{}^{2} + 
\tilde{M}_{2}^o{}^{2}}{2 \mu_{o}^{2}}\right)
+\frac{1}{3} \left(\Delta_{\epsilon} - \log \frac{\tilde{M}_{2}^+{}^{2} + 
\tilde{M}_{3}^o{}^{2}}{2 \mu_{o}^{2}}\right) \right. \nonumber \\
\displaystyle &+& \left. \left.
\frac{1}{3} \left(\Delta_{\epsilon} - \log \frac{\tilde{M}_{2}^+{}^{2} + 
\tilde{M}_{4}^o{}^{2}}{2 \mu_{o}^{2}}\right) \right] \right\}, \\
\nonumber \\
\label{eq:RFcnWW}
\nonumber \\
\displaystyle \Sigma_{{\scriptscriptstyle L}}^{{\scriptscriptstyle WW}}(k)_{\tilde{\chi}}  &=& 
\frac{e^{2}}{16 \pi^{2}}  \frac{1}{s_{{\scriptscriptstyle W}}^{2}} 
\left\{  2 (\tilde{M}_{1}^{+} - \tilde{M}_{2}^{o})^{2} 
\left(\Delta_{\epsilon} - \log \frac{\tilde{M}_{1}^+{}^{2} + 
\tilde{M}_{2}^o{}^{2}}{2 \mu_{o}^{2}}\right) \right. \nonumber \\
\displaystyle &+& \left. \frac{1}{2} (\tilde{M}_{2}^{+} - \tilde{M}_{3}^{o})^{2} 
\left(\Delta_{\epsilon} - \log \frac{\tilde{M}_{2}^+{}^{2} + 
\tilde{M}_{3}^o{}^{2}}{2 \mu_{o}^{2}}\right)+\frac{1}{2} (\tilde{M}_{2}^{+}-\tilde{M}_{4}^{o})^{2} 
\left(\Delta_{\epsilon} - \log \frac{\tilde{M}_{2}^+{}^{2} + 
\tilde{M}_{4}^o{}^{2}}{2 \mu_{o}^{2}}\right)\right\}, \nonumber \\
\end{eqnarray}

We also include here the large masses limit of the frequently 
appearing function $h({m}_{1}^{2}, {m}_{2}^{2})$:

$\bullet$ If \hspace*{0.5cm} $|{m}_{1}^{2} - {m}_{2}^{2}| \ll 
|{m}_{1}^{2} + {m}_{2}^{2}|$ ,\hspace*{0.4cm} 
${m}_{1}^{2}; {m}_{2}^{2}$ large,
\begin{equation}
h({m}_{1}^{2}, {m}_{2}^{2}) \longrightarrow
\frac{{m}_{1}^{2} - {m}_{2}^{2}}{2} \left[
\frac{({m}_{1}^{2} - {m}_{2}^{2})}{({m}_{1}^{2} + {m}_{2}^{2})} +
O{\left(\frac{{m}_{1}^{2} - {m}_{2}^{2}}{{m}_{1}^{2} + {m}_{2}^{2}}\right)}^{2}
\right]
\end{equation}


\newpage
{\Large {\bf Figure Captions}}\\
\begin{itemize}
\item[Fig.{\bf 1a}] Feynman diagrams with scalars in the loops that contribute to the 
two-point functions of the electroweak gauge bosons. The diagrams shown for neutral gauge
bosons are of stop loops. Not shown are the similar diagrams of sbottom loops. For the 
sleptons sector, $\nu, \tauu$ and $\taud$, analogous diagrams are obtained.
\item[Fig.{\bf 1b}] Feynman diagrams with charginos and neutralinos in the loops that 
dominate the two-point functions of electroweak gauge bosons in the asymptotic limit of
very large ino masses.
\end{itemize}
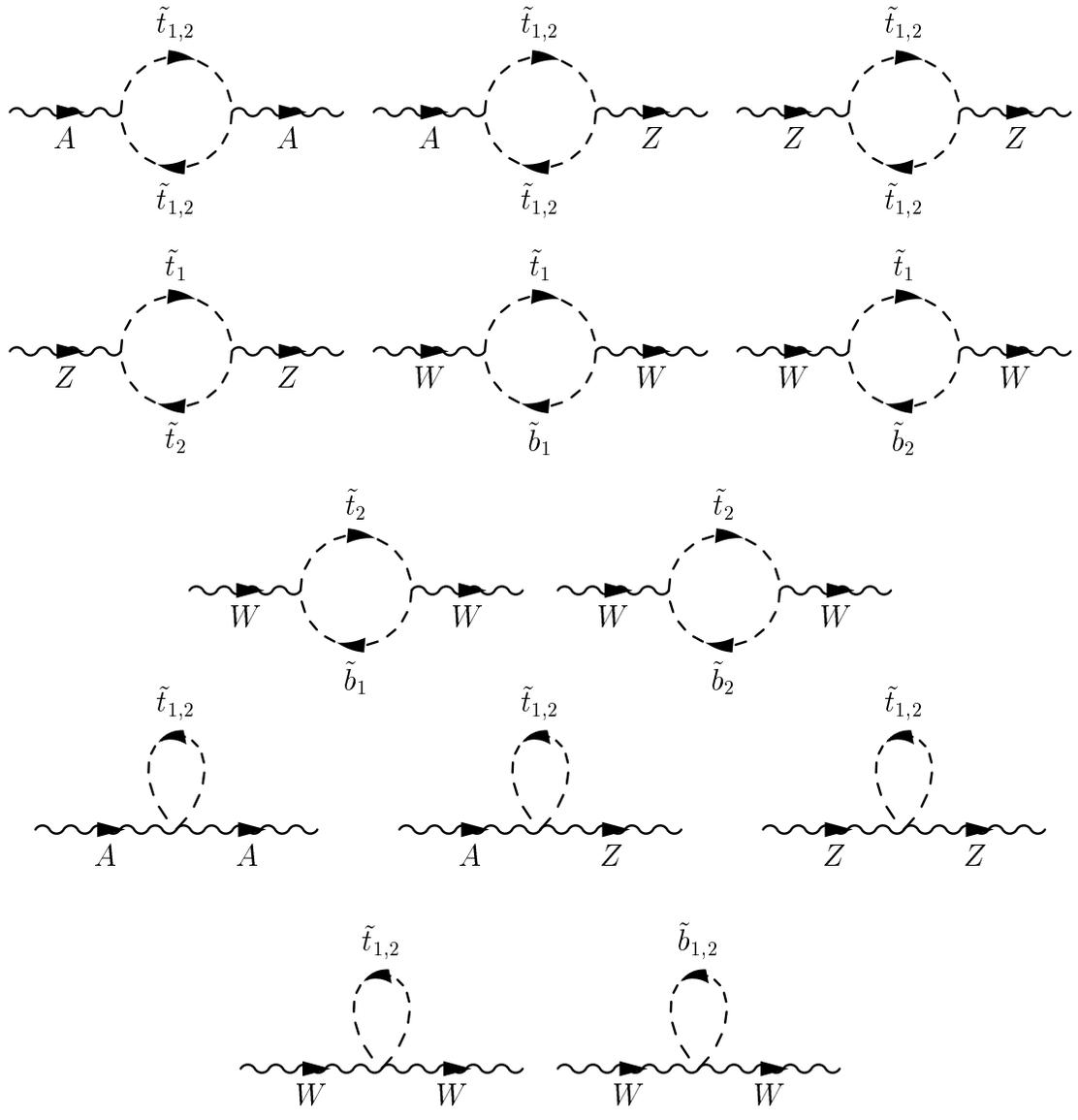
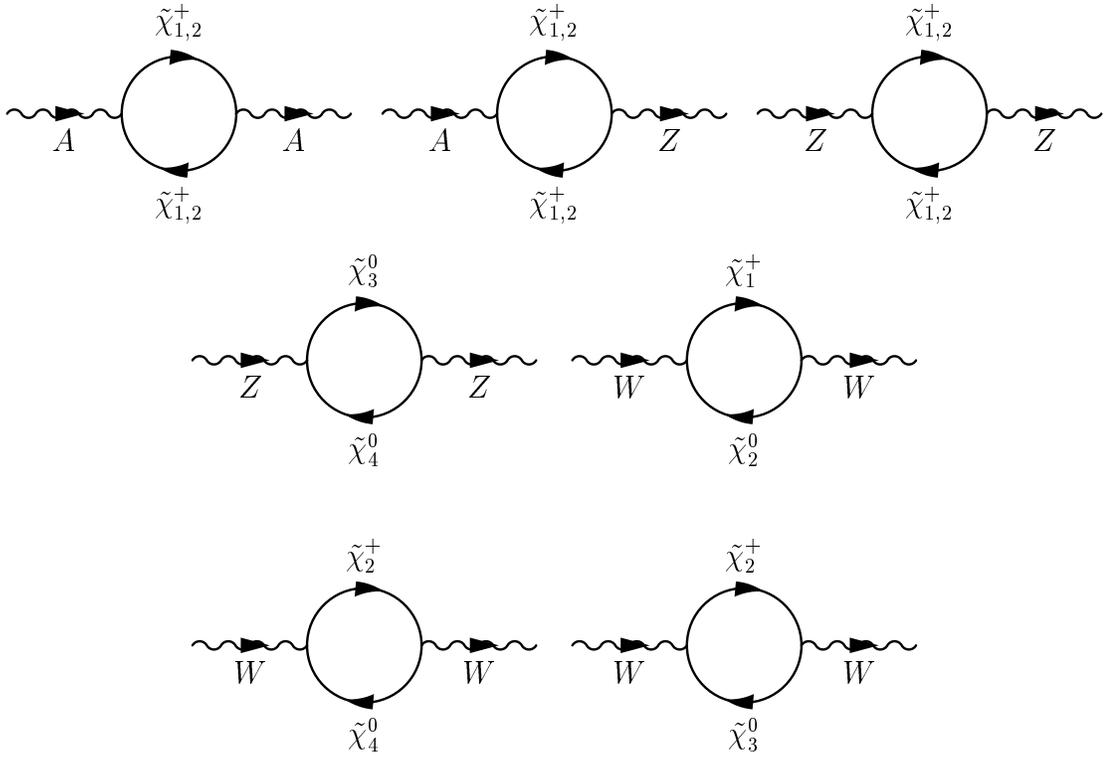
\end{document}